\documentclass[manuscript]{aastex}

\usepackage{lscape}
\usepackage{pdflscape}
\usepackage{epsfig}
\usepackage{epstopdf}

\newcommand{\lapprox }{{\lower0.8ex\hbox{$\buildrel <\over\sim$}}}
\newcommand{\gapprox }{{\lower0.8ex\hbox{$\buildrel >\over\sim$}}}

\newcommand{\ha}{H$_\alpha$ }
\newcommand{\vsini}{$v$ sin $i$}
\newcommand{\nai}{\ion{Na}{1} }
\newcommand{\cai}{\ion{Ca}{1} }
\newcommand{\feh}{[Fe/H]}

\newcommand{\noprint}[1]{}
\newcommand{\figsetstart}{{\bf Fig. Set} }
\newcommand{\figsetend}{}
\newcommand{\figsetgrpstart}{}
\newcommand{\figsetgrpend}{}
\newcommand{\figsetnum}[1]{{\bf #1.}}
\newcommand{\figsettitle}[1]{ {\bf #1} }
\newcommand{\figsetgrpnum}[1]{\noprint{#1}}
\newcommand{\figsetgrptitle}[1]{\noprint{#1}}
\newcommand{\figsetplot}[1]{\noprint{#1}}
\newcommand{\figsetgrpnote}[1]{\noprint{#1}}

\slugcomment{DRAFT \today}
\shorttitle{Metallicity and Temperature Indicators in M dwarf K band Spectra}
\shortauthors{Rojas-Ayala et al.}

\begin{document}

\title{Metallicity and Temperature Indicators in M dwarf K band Spectra: Testing New \& Updated Calibrations With Observations of 133 Solar Neighborhood M dwarfs}

\author{B{\'a}rbara~Rojas-Ayala\altaffilmark{1,2,3}, Kevin~R.~Covey\altaffilmark{2,4,5}, Philip~S.~Muirhead\altaffilmark{6}, James~P.~Lloyd\altaffilmark{2}}

\altaffiltext{1}{Department of Astrophysics, American Museum of Natural History, Central Park West at 79th Street, New York, NY 10024, USA; babs@amnh.org}
\altaffiltext{2}{Department of Astronomy, Cornell University, 122 Sciences Drive, Ithaca, NY 14853, USA}
\altaffiltext{3}{Z. Carter Patten '25 Fellow}
\altaffiltext{4}{Hubble Fellow}
\altaffiltext{5}{Visiting Researcher, Department of Astronomy, Boston University, 725 Commonwealth Ave, Boston, MA 02215, USA.}
\altaffiltext{6}{Current Address: Department of Astronomy, California Institute of Technology, 1200 East California Boulevard, MC 249-17, Pasadena, CA 91125, USA.}

\begin{abstract}
We present K band spectra for 133 nearby (d $<$ 33 parsecs) M dwarfs, including 18 M dwarfs with reliable metallicity estimates (as inferred from an FGK type companion), 11 M dwarf planet hosts, more than 2/3 of the M dwarfs in the Northern 8 pc sample, and several M dwarfs from the LSPM catalog. From these spectra, we measure equivalent widths of the Ca and Na lines, and a spectral index quantifying the absorption due to H$_2$O opacity (the H$_2$O-K2 index).  Using empirical spectral types standards and synthetic models, we calibrate the H$_2$O-K2 index as an indicator of an M dwarf's spectral type and effective temperature.  We also present a revised relationship that estimates the [Fe/H] and [M/H] metallicities of M dwarfs from their Na I, Ca I, and H$_2$O-K2 measurements. Comparisons to model atmosphere provide a qualitative validation of our approach, but also reveal an overall offset between the atomic line strengths predicted by models as compared to actual observations. Our metallicity estimates also reproduce expected correlations with Galactic space motions and H$\alpha$ emission line strengths, and return statistically identical metallicities for M dwarfs within a common multiple system. Finally, we find systematic residuals between our H$_2$O-based spectral types and those derived from optical spectral features with previously known sensitivity to stellar metallicity, such as TiO, and identify the CaH1 index as a promising optical index for diagnosing the metallicities of near-solar M dwarfs.

\end{abstract}


\section{Introduction}

Contrary to their dimness, M dwarf stars hold significant promise for illuminating the processes that govern the formation and evolution of stars, planets, and the Milky Way.  M dwarfs possess masses of 0.6 M$_{\odot} >$ M$_* >$ 0.08 M$_{\odot}$ \citep{Delfosse2000}, straddling the peak of the stellar initial mass function and dominating stellar populations by number \citep{Bastian2010}.  With main sequence lifetimes that exceed a Hubble time, Galactic M dwarfs also represent a complete archeological record of the chemical evolution and star formation history of the Milky Way \citep[e.g.][]{2007AJ....134.2418B}.  Moreover, M dwarfs are of great interest as potential exoplanet host stars, as all of a planet's observable signals will be significantly easier to detect if it orbits an M dwarf compared to a similar planet in orbit around a G dwarf \citep[e.g., ][]{2008PASP..120..317N}.  

Calibrating the fundamental parameters of M dwarfs, however, is a difficult challenge from both an observational and theoretical perspective.  Empirical measurements of M dwarf masses, luminosities, temperatures, and radii are anchored primarily by the information extracted from the orbits of M dwarf binaries, either the very rare eclipsing systems \citep[$\sim$15 systems known; e.g., ][]{Kraus2011,2011ApJ...742..123I} or less rare (but also somewhat less informative) spectroscopic/astrometric systems \citep[e.g., ][]{Shkolnik2010,Martinache2007}.  Overcoming the intrinsic faintness of these low-luminosity systems, however, makes their characterization a taxing observational challenge, and analyses of known binaries reveal systematic offsets in inferred temperatures and radii that correlate with both orbital period and tracers of magnetic activity \citep{Ribas2006,Lopez-Morales2007}, suggesting that more wide binaries are needed to accurately infer the parameters typical isolated M dwarfs \citep{Kraus2011}.  Theoretical constraints on M dwarf parameters have been similarly difficult to achieve: accurately modeling the deep convective zones in M dwarf interiors \citep{Mullan2001, 2008ApJ...676.1262B} and the mixture of molecules and grains that dominate M dwarf atmospheres \citep{Tsuji1996,Allard2000} requires significant computational resources, as well as extensive databases of oscillator strengths and opacities measured from laboratory experiments.  Confronting theoretical predictions with empirical measurements have also identified significant offsets even for isolated field stars: empirical effective temperature scales appear $\sim$200-300 K cooler than would be inferred from theoretical model atmospheres \citep[e.g.][]{2005MNRAS.358..105J}, for example.

While M dwarf masses, temperatures, and radii have proven challenging to calibrate, those parameters appear strikingly pedestrian when contrasted with the difficulty of inferring M dwarf metallicities.  \citet{1976A&A....48..443M} was the first to systematically assess the strengths of M dwarf spectral features as a function of metallicity, constructing a grid of synthetic model atmospheres that included molecular opacities and spanned temperatures of 4250--3000K and metallicities of \feh = -2.0 -- 0.0 dex.  With these models, Mould established several foundations of subsequent investigations of M dwarf metallicities, such as the metallicity sensitivity of TiO absorption, the gravity sensitivity of CaH absorption, and the rough metallicities of Galactic M subdwarfs \citep[ \feh $\sim$ -1.0 dex; ][]{Mould1976b}.  These studies enabled the subsequent photometric and spectroscopic identification of M subdwarfs \citep{Stauffer1986,Ruiz1993}, the development of increasingly detailed models of M dwarf atmospheres \citep{Allard1990}, and represent the fundamental origin of the TiO and CaH-based indices used today to identify metal-poor M subdwarfs \citep{1997AJ....113..806G, 2005MNRAS.356..963W, 2007ApJ...669.1235L}.  

While techniques to identify metal-poor M dwarfs have been developing for more than 35 years, methods for identifying metal-$rich$ M dwarfs have received significantly less attention.  Interest in identifying M dwarfs with super-solar metallicities was only recently spurred by the realization that such stars could be highly promising targets for searches for Earth-like planets.  The existence of a correlation between (gas giant) planet frequency and host-star metallicity is well established for FGK-dwarfs \citep{1997MNRAS.285..403G, 2004A&A...415.1153S, 2005ApJS..159..141V}, but the first investigations to examine if the planet-metallicity correlation extended to M dwarf hosts returned mixed results \citep{2005A&A...442..635B,2006ApJ...652.1604B}.  Bonfils et al. (2005, B05 hereafter) analyzed V and K photometry for 20 wide M dwarf companions to FGK-dwarfs whose metallicities could be determined reliably via standard spectroscopic techniques. Assuming that both binary components inherit the same metallicity from their parent molecular cloud material, B05 assigned the metallicity measurements for each primary to the M dwarf secondary, and derived iso-metallicity contours in the M$_V$ vs. V-K$_s$ color-magnitude plane.  This photometric metallicity calibration suggested that nearby M dwarfs, including the planet hosts Gl 876 and Gl 436, were slightly metal-poor compared to the mean metallicity of the Galactic disk.  This conclusion was reinforced by the work of \citet{2006ApJ...652.1604B}, whose spectroscopic analysis indicated sub-solar metallicities for Gl 436, Gl 581, and Gl 876.

The persistence of the planet-metallicity correlation into the M dwarf regime has been suggested, however, by the recent re-calibration of the photometric metallicity calibration by \citet{2009ApJ...699..933J}.  Using six M dwarfs with wide, metal-rich FGK companions to update the M$_V$ vs. V-K metallicity contours, Johnson \& Apps (2009, JA09 hereafter) found that the B05 calibration systematically underestimated the compositions of these metal-rich stars.  JA09 used their revised calibration to estimate the metallicities of six M dwarfs with planetary mass companions, concluding that M dwarf planet hosts are indeed preferentially metal rich, just like FGK hosts.  This conclusion was supported by the subsequent work of \citet{2010A&A...519A.105S}, who updated the B05 and JA09 calibrations by using theoretical models to inform the functional form of the M$_K$ vs. V-K relation, and by inferring the mean metallicity of M dwarfs in the Solar Neighborhood from a volume-limited and kinematically-matched sample of F \& G dwarfs in the Geneva-Copenhagen Survey \citep{Nordstrom2004}.

The photometric metallicity calibration developed by B05 and extended by JA09 and SL10 has proven to be a valuable resource, but its dependence on precise V magnitudes and trigonometric parallaxes limits its utility to M dwarfs in the immediate solar neighborhood, at least for the remainder of the pre-Gaia era.  In a recent contribution, we presented an alternative, spectroscopic technique for estimating the metallicities of near-solar metallicity M dwarfs \citep[][hereafter RA10]{2010ApJ...720L.113R}. This technique is capable of providing metallicity estimates with an accuracy comparable to that of the photometric technique, and requires only a moderate resolution K band spectrum of the M dwarf target, providing a significantly lower observational burden for empirically estimating the metallicities of distant, near-solar or super-solar M dwarfs. An example of the utility of this technique is provided by a recent paper on M dwarf planet hosts identified in the most recent Kepler data release by Muirhead et al. (2011; submitted) 

While RA10 provided a concise introduction to the spectroscopic K band metallicity indicator, space limitations prevented a full exploration of the technique.  In this paper, we provide a full analysis of the motivation, calibration, demonstration and application of the technique using spectra we obtained for 133 Solar Neighborhood M dwarfs.  In Section 2, we describe the composition of our sample, and the acquisition and reduction of our spectroscopic observations.  In Section 3, we describe our measurements of the new H2O-K2 (a modified version of the H2O-K index utilized by RA10), \nai and \cai spectral features upon which the K band metallicity technique depends. In Section 4, we analyze model atmospheres to demonstrate the H2O-K2 index's insensitivity to metallicity, and use the H2O-K2 measurements from our sample to calibrate the index as a spectral type and T$_{eff}$ indicator.  In Section 5, we use the H2O-K2 index as the basis for a revised K band metallicity calibration; by incorporating an additional M dwarf binary into our sample of metallicity calibrators, and adopting a modified functional form to better normalize the temperature dependence of the Na and Ca features, we derive an updated K band \feh  metallicity indicator, as well as a new calibration for overall metallicity ([M/H]), along with robust estimates of the uncertainty in each calibration.  In Section 6 we perform several sanity checks of the K band metallicity technique, demonstrating that the K band metallicity estimates preserve expected correlations between metallicity and Galactic kinematics as well as signatures of chromospheric activity.  We conclude in Section 7 by using the K band metallicity estimates, in combination with other stellar parameters such as rotational velocity ($v$ sin $i$), magnetic activity (as diagnosed by H$_{\alpha}$ emission) and optical brightness, to identify a sample of nearby M dwarfs with particular promise for exoplanet surveys.  We summarize our findings in Section 8, and present our full K band spectral atlas in an appendix (we also make our spectra available to the community online).

\section{Sample Selection \& Observations}\label{sec:Obs}

We observed 133 nearby M dwarfs with declinations higher than -30$^o$. Our sample consists of:

\begin{itemize}

\item{Eighteen M dwarfs with wide ($>$ 5$''$ separation), common-proper-motion solar-type companions to serve as metallicity calibrators. The FGK-primaries have spectroscopic metallicity measurements by \citet[SPOCS Catalogue]{2005ApJS..159..141V}, obtained by fitting synthetic atmospheric spectra to their high-resolution, high signal-to-noise echelle spectra. The SPOCS [Fe/H] and [M/H] values for the FGK-primaries have been assumed to also describe the metallicities of their M dwarf companions. This assumption is justified if both binary components formed together, from the same well-mixed molecular cloud, and no mass transfer or dredge-up has occurred in the system. This assumption is supported by the measurements of the metallicities of binaries with two FGK components:  \citet{2004A&A...420..683D,2006A&A...454..581D} find typical metallicity differences of $\le$0.02 dex between components of binary systems. The binary systems were selected from the \citet{1991STIA...9233932G} catalogue of nearby stars, the \citet{1994RMxAA..28...43P} catalogue of nearby wide binary and multiple systems, and the list of new HIPPARCOS binaries by \citet{2004ApJS..150..455G}. The SPOCS metallicity values are shown in Table \ref{caltable}. The stars in the calibration sample have [Fe/H] metallicities from -0.69 dex to +0.31 dex, [M/H] metallicities from -0.49 dex to +0.25 dex, and spectral types from M1 to M6.}

\item{One hundred and fifteen M dwarfs within 33 parsecs, including 11 M dwarf planet hosts observable from Palomar Mountain: Gl 876 \citep{1998ApJ...505L.147M,1998A&A...338L..67D}, Gl 436 \citep{2004ApJ...617..580B}, Gl 581 \citep{2005A&A...443L..15B}, Gl 849 \citep{2006PASP..118.1685B}, Gl 176 \citep{2009A&A...493..645F}, GJ 1214 \citep{2009Natur.462..891C}, Gl 649 \citep{2010PASP..122..149J}, HIP 57050 \citep[aka GJ 1148,][]{2010ApJ...715..271H}, HIP 79431 \citep{2010PASP..122..156A}, Gl 179 \citep{2010ApJ...721.1467H}, and HIP 12961 \citep{2011A&A...526A.141F}. More than half are members of the Northern 8 parsec sample; the remainder were selected from the LPSM catalog \citep{2005AJ....129.1483L}}
\end{itemize}

Near-infrared spectra of these stars were obtained with the TripleSpec spectrograph on the Palomar 200-inch Hale Telescope \citep{2008SPIE.7014E..30H} during several observing runs between 2007 and 2010. TripleSpec at Palomar has no moving parts and simultaneously acquires 5 cross-dispersed orders covering 1.0-2.4 $\mu$m at a resolution of $\lambda/\Delta\lambda$$\approx$2700. 

The observations were carried out using two different methods:

1) The star was placed at 2 different positions along the slit, A and B. Four exposures were taken with an ABBA slit-nodding pattern. Sky-subtraction was performed by differencing A and B exposures in each paired nod.

2) The star was placed at 5 different positions along the slit. This type of observation pattern was preferred since the array showed a significant number of bad pixels. Each of the 5 exposures were sky-subtracted using a sky-frame made by median combining the 5 exposures of the star. 

These two types of sky-subtraction methods also remove the emission from the instrument and the dark signal from the detector.

Calibration data were taken at the beginning of each night to correct for the non-uniform response between detector pixels. Several exposures of 30 seconds were taken with the dome dark (lamp-off frames), and while illuminating the instrument with a white light source (lamp-on frames). A master-flatfield was obtained by subtracting the median of the lamp-off frames from the median of the lamp-on frames. 

The spectra were reduced using a modified version of the TEDI reduction software \citep{2011arXiv1103.0004M}\footnote{http://www.astro.cornell.edu/$\sim$muirhead/\#Downloads}. Each sky-subtracted exposure was divided by the normalized master-flatfield, wavelength calibrated and optimally extracted \citep{1986PASP...98..609H}. The IDL pipeline walks through the stellar spectrum in each order of the calibrated star image, finds the maximum every 10th pixel, and fits a Gaussian to the slit profile. The Gaussian fit therefore contains all of the stellar signal. The spectral orders of TripleSpec are substantially curved, so the pipeline interpolates onto a rectilinear grid with a thin-plate-spline to correct for the slit tilt, using the spatial and wavelength solutions for the detector by Terry Herter\footnote{herter@astro.cornell.edu}. 

The 1-D M dwarf spectra were telluric corrected using observations of an A0V star with the IDL-based code $xtellcor$\_$general$ by \citet{2003PASP..115..389V}. The spectrum of an A0V star is almost free of metal lines: its observed near-infrared spectrum consists mainly of a featureless and smooth continuum superposed by \ion{H}{1} absorption from the star's atmosphere, and telluric absorption features contributed by the Earth's atmosphere. A telluric spectrum is obtained by removing the hydrogen lines of the A0V star using a high-resolution model of Vega. The target spectrum is then divided by the telluric spectrum constructed from observations of a A0V star obtained near in time and close in airmass to the target.

Finally, the telluric-corrected spectra were flux-calibrated using their 2MASS K band photometry. The flux density of the target star is proportional to the data number count (D$_{\lambda}$) after the telluric calibration:

\begin{eqnarray}
\mathrm{F_{\lambda}} = \mathrm{C~D_{\lambda}} 
\label{eq3}
\end{eqnarray}

Then, the star's spectrum can be flux calibrated from its m$_{K_{s}}$ magnitude, the K$_{s}$ spectral response function R$_{K_{s}}$, and the K$_{s}$-band average flux density of a m$_{K_{s}}$$=$0 star F$_{K_{s}}^{o}$, by finding C :

\begin{eqnarray}
\mathrm{C} = \frac{\int\! \mathrm{R_{K_{s}}}\, \mathrm{d}\lambda}{\int\! \mathrm{D_{\lambda}~R_{K_{s}}}\,\mathrm{d}\lambda} \mathrm{F_{K_{s}}^{o}~10^{-0.4 m_{K_{s}}}}
\label{eq2}
\end{eqnarray}
 
The K$_{s}$ magnitudes for all the targets were obtained from the 2MASS All-Sky Catalog of Point Sources\footnote{http://vizier.u-strasbg.fr/viz-bin/VizieR?-source=II/246} \citep{2006AJ....131.1163S}. The 2MASS K$_{s}$ spectral response curve and F$_{K_{s}}^{o}$ are available online\footnote{http://www.ipac.caltech.edu/2mass/releases/allsky/doc/sec6$\_$4a.html}. 

\begin{figure}
\begin{center}
\includegraphics[scale=0.6,angle=90]{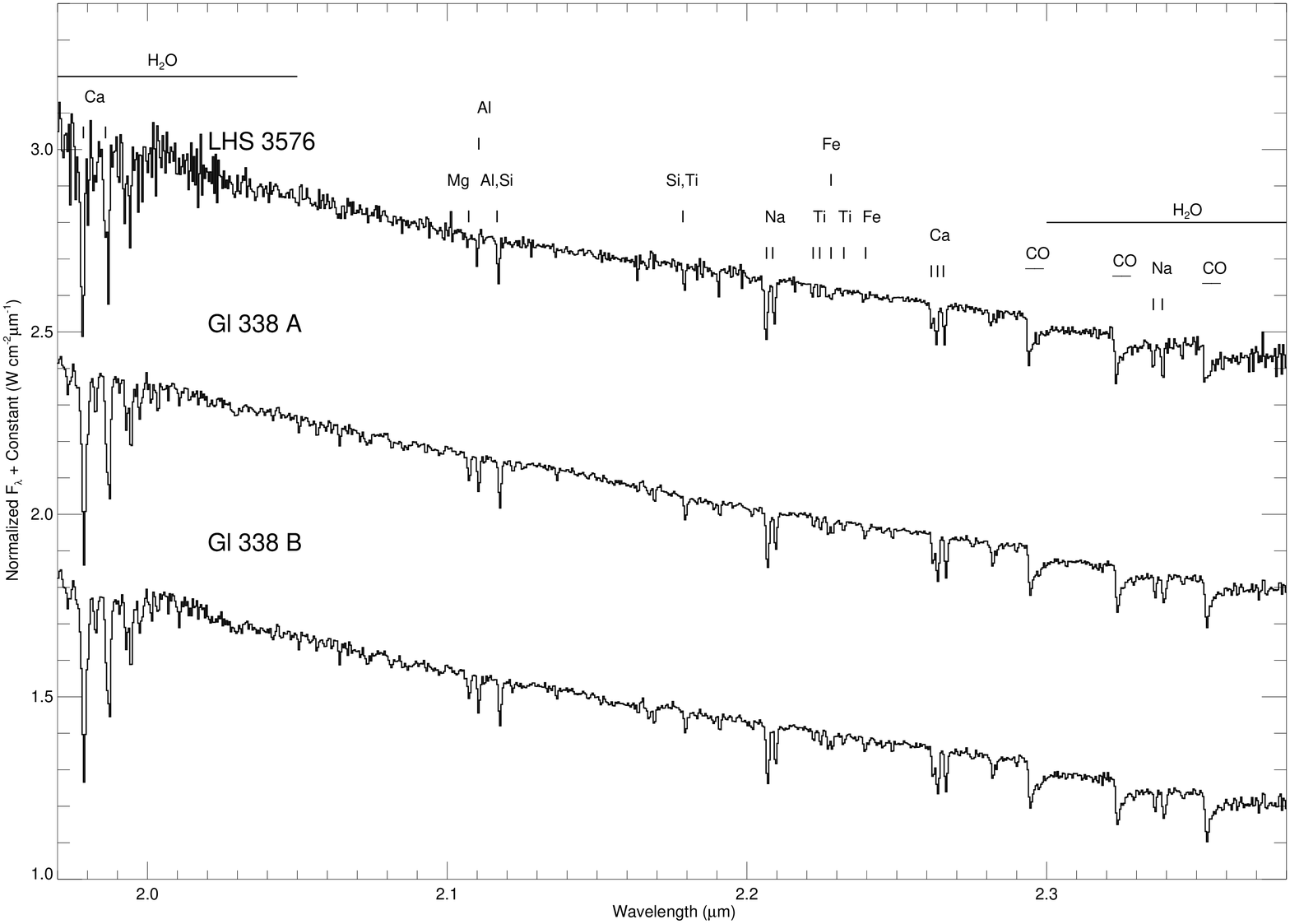}
\caption{\footnotesize{K band spectra of LHS 3576, Gl 338 A, and Gl 338 B. The main spectral features are indicated.}}
\label{FirstGoodSpectra}
\end{center}
\end{figure}

\begin{figure}
\begin{center}
\includegraphics[scale=0.6,angle=90]{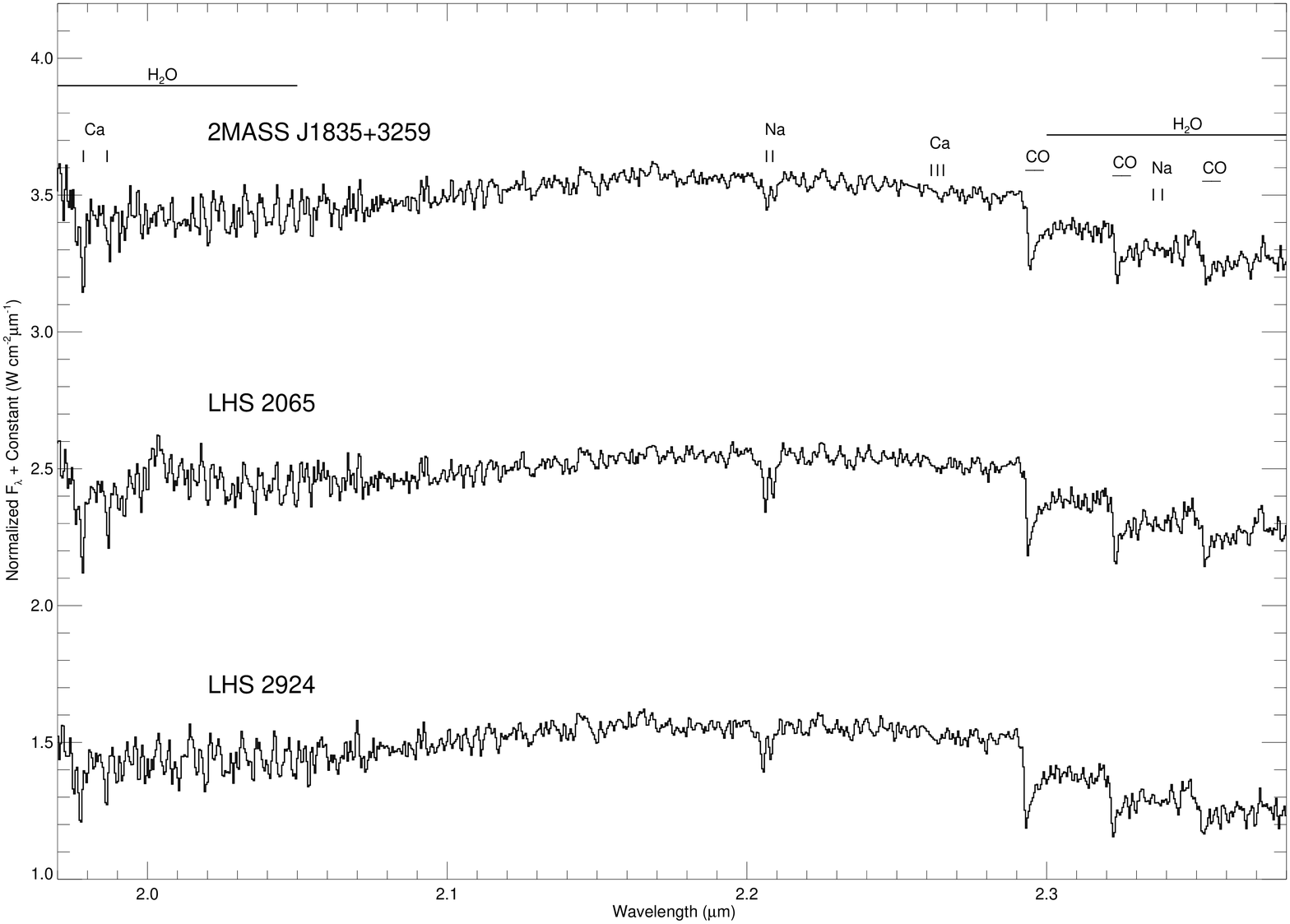}
\caption{\footnotesize{K band spectra of 2MASS J1835+3259, LHS 2065, and LHS 2924. The main spectral features are indicated.}}
\label{LastGoodSpectra}
\end{center}
\end{figure}

\begin{figure}
\begin{center}
\includegraphics[scale=0.6,angle=90]{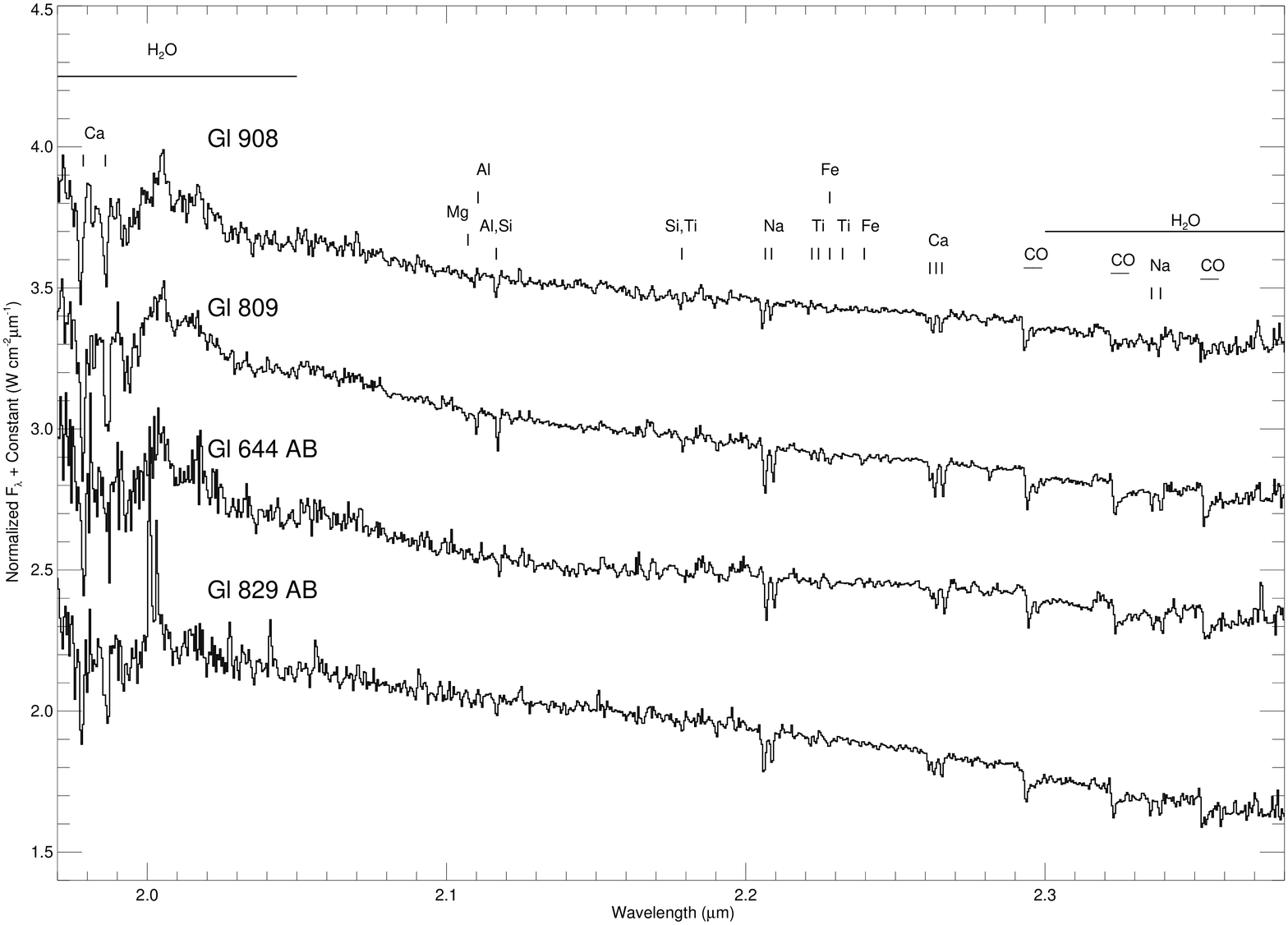}
\caption{\footnotesize{K band spectra of Gl 908, Gl 809, Gl 644 B and Gl 829 AB.  The main spectral features are indicated.}}
\label{dodgyTelluricSpectra}
\end{center}
\end{figure}

Table \ref{SampleTable} presents the properties of each star in the sample, including the star's distance, V and K$_s$ magnitudes, the date the spectrum was observed, and the average signal-to-noise ratio (SNR) obtained in the K band continuum.  The full sample of K band spectra are presented in Figures \ref{FirstGoodSpectra}-\ref{LastGoodSpectra}; the sample is ordered in Table \ref{SampleTable} and in Figures \ref{FirstGoodSpectra}-\ref{LastGoodSpectra} in order of increasing H$_2$O-K2 index (see Section \ref{H2O-K2}).  Four stars appear to have sub-optimal telluric corrections (Gl 644AB, 829AB, 809, \& 908); these stars are flagged in Table \ref{SampleTable} and their spectra are presented separately in Figure \ref{dodgyTelluricSpectra}.

\section{Spectroscopic Analysis}

\subsection{Na \& Ca EWs}

RA10 demonstrated that the 2.205 $\mu$m \nai and 2.263 $\mu$m \cai lines can be used to estimate the metallicities of M dwarf stars.  We measured the equivalent widths (EWs) of each of these lines for every star in our sample. The standard definition of the EW of a line is given by the following equation
 
\begin{eqnarray}
\mathrm{EW_\lambda}&=&  \int _{\lambda_1}^{\lambda_2}  ~ \left [ 1 - \frac {F(\lambda)}{F_c(\lambda)} \right ] {\mathrm{d}\lambda}
\label{eweq}
\end{eqnarray}

Here, F($\lambda$) represents the flux across the wavelength range of the line ($\lambda_2$-$\lambda_1$), and F$_c$($\lambda$) represents the estimated continuum flux on either side of the absorption feature. 

The EWs of the \nai and \cai features were calculated with an IDL pipeline using the following approximation:

\begin{eqnarray}
\mathrm{EW_\lambda} &\simeq& \sum _{i=0}^{n} {\left [ 1 - \frac {F(\lambda_{i})}{F_c(\lambda_{i})} \right ] {\Delta\lambda_{i}} } 
\label{ewreiman}
\end{eqnarray}

Equation \ref{ewreiman} is the Riemann sum expression of the integral in Equation \ref{eweq}, where the F($\lambda_{i}$) and F$_c$($\lambda_{i}$) are the line flux and the estimated continuum flux of the wavelength interval $\Delta\lambda_{i}$, respectively, and n is the number of intervals. The integration limits adopted for the \nai doublet  and \cai triplet features are shown in Table \ref{ewtable}.

Since the continuum of the M dwarf spectra is affected greatly by molecular/broad absorption features, a pseudo-continuum is calculated instead using regions adjacent to the feature of interest and free of any other atomic features.  The pseudo-continuum for each feature is estimated from a linear fit to the median flux within 0.003 $\mu$m ($\sim$ 7 pixels) wide regions centered on the continuum points listed in Table \ref{ewtable}; the median flux at each continuum point is also weighted by its uncertainty in performing the fit. The continuum points in Table \ref{ewtable} were originally defined by \cite{2000AJ....120.2089F} for use in modest resolution spectra of late-type giants.

To estimate the uncertainties on the EW measurements, a Monte-Carlo approach was used. An IDL pipeline added random Gaussian noise (based on the signal-to-noise of the spectrum) to the star's spectrum and then calculated the EW of each feature using Equation \ref{ewreiman}. The entire procedure was repeated 1000 times and the adopted EW uncertainties of the \nai and \cai features for each star correspond to the standard deviations of the 1000 EW measurements of each feature.

The values of the Na I and Ca I EWs measured for each star, along with their errors, are presented in Table \ref{spectral_table}.

\subsection{H$_2$O-K2 Index}\label{H2O-K2}

\begin{figure}[htp]
\begin{center}
\includegraphics[scale=0.6]{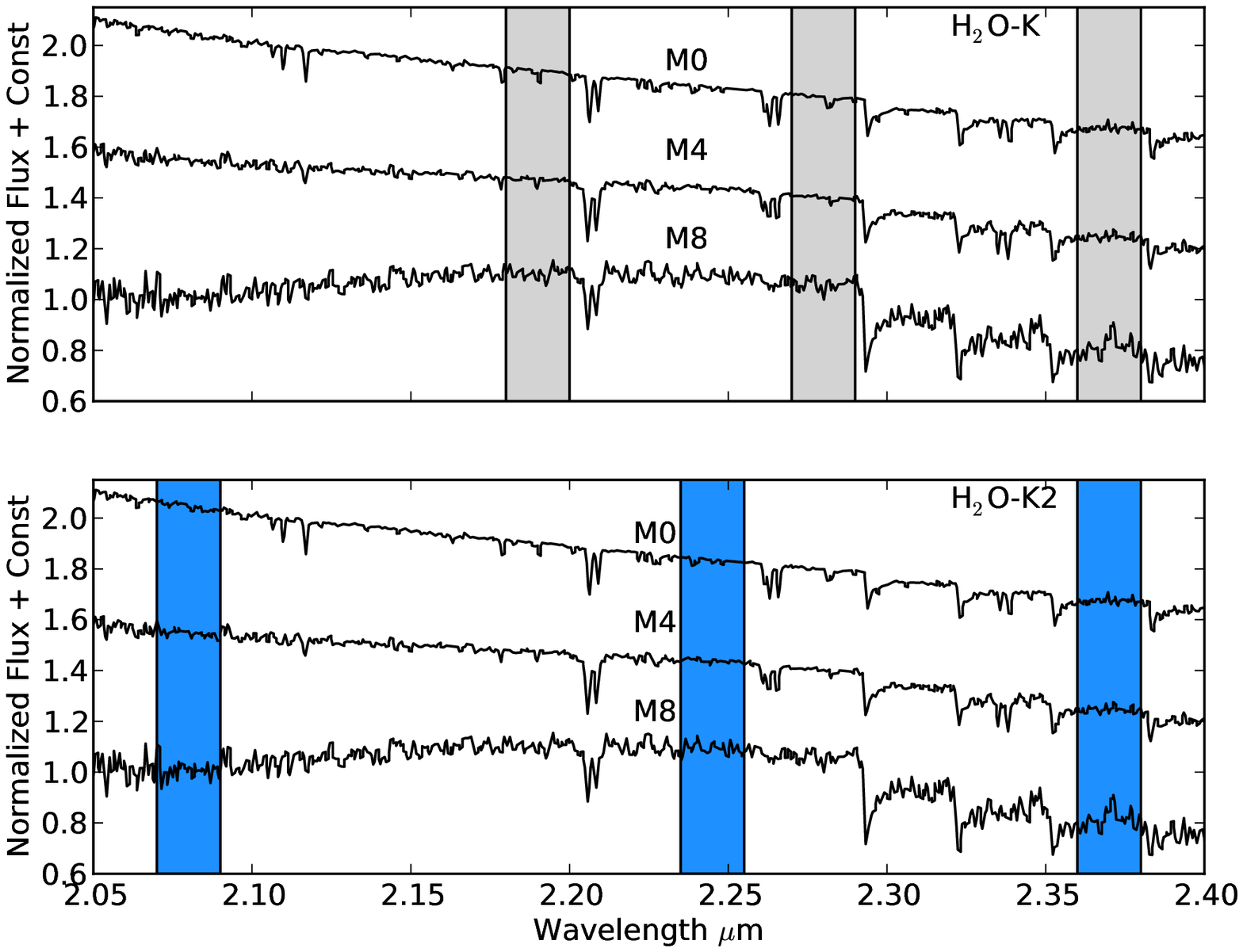}
\caption{The H$_2$O-K index by \citet{2010ApJ...722..971C} versus the new H$_2$O-K2 index in Equation \ref{waterindex}. The portions of the spectrum used by the H$_2$O-K2 index are almost free of atomic features and better sample the change in the overall shape of the spectra of M dwarfs due to the water absorption from 2.07 $\mu$m to 2.38 $\mu$m.}
\label{figh2oK2}
\end{center}
\end{figure}

In RA10, the H$_2$O-K index by \citet{2010ApJ...722..971C} was used to account for the influence of temperature on the strengths of the Na and Ca lines being used to diagnose each star's metallicity. \citet{2010ApJ...722..971C} adopted this index to characterize the spectral types of highly reddened young stars from moderate S/N spectra, and optimized the H$_{2}$O-K index to sample the redder portion of the K band and ignored weak atomic features (\ion{Mg}{1}, \ion{Ti}{1}) that were not typically visible in their spectra.  Stars in our sample, by contrast, possess low extinctions and are relatively bright, such that two of the regions used to calculate the H$_2$O-K index show obvious atomic absorption features in our high-S/N spectra of early M dwarfs.  To ensure our water-index measurements are not affected or biased by these features, we developed a modified H$_2$O-index using two new regions that do not show any noticeable atomic lines. This H$_2$O-K2 water index is defined as:

\begin{eqnarray}
\mathrm{H_{2}O\!\!-\!\!K2} &=& \frac{\langle \mathcal{F}(2.070-2.090) \rangle / \langle \mathcal{F}(2.235-2.255) \rangle}{\langle \mathcal{F}(2.235-2.255) \rangle / \langle \mathcal{F}(2.360-2.380) \rangle}
\label{waterindex}
\end{eqnarray}

\noindent where $\langle \mathcal{F}(a-b) \rangle$ denotes the median flux level in the wavelength range defined by $a$ and $b$, in microns. This new index represents the change in the overall shape of the spectra of M dwarfs due to water absorption from 2.07 $\mu$m to 2.38 $\mu$m, with smaller values of the H$_2$O-K2 index corresponding to greater amounts of H$_2$O opacity.  The difference between the H$_2$O-K index defined by \citet{2010ApJ...722..971C} and the H$_2$O-K2 index is shown in Figure \ref{figh2oK2}. Uncertainties in the H$_2$O-K2 index measurements were computed using the same Monte-Carlo approach used to estimate uncertainties in the \nai and \cai lines; each H$_2$O-K2 error estimate represents the standard deviation of 1000 H$_2$O-K2 measurements after adding synthetic noise to the spectrum consistent with the S/N value of the observed spectrum.

The values of the H$_2$O-K2 index measured for each star are included in Table \ref{spectral_table} along with their uncertainties.

\subsection{Expectations from theoretical models}

The metallicity calibration presented in RA10 was empirically derived, with the temperature and metallicity dependences of the \nai, \cai, and H$_2$O features inferred largely from visual inspection of the calibrator stars.  While RA10 used PHOENIX model atmospheres to qualitatively demonstrate the metallicity sensitivity of \nai and \cai, and the reduced metallicity sensitivity of the H$_2$O feature, no detailed exploration of the behavior of the \nai, \cai, and H$_2$O features as functions of temperature or metallicity were performed. 

The lack of a quantitative exploration of the line strengths predicted by model atmospheres is partially due to the fact that, until recently, synthetic models struggled to match the infrared spectrum of M dwarfs, which is predominately dominated by water opacity \citep{1995ApJ...445..433A}. The discrepancy was believed to be due to incomplete water vapor lists; several water opacity profiles were used through the years, all of which over-predicted the K band opacity and resulted in a lack of flux in the synthetic model when compared to observed spectra \citep{2010arXiv1011.5405A}. \citet{2009ARA&A..47..481A} recently derived a new estimate of the solar oxygen abundance, however, which is a factor of 2 lower than previous estimates and seems to have solved the water opacity discrepancy. The new BT-Settl-2010 synthetic spectra by \citet{2010arXiv1011.5405A} incorporate the new solar abundances, along with updated molecular line lists, and provide a better match to the spectral distribution of M dwarfs across the near-infrared than previous models. The BT-Settl-2010 models are available online\footnote{http://phoenix.ens-lyon.fr/simulator}; using this interface, we constructed a grid of BT-Settl-2010 synthetic spectra with effective temperatures from 2200K to 4100K, and overall metallicities [M/H] equal to +0.5, +0.3, +0.0, -0.5, and -1.0 dex. This effective temperature range covers the whole M dwarf sequence, and the model grid covers a range of metallicities larger than that spanned by our calibration stars, or the metallicities estimated with Equation \ref{overmet} for any of our target stars. All the stars analyzed here are known to be dwarfs and none show any spectral features indicating otherwise, so a value of log g=5.0 was selected to provide the best agreement with empirical M dwarf gravity determinations \citep{2009ApJ...701..764F, 2009A&A...505..205D}.

To provide a quantitative indication of the feature strengths predicted by the theoretical models, we degraded the resolution of each theoretical spectrum to match that of our observed sample and measured the EWs of the \nai doublet and \cai triplet, as well as the H$_2$O-K2 index, using the same routines applied to our observed spectra.

\begin{figure}[htp]
\begin{center}
\includegraphics[scale=0.45]{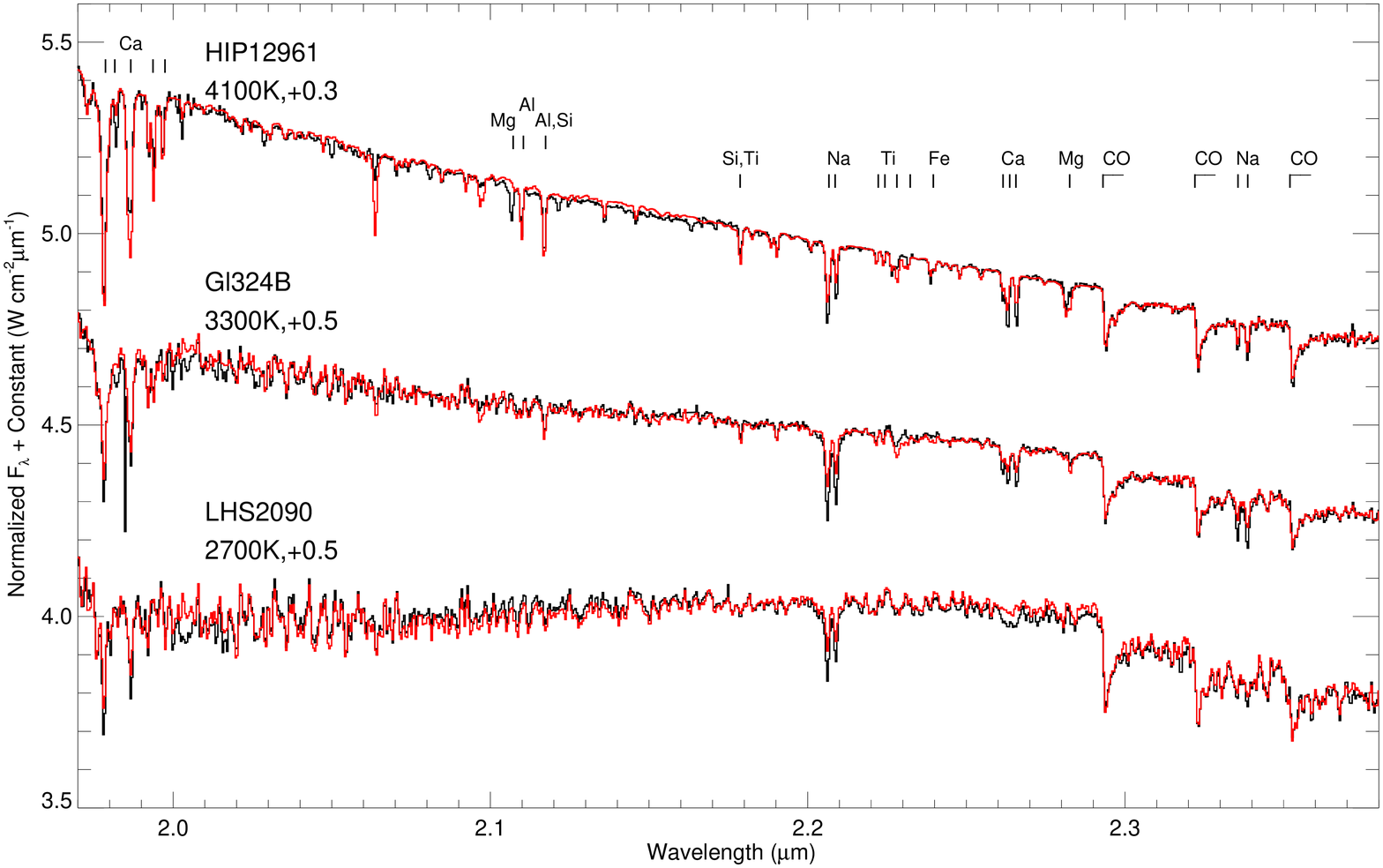}
\caption{Comparison of BT-Settl-2010 synthetic spectra by \citet{2010arXiv1011.5405A} vs. TripleSpec K band spectra of HIP 12961, Gl 324 B and LHS 2090. The most prominent molecular and atomic features are indicated. The overall shape of the observed spectra, dominated mostly by water absorption, are well matched by the BT-Settl-2010 models. The strong \nai doublet and the \cai triplet features in these stars cannot be reproduced by any of the supersolar metallicity models. Perhaps the best match is for HIP 12961, where the [M/H] =+0.3 and T$_{\mathrm{eff}}$= 4100K BT-Settl-2010 model matches the strengths of the CO bands and the neutral Ti and Al lines quite well.  The BT-Settl-2010 model modestly underpredicts the strengths of the 2.205 \nai and 2.26 \cai lines, however, and also shows stronger \cai features at $\sim$1.97 $\mu$m, and no Mg {\footnotesize I} line at $\sim$2.11 $\mu$m. Uncertain oscillator strengths and missing opacity sources could explain these discrepancies \citep{2010sf2a.conf..275R}.}
\label{syn_spec}
\end{center}
\end{figure}

Figure \ref{syn_spec} shows the spectra of three stars in our sample, ranging in spectral type from M1 to M7, superimposed by BT-Settl-2010 synthetic spectra degraded to the resolution of the TripleSpec data. The overall shape of the observed spectra, dominated mostly by water absorption, are well matched by the BT-Settl-2010 models.
The \nai and \cai atomic features used by the metallicity calibration fits, however, are noticably weaker in the BT-Settl-2010 models than in the observed spectra: even the highest metallicity synthetic spectra cannot reproduce the strengths of the \nai doublet and \cai triplet in the spectra of Gl 324 B and LHS 2090. The [M/H] =+0.3 and T$_{\mathrm{eff}}$= 4100K BT-Settl-2010 model matches the strengths of the CO bands and the neutral Ti and Al lines for HIP 12961 quite well, but it shows stronger \cai features at $\sim$1.97 $\mu$m, and no \ion{Mg}{1}  line at $\sim$2.11 $\mu$m. These discrepancies could be explained by uncertainties in the oscillator strengths and opacities adopted for these lines, as suggested by \citet{2010sf2a.conf..275R}.  Another explanation could be the size of the wavelength interval (resolution) chosen to calculate the synthetic spectrum. If the lines in question are not well-sampled at the resolution used to calculate the synthetic spectrum, convolving the synthetic spectrum with a Gaussian to degrade the resolution to match TripleSpec could cause them to be artificially `weakened'. 

Figure \ref{ew_synspec} shows the H$_2$O-K2 index and EWs of the \nai doublet and \cai triplet measured from the BT-Settl-2010 models, after degrading their spectral resolution to match that of TripleSpec, as a function of effective temperature and color-coded by metallicity. The \cai triplet behaves as expected from observed data: its strength increases as both temperature and metallicity increase.  However, the \cai triplet disappears from the synthetic models at effective temperatures lower that $\sim$2900K, even for the supersolar models. In our high S/N empirical spectra, the \cai triplet remains visible even at spectral type M8.  

The water absorption features predicted by the models also agree reasonably well with expectations from empirical spectra, with water absorption increasing monotonically with decreasing temperature, in a manner independent of metallicity, down to T$_{\mathrm{eff}} \sim$3000K. Below 3000K, however, the models show a sharp metallicity difference: for the subsolar metallicity models, the H$_2$O absorption remains almost constant from 3000K to 2200K, while water absorption in solar and supersolar models continues to increase with decreasing temperature. The stark difference in the H$_2$O opacity between solar and subsolar metallicity models suggests the discrepancy in H$_2$O absorption may be more likely a computational artifact than an astrophysical difference.  

The most puzzling behavior is shown by the EWs for the \nai doublet. Observationally, the strength of the \nai doublet increases as temperature decreases, at least until effective temperatures equivalent to an M6V star \citep[e.g.][]{2005ApJ...623.1115C}. Strong \nai doublets are also seen in the late-type stars studied in this work. The only synthetic spectra that show this trend, however, are the solar metallicity spectra. The most metal-poor ([M/H] = -1.0) synthetic spectra, by contrast, show the strongest \nai absorption near $\sim$3000K, with the \nai EWs decreasing to both hotter and cooler temperatures; the moderately metal poor ([M/H] = -0.5) models show a similar behavior but with a strong jump near T$_{\mathrm{eff}}$$=$2900K.  The super-solar models, by contrast, possess the strongest \nai lines at warm temperatures, but the \nai lines in these models begin to weaken at T$_{\mathrm{eff}} \leq$ 3600K, ultimately predicting \nai line strengths that are \textit{anti-correlated} with metallicity (in the solar/super-solar regime) below T$_{\mathrm{eff}}$=3000K. 

The behavior of the \nai doublet in the solar and supersolar synthetic models is not reproduced in our empirical spectra. The metal-rich calibrator, Gl 376 B ([M/H]$_{SPOCS}$ = +0.11, [Fe/H]$_{SPOCS}$=+0.20), is a late-type star (M6) and it exhibits one of the strongest  \nai doublets of the 18 calibrators. Since low temperatures favor the formation of molecular species, the anomalous \nai behavior in the synthetic models could be an effect of the treatment of molecular formation at these temperatures. The most recent study of alkali chemistry in cool atmospheres was conducted by \citet{1999ApJ...519..793L}. \citet{1999ApJ...519..793L} calculated the mole fraction, the amount of an element divided by the total amount of the elements, of neutral Na, ionized Na, and the molecules NaCl, NaH, and NaOH as function of temperature at 1 bar total pressure for cool dwarf atmospheres (3300K $<$ T $<$ 1500K). The computations performed by \citet{1999ApJ...519..793L} indicated that the mole fraction of ionized Na decreases towards cooler temperatures, while the mole fractions of NaCl and NaOH increase; the mole fractions of neutral Na and NaH remain almost constant over this temperature range. Indeed, \citet{1999ApJ...519..793L} found that neutral Na is the most abundant Na-based species over this temperature range, with the abundance peaking near $\sim$2000K and then beginning to decline as Na$_{2}$S condensation weakens the Na features in brown-dwarf spectra \citep{2002ApJ...577..974L}.  It is possible that these complex chemical networks could play a role in the anomalous behavior of the \nai feature in the synthetic models: perhaps the most metal-rich models favor the formation of Na-based molecules toward cooler temperatures, for example, such that the neutral Na responsible for the \nai 2.205 feature depletes more rapidly in metal-rich models, or perhaps Na$_{2}$S condensation occurs at higher temperatures in the super-solar models.  These suggestions are highly speculative, however, and based on the number of odd behaviors exhibited for models cooler than 3000K, we suspect the root cause is as likely to be computational as astrophysical in nature.

\section{Calibrating The H$_2$O-K2 index as a spectral type and T$_{eff}$ indicator}

Water is a heteroatomic molecule and the appearance of the absorption band reflects the pressure and temperature sensitivity of the formation of these molecules \citep{1979ARA&A..17....9M}.   Numerous investigators have leveraged this fact to use near-infrared water absorption bands to diagnose M dwarf effective temperatures and spectral types.  \citet{1986ApJS...62..501K} noted that the depression at 4778 cm$^{-1}$ ($~$2.096 $\mu$m) due to water absorption was largest for dwarf stars, and increases with decreasing spectral type.   \citet{1994MNRAS.267..413J} placed this calibration on an absolute T$_{eff}$ scale, deriving effective temperatures for stars with spectral types ranging from M2V to M9V by comparing the observed H$_2$O bandhead at 1.34 $\mu$m to laboratory data for the water absorption coefficient.  \citet{1995AJ....110.2415A} similarly created a T$_{eff}$ indicator using an index sampling the K band H$_2$O feature in low resolution spectra at $\sim$2.15 and $\sim$2.20 $\mu$m, deriving a linear relationship between effective temperature from low-resolution spectra of MK-dwarf standards.  More recently, \citet{2003ApJ...596..561M} calculated linear relationships between stellar spectral type and the depth of water absorption bands present in $R\sim$2000 near infrared spectra of M, L and T dwarfs.  \citet{2003ApJ...596..561M} found that their K band water index, H$_2$O-D, was constant for spectral types M0-M4, at which point the index decreased (due to increased H$_2$O absorption) towards cooler spectral types.  

\begin{figure}[htp]
\begin{center}
\includegraphics[scale=0.75]{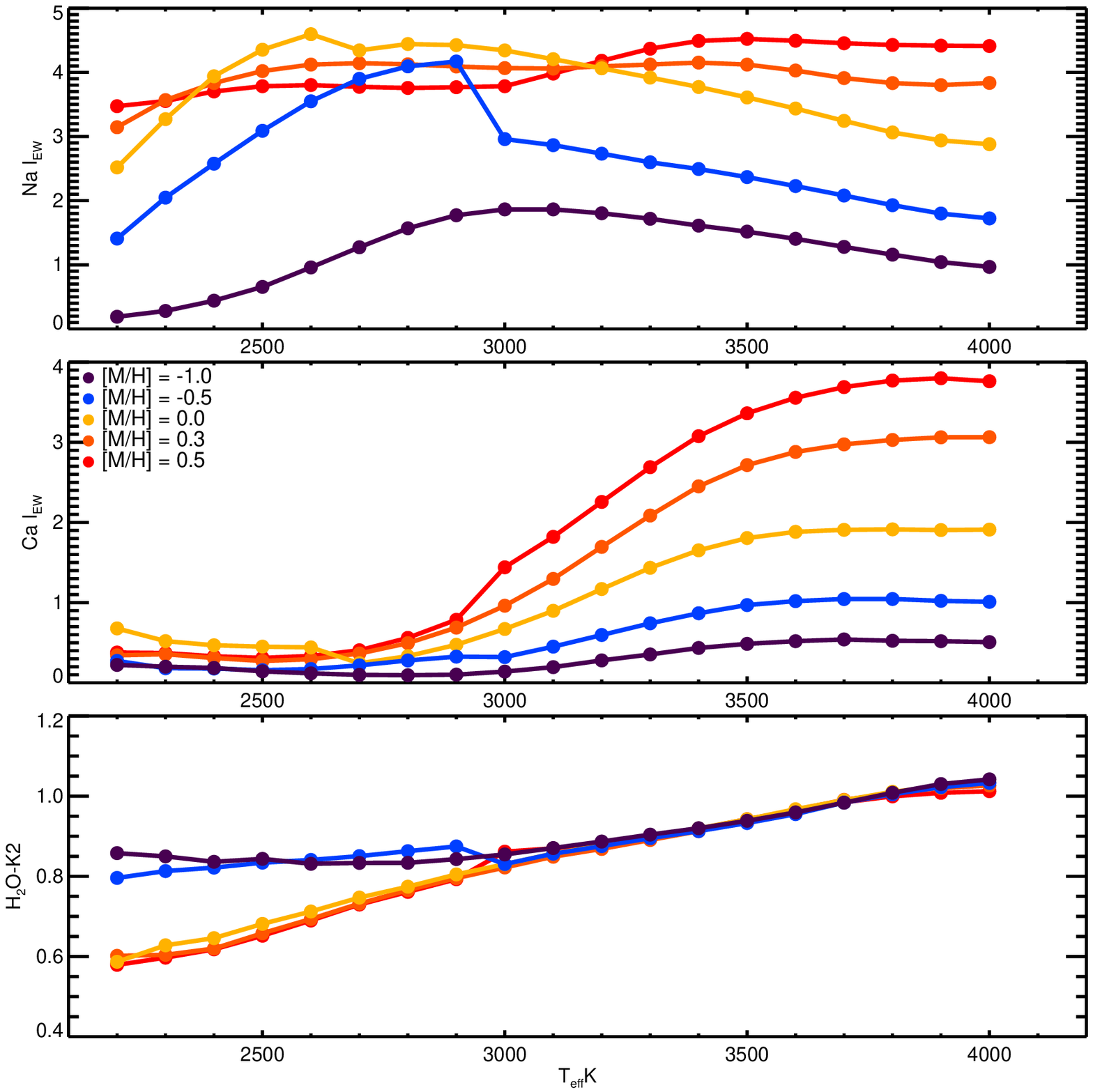}
\caption{Equivalent widths of the \nai doublet (top) and \cai triplet (middle), and H$_2$O-K2 index (bottom), measured from the BT-Settl-2010 synthetic spectra computed by \citet{2010arXiv1011.5405A} and shown as a function of model T$_{eff}$. The strengths of the \nai and \cai features are somewhat weaker in the synthetic spectra than in observed data, but the qualitative behavior of the \cai feature as a function of both temperature and metallicity is consistent. Water absorption appears to be a monotonic function of temperature, independent of metallicity, for models with T$_{\mathrm{eff}} \geq$ 3000K.  Below 3000K, however, the H$_2$O-K2 indices measured from the [M/H] $\leq$ -0.5 dex spectral grids diverge from the [M/H] $>$ -0.5 models; the stark divergence between the two domains suggest the difference may be a computational artifact rather than a true astrophysical difference.  The \nai doublet shows the oddest behavior in these spectra, with significant structure as a function of both metallicity and temperature.}
\label{ew_synspec}
\end{center}
\end{figure}

\begin{figure}[htp]
\begin{center}
\includegraphics[scale=0.5]{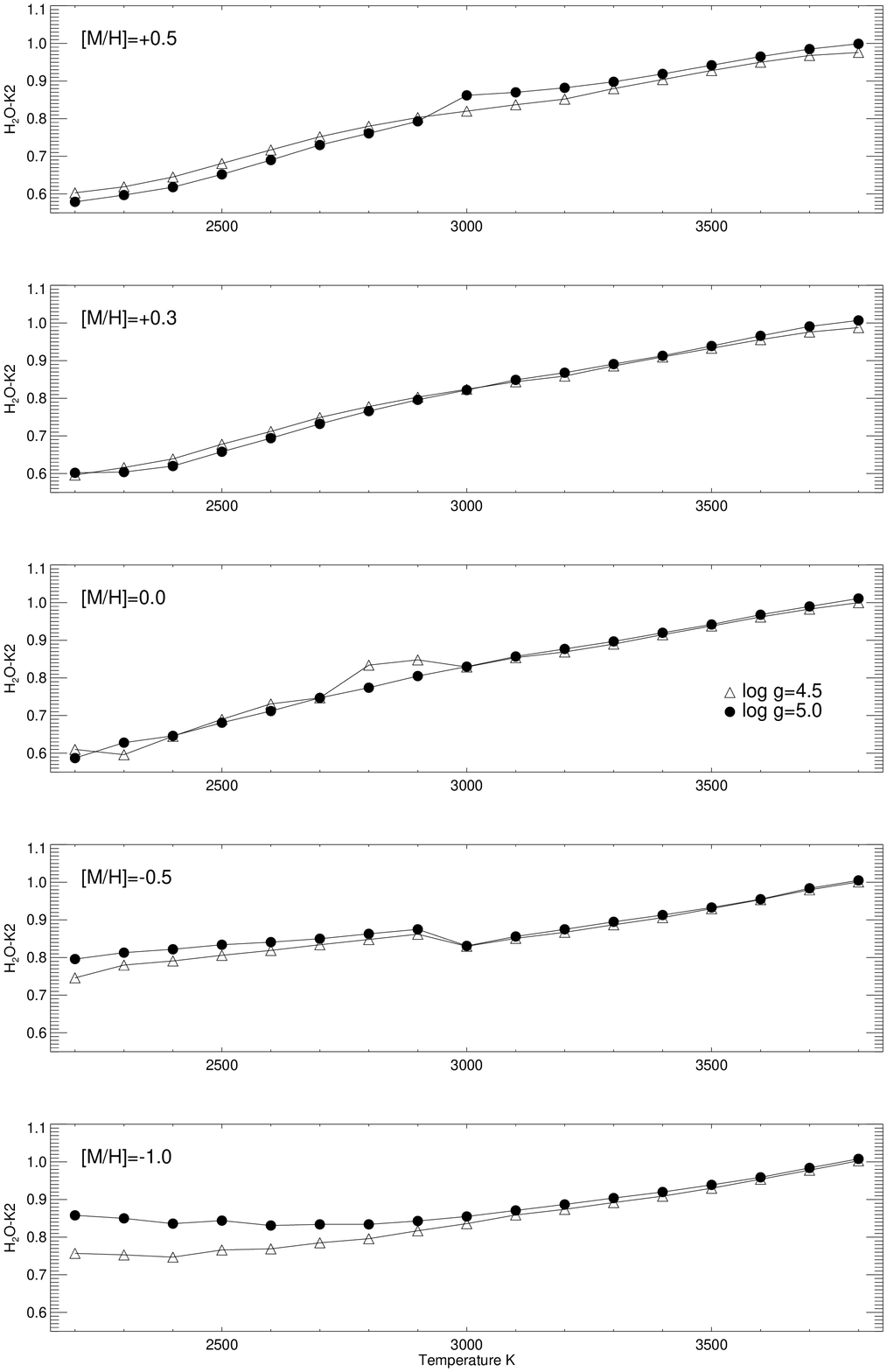}
\caption{The H$_2$O-K2 index measured from the BT-Settl-2010 synthetic spectra shown as a function of model T$_{eff}$. Triangles represent models with $log$ g = 4.5 and black dots represent models with $log$ g= 5.0. The H$_2$O-K2 index shows negligible sensitivity to surface gravity in all models with T$_{eff}$$\ge$ 3000 K. Differences in H$_2$O-K2 index due to surface gravity are small in the solar and super-solar models throughout the whole effective temperature range. The largest discrepancies due to surface gravity are found at lower temperatures in the subsolar [M/H] models. The largest discrepancy between the two surface gravities in the [M/H]=0.0 dex models is believed to be a computational artifact.}
\label{figgrav}
\end{center}
\end{figure}

To calibrate the H$_2$O-K2 index as a well-defined T$_{eff}$ indicator, we provide in Table \ref{H2O-K2Table} the values of the H$_2$O-K2 indices measured from the BT-Settl-2010 models and originally presented in Figure \ref{ew_synspec}.  As Figure \ref{ew_synspec} demonstrates, the BT-Settl-2010 models predict that the H$_2$O-K2 index is a monotonic function of temperature, but independent of metallicity, for stars in the 3000 K $<$ T$_{eff} <$ 4000 K temperature range.  As noted earlier, the lower bound of this metallicity-independent behavior is defined by the clear difference in the behavior of the H$_2$O-K2 index as a function of T$_{eff}$ for metal-poor and solar/metal-rich stars, respectively.  While this low-temperature behavior may be a computational artifact, the metallicity dependent differences at T$_{eff}$s $>$ 4000 K are much more likely to represent a true astrophysical behaviors.  That is, while the H$_2$O opacity appears to `saturate' for even very low metallicities, such that the most metal-poor stars in the model grid do not appear to possess any less H$_2$O opacity than the most metal-rich stars of the same temperature, this is not true at the highest temperatures, near the onset of the H$_2$O feature in the NIR spectral sequence.  Rather, the H$_2$O index values tabulated in Table \ref{H2O-K2Table} appears to indicate that the H$_2$O feature appears at higher temperatures for more metal-rich stars.  Care must therefore be exercised in inferring the temperature of stars with very weak H$_2$O features; in addition to the greater difficulty of ensuring a high S/N measurement of a weak feature, the BT-Settl-2010 models suggest that even if high quality H$_2$O-K2 measurements can be obtained for stars with weak H$_2$O absorption, there will be large astrophysical uncertainties introduced into the resultant T$_{eff}$ estimate due to the strong metallicity dependence of those features at relatively high $T_{eff}$s. The gravity dependence of the H$_2$O-K2 index, as a function of model effective temperature and fixed metallicity, is shown in Figure \ref{figgrav}. The gravity values of log g= 4.5 and 5.0 are the standard values adopted for M dwarf models. Within the metallicity range explored with the BT-Settl-2010 models, the sensitivity of the H$_2$O-K2 index to surface gravity is negligible for T$_{eff}$$\ge$3000 K, and throughout the whole effective temperature range studied for the solar and super solar metallicity models.  The largest discrepancies in H$_2$O-K2 index due to surface gravity are found at lower temperatures in the [M/H]=-1.0 models.    
Therefore, according to the BT-Settl-2010 models, the H$_2$O-K2 index values are metallicity and gravity insensitive for 3000K $\le$T$_{eff}$$\le$ 3800K.

Effective temperatures for the 133 stars were inferred from their H$_2$O-K2 index using the solar ([M/H]=0.0) BT-Settl-2010 models. For each star, we interpolated the measured H$_2$O-K2 value onto the solar metallicity model [H$_2$O-K2,T$_{eff}$] grid to get a T$_{eff}$ estimate and error from its H$_2$O-K2 uncertainty. The values used for the solar metallicity [H$_2$O-K2,T$_{eff}$] grid are shown in Table \ref{H2O-K2Table}. The effective temperatures derived should not be considered extremely accurate or reliable, especially for the late type M dwarfs in the sample, since the BT-Settl-2010 models of T$_{eff}$$<$3000K show H$_2$O absorption discrepancies due to [M/H] that are more likely to be a computational artifact that an astrophysical difference. Due to the finite resolution of the model grid, where models have only been calculated on a grid with a spacing of $\Delta$T$_{eff}$$=$100K, systematic errors of $\sim$100K are expected. 

The dominant spectral sequence for M dwarf stars was established two decades ago by \citet[][hereafter KHM]{1991ApJS...77..417K}, who used red-optical spectra covering 6300-9000 \AA to define a set of primary and secondary M dwarf spectral standards. KHM derived their classifications from a least-squares minimization technique that depends on individual spectral features as well as the overall spectral slope. In the KHM technique, the low resolution spectrum  (~18 \AA) is only normalized using almost featureless region near 7500 \AA before performing the least-squares minimization, so the overall shape of the spectra is conserved.  

Sixty eight of the stars in our sample have KHM spectral types reported by the Research Consortium On Nearby Stars \citep[RECONS,][]{2006AJ....132.2360H}, enabling us to calibrate the H$_2$O-K2 index as a proxy for KHM spectral type. We exclude from this calibration sample 14 objects with potentially ambiguous H$_2$O-K2 indices: eleven are members of binaries or multiple systems in very tight orbits, such that the final K band spectrum mixes emission from all the stars in the system; we exclude the remaining three stars on the basis of their sub-standard telluric/flux calibration.   For the fifty four single stars in our sample with high-quality spectra and KHM spectral types, we performed a linear regression between the measured H$_2$O-K2 indices and KHM spectral types using a Bayesian approach that takes into account the errors associated with each of the variables \citep{2007ApJ...665.1489K}. According to \citet{1991ApJS...77..417K}, the KHM spectral types have $\sim$ 0.5 subtype errors; the errors associated with each H$_2$O-K2 index are listed in Table \ref{spectral_table}. This regression analysis indicated that a star's KHM spectral type, expressed as a numerical M subtype, can be derived from its H$_2$O-K2 indices as:

\begin{eqnarray}
\label{fitspec}
\mathrm{M~subype} &=& ~\mathrm{A}~+~\mathrm{B}~(\mathrm{H_{2}O\!\!-\!\!K2}) \\
\mathrm{A} &=&~24.699~\pm~0.930\nonumber \\
\mathrm{B} &=&~- 23.788~\pm~1.067\nonumber \\
\mathrm{RMSE(M~ subtype)} &=&0.624\nonumber  
\end{eqnarray}

\begin{figure}[htp]
\begin{center}
\includegraphics[scale=0.7]{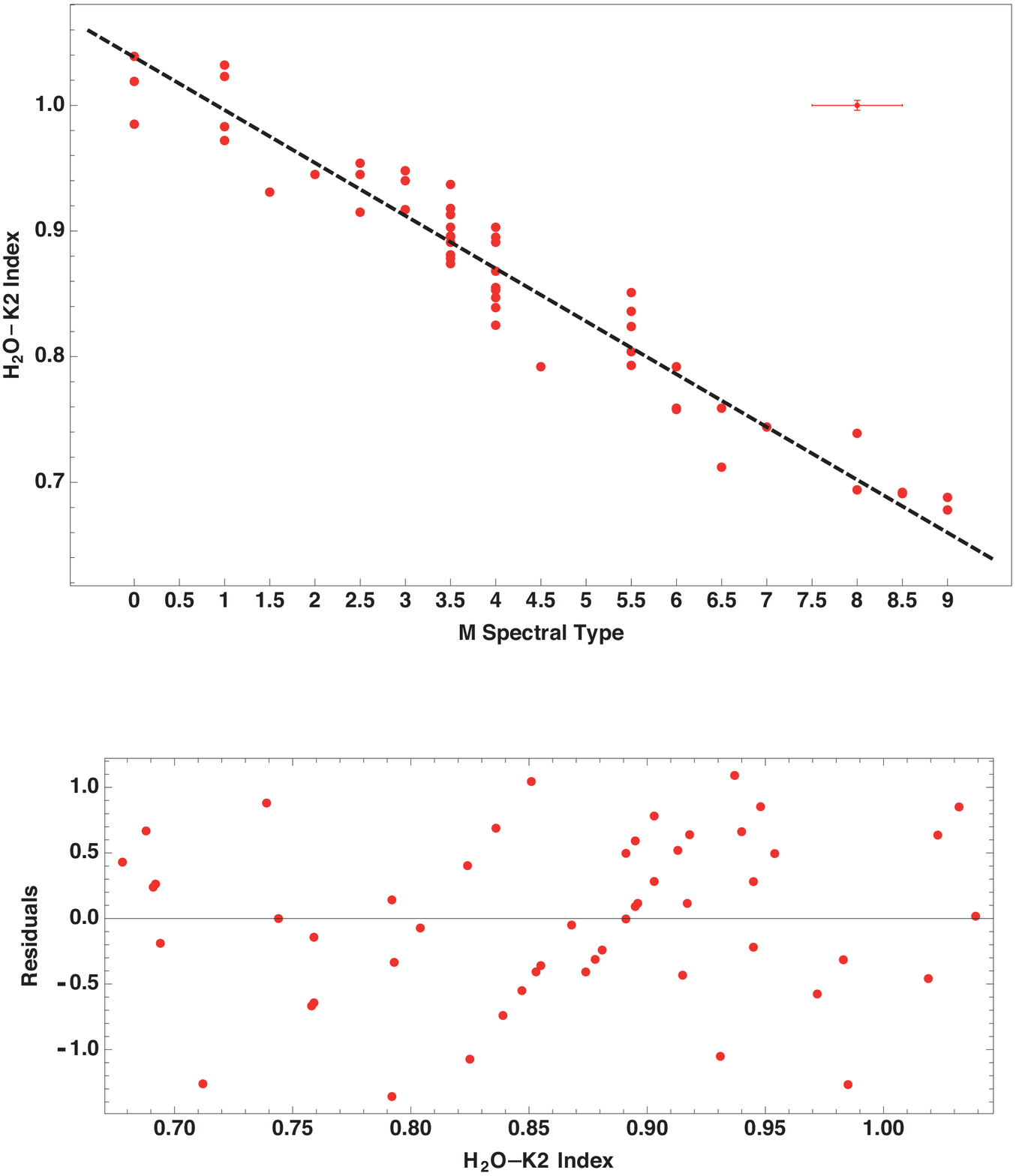}
\caption{Top: The H$_2$O-K2 index versus the Spectral Type of M dwarfs. The red dots represent the fifty four stars with optical KHM spectral types by RECONS. The dashed black line is the linear relationship  between KHM spectral types and the H$_2$O-K2 index in Equation \ref{fitspec}, obtained using a bayesian approach that includes the errors in both coordinates of the objects. Bottom: Spectral type residuals vs H$_2$O-K2 index. }
\label{figfitspec}
\end{center}
\end{figure}

The correlation between KHM spectral type, H$_2$O-K2 index, and the residuals of the fit for our 54 spectral type calibrators, are shown in Figure \ref{figfitspec}. The residuals indicate that the H$_2$O-K2 index can predict each star's spectral type with a typical accuracy of 0.6 sub-types.  As Figure \ref{figfitspec} shows, however, some stars possess KHM spectral types that differ from the type predicted from their H$_2$O-K2 index by $\gtrsim$1 subtype.  A careful inspection of each of the fifty four objects rules out poor quality spectra as the cause of the outliers, but reveals that for the more extreme cases, the overall shape of their K band spectra disagrees with stars of the same KHM spectral type. These discrepancies could be explained by the influence of metallicity on the strengths of double-metal molecular bands, a topic we return to in \S \ref{optical_metallicity_indicators}. 

\begin{figure}[htp]
\begin{center}
\includegraphics[scale=0.6]{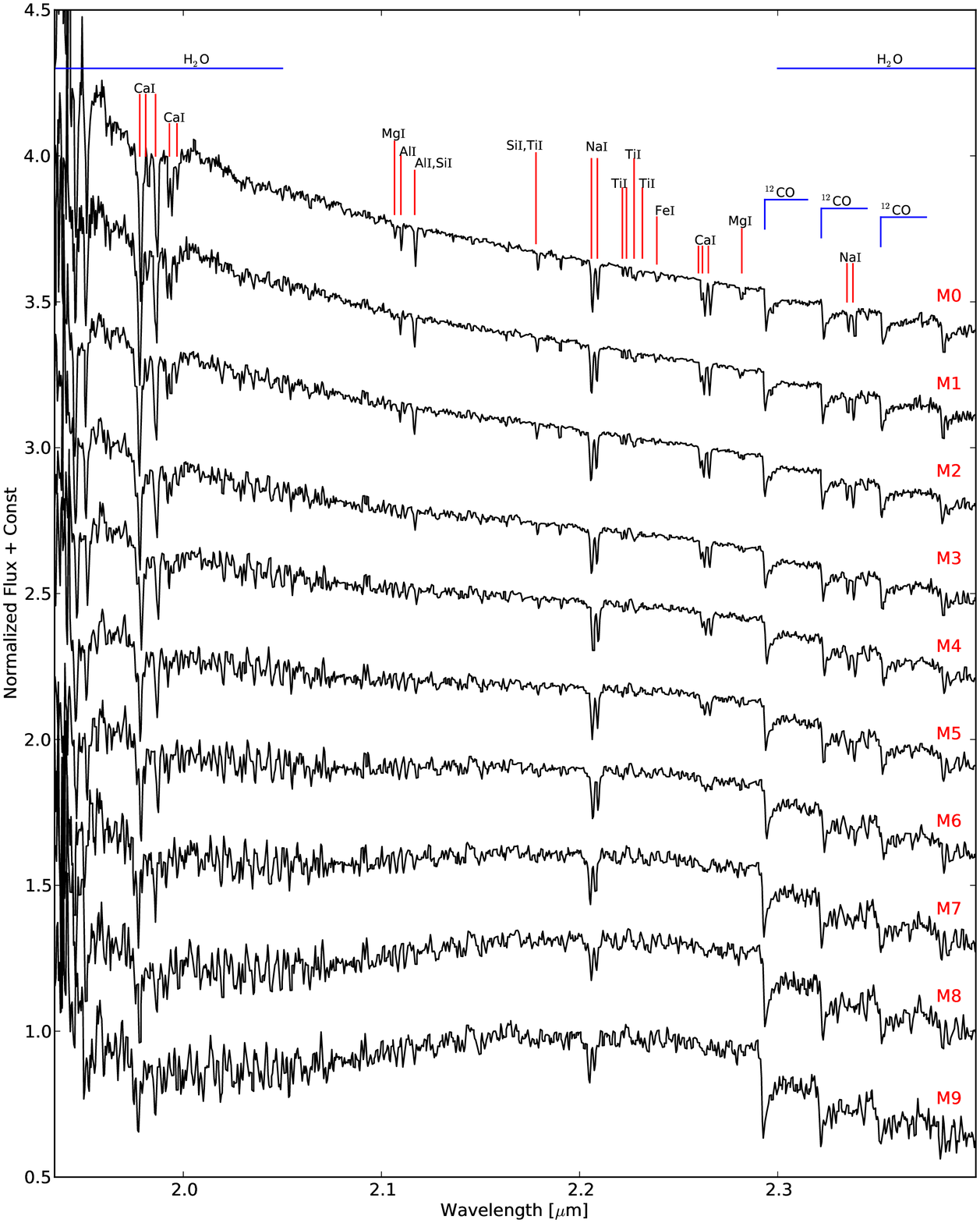}
\caption{Mean TripleSpec K band spectral type templates. The stars in our sample were organized by spectral type and a template was constructed by combining the spectra in each subtype. All of the subtypes are constituted by at least three stars, and up to as many as thirty, with the exception of the M9 subtype, for which LHS 2924 is the only prototype. The most prominent molecular and atomic features are indicated.}
\label{templates}
\end{center}
\end{figure}

We have used this relationship to infer the KHM spectral type for each star in our sample from its H$_2$O-K2 index; these types are listed in Table \ref{spectral_table}, and were used to order the sample from earliest to latest spectral types. 
To provide high quality template spectra for subsequent spectral type measurements, we also combined all the spectra in our sample in bins according to their decimal H$_2$O-K2 types. For each subtype bin, each spectrum was normalized by its mean flux between 2.16 and 2.2 $\mu$m. The spectra then were all cross-correlated with the spectrum of one of the stars in the subtype bin to make their spectral features coincide. Finally, the wavelengths offsets were applied to each spectrum.  At least three stars, and as many as thirty, were then median combined to make the master template for each subtype, with the exception of the M9 template, for which LHS 2924 is the only prototype in our sample. The templates from M0 to M9 are shown in Figure \ref{templates}.

\section{Updated and Extended K band Metallicity Calibrations}

\subsection{A Revised Empirical [Fe/H] Calibration}

RA10 presented a linear equation for estimating an (early-to-mid-) M star's [Fe/H] value based on the strengths of its \nai and \cai lines, and its temperature as diagnosed via its H$_{2}$O-K index.  We present here a revised formalism to describe this relationship, where the revision has been motivated by three factors:

\begin{itemize}
\item{The adoption of the revised H$_{2}$O-K index, as introduced above;}
\item{The addition of a new M dwarf star, Gl 166C, to the metallicity calibrator sample (Table \ref{caltable}) following the publication of RA10.}
\item{The adoption of a new functional form to estimate the star's metallicity from its K band spectral features.}
\end{itemize}

The adoption of a new functional form for the K band metallicity calibration is perhaps the most meaningful change, and deserves further explanation.  The linear expression in RA10 produced unrealistically high [Fe/H] values for some of the more metal-rich M dwarfs in this sample (e.g., LHS 3799; [Fe/H] $\sim +$0.64 according to the RA10 calibration). While it is necessary to include the water index in the relationship to remove the temperature dependence of the \nai and \cai features, the assumption that water absorption should be included as an independent variable is not necessarily correct. Extensive experimentation with various functional forms revealed that a better [Fe/H] calibration fit could be obtained if water absorption is used as a ``weight" for the \nai and \cai strengths rather than as a fully independent variable.

We conducted multivariate linear regressions on the \nai and \cai EWs, weighted by each star's H$_{2}$O-K2 index, for the 18 metallicity calibrators to identify the best fit relationship to predict each star's [Fe/H] value.  The resulting calibration is:

\begin{eqnarray}
\label{ironmet}
\mathrm{[Fe/H]} &=&~\mathrm{A}~+~\mathrm{B}~ \frac{\mathrm{Na~{\footnotesize I}_{EW}}}{\mathrm{H_{2}O\!\!-\!\!K2}}~+~\mathrm{C}~\frac{\mathrm{Ca~{\footnotesize I}_{EW}}}{\mathrm{H_{2}O\!\!-\!\!K2}}\\
\mathrm{A} &=&~-1.039~\pm~0.170\nonumber \\
\mathrm{B} &=&~0.092~\pm~0.023\nonumber \\
\mathrm{C} &=&~0.119~\pm~0.033\nonumber \\
\mathrm{RMSE([Fe/H])} &=&0.141 \nonumber 
\end{eqnarray}

For completeness, we summarize in Appendix \ref{statistics_appendix} the Residual Mean Square ($RMSp$), the Root Mean Squared Error ($RMSE$) and adjusted square for the multiple correlation coefficient ($R$$^2_{ap}$) statistics for evaluating the uncertainties and predictive power associated with a functional fit to a given set of calibration data. These quantities were introduced by \cite{2010A&A...519A.105S} to compare the quality of the photometric [Fe/H] calibrations.  The [Fe/H] fit in RA10 has a $RMSp$([Fe/H])=0.02 and an $R$$^2_{ap}$([Fe/H])=0.63. The addition of another binary system and the new independent variables maintained or improved each of these statistics with respect to the RA10 calibration: the RA10 [Fe/H] calibration and the revision presented here possess equivalent $RMSp$ values, but the new [Fe/H] fit has a higher $R$$^2_{ap}$ ($R$$^2_{ap}$([Fe/H]$_{RA11}$)$=$0.67). 

For the [Fe/H] model in Equation \ref{ironmet}, the Root-Mean-Squared Error of Validation ($RMSE$$_{V}$, Appendix \ref{statistics_appendix}) is equal to 0.161 dex. The RMSE$_{V}$ value is equivalent to a $\sim$ 70$\%$ confidence interval or 1-$\sigma$, and is considered a sensible estimate of average prediction error. The 95$\%$ confidence interval is equal to $\pm$ 0.344 dex.

\begin{figure}[htp]
\begin{center}
\includegraphics[scale=0.55]{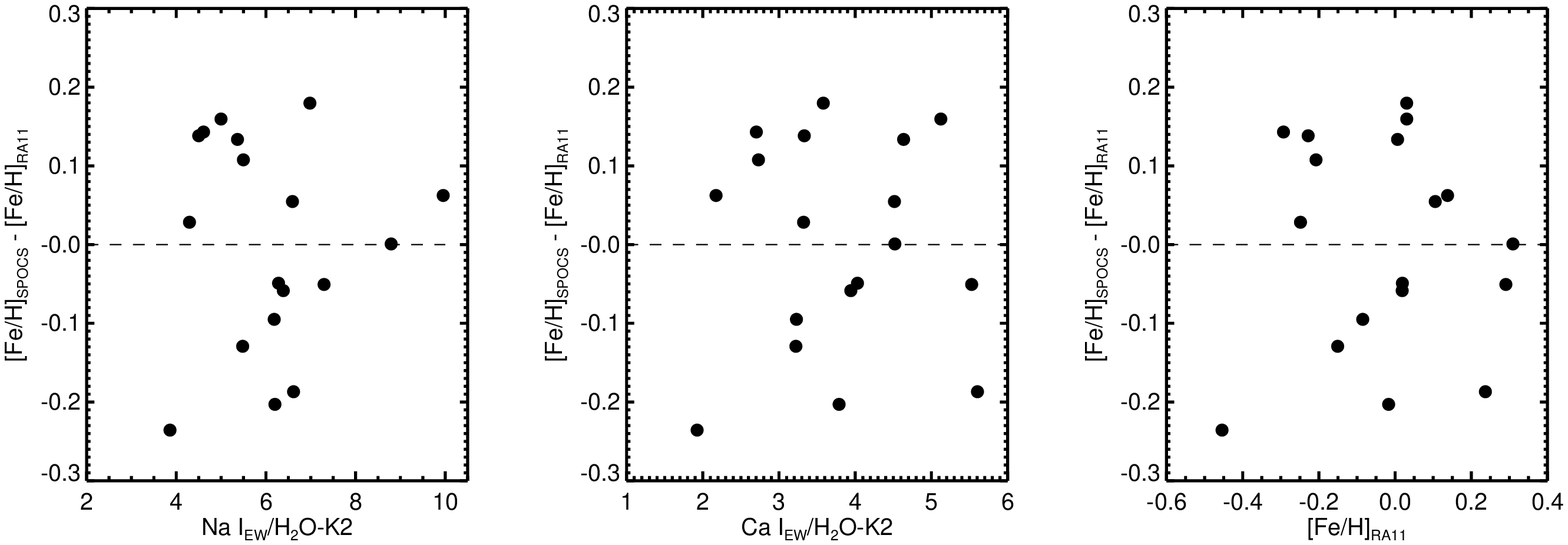}
\caption{[Fe/H] residuals versus the independent variables of the [Fe/H] fit in Equation \ref{ironmet} (\nai$/$H$_2$O-K2, left; \cai$/$H$_2$O-K2, middle), and as a function of the predicted metallicity (right).}
\label{resironmet}
\end{center}
\end{figure}

The [Fe/H] residuals versus the dependent variables of the fit in Equation \ref{ironmet} and the SPOCS [Fe/H] values are shown in Figure \ref{resironmet}. The residuals show mostly random scatter, and no evident structure can be seen as a function of the weighted line strengths (i.e. left and middle panels), and the computed [Fe/H] from Equation \ref{ironmet} (right panel). The "outlier"  in the bottom left of the [Fe/H] residuals vs the SPOCS [Fe/H] values plot is the lowest metallicity star in the calibration sample, Gl 611 B, however, and demonstrates the pressing need for an expanded sample of metallicity calibrators to establish if any systematic errors are present, and if so, remove them.

\begin{figure}[htp]
\begin{center}
\includegraphics[scale=0.8]{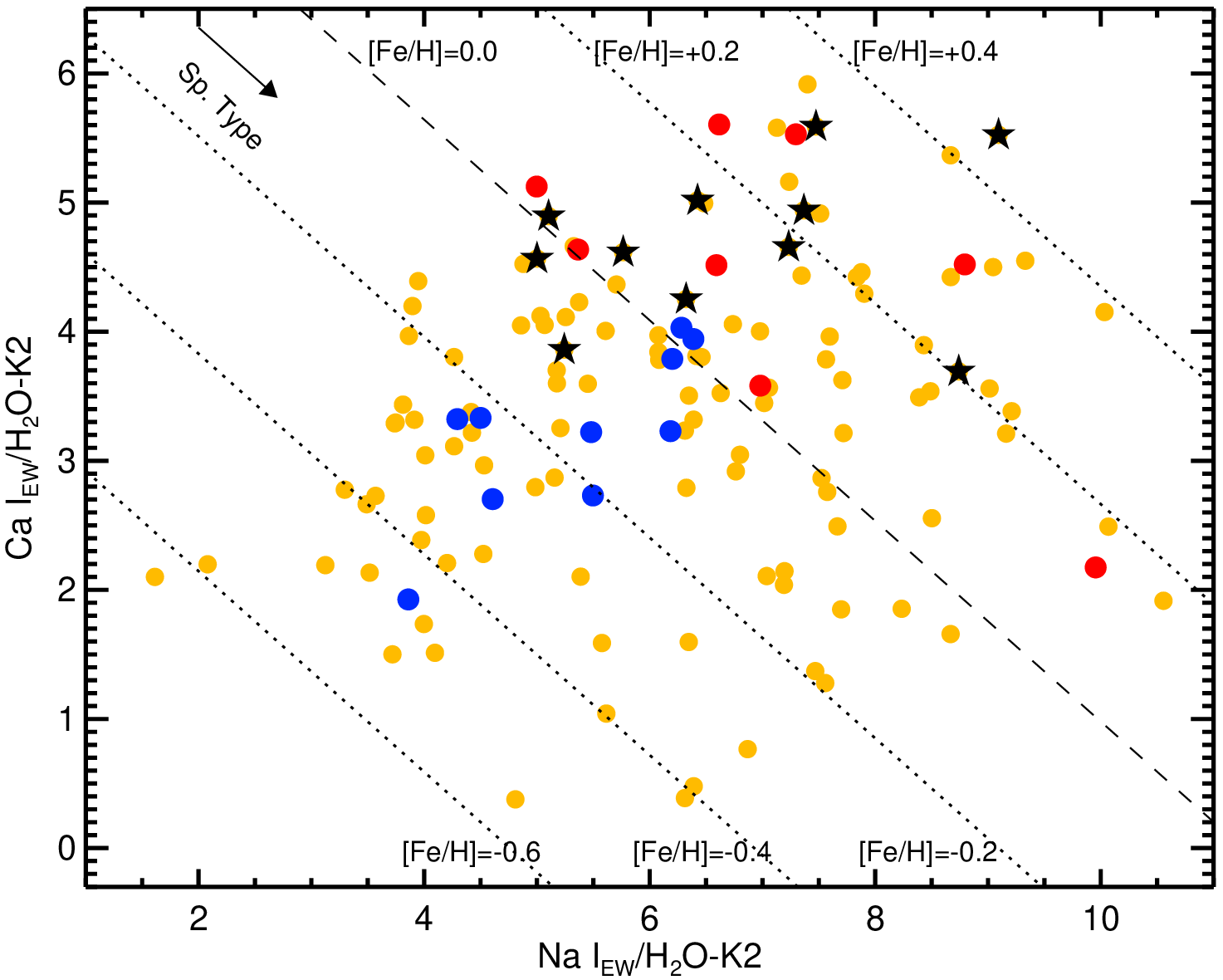}
\caption{The EW of the \cai vs the EW of the \nai, both weighted by the H$_2$O-K2 index, for the M dwarfs in our sample. The red and blue dots are the 18 M dwarfs in the metallicity calibration sample with [Fe/H]$\geq$0.0 and [Fe/H]$<$0.0, respectively. The black stars represent the M dwarf planet hosts, and the yellow dots the rest of the M dwarfs studied in this work. The dashed line is an iso-metallicity contour for [Fe/H] $=$0.0. The dotted lines are iso-metallicity contours for [Fe/H] values +0.4, +0.2, -0.2, -0.4, and -0.6 dex from the top right corner to the bottom left corner, respectively. The isometallicity contours were calculated from Equation \ref{ironmet}. In this plane, there is a clear distinction between the M dwarfs in the calibration sample with metal-rich and metal-poor FGK-dwarf companions. The M dwarf planet hosts all have [Fe/H]$>$0.0 dex, except Gl 581 and Gl 649, whose [Fe/H] are equal to -0.10 dex and -0.04 dex, respectively. The most iron-rich star in the sample is the planet host HIP 79431(+0.46 dex) and the least iron-rich is the flare star V$*$ V1513 Cyg (-0.64 dex).}
\label{scatteriron}
\end{center}
\end{figure}

The EWs of the \cai triplet and the \nai doublet features, weighted by the H$_{2}$O-K2 index, are plotted in Figure \ref{scatteriron} for all the M dwarfs analyzed here. Iso-metallicity [Fe/H] contours calculated from Equation \ref{ironmet} are shown as dashed and dotted lines. In this plane, there is a clear distinction between the M dwarfs in the calibration sample with metal-rich and metal-poor FGK-dwarf companions. The M dwarf planet hosts also all have [Fe/H]$>$0.0 dex according to this relationship, with the exception of Gl 581 and Gl 649, whose [Fe/H] are equal to -0.10 dex and -0.04 dex, respectively. The [Fe/H] values predicted by Equation \ref{ironmet} for the full sample of stars analyzed here are listed in Table \ref{spectral_table}: the most iron-rich star in the sample is the planet host HIP 79431(+0.46 dex) and the least iron-rich is the flare star V$*$ V1513 Cyg (-0.64 dex).

\subsection{A New K band Overall Metallicity ([M/H]) Calibration}

In addition to iron abundance estimates, the SPOCS catalogue also provides overall metallicity [M/H] values for the FGK-companions of stars in the calibration sample. In the fitting procedure described in \citet{2005ApJS..159..141V}, the [M/H] value is an individual model parameter, rather than a quantity constructed from individual element abundances.  We performed a linear regression as described above, assuming the same functional form as adopted for the iron metallicity estimate, to calibrate the relationship between the water weighted \nai doublet and \cai triplet and [M/H]:

\begin{eqnarray}
\label{overmet}
\mathrm{[M/H]} &=&~\mathrm{A}~+~\mathrm{B}~ \frac{\mathrm{Na~{\footnotesize I}_{EW}}}{\mathrm{H_{2}O\!\!-\!\!K2}}~+~\mathrm{C}~\frac{\mathrm{Ca~{\footnotesize I}_{EW}}}{\mathrm{H_{2}O\!\!-\!\!K2}}\\
\mathrm{A} &=&~-0.731~\pm~0.120\nonumber \\
\mathrm{B} &=&~0.066~\pm~0.016\nonumber \\
\mathrm{C} &=&~0.083~\pm~0.023\nonumber \\
\mathrm{RMSE([M/H])} &=&0.100 \nonumber 
\end{eqnarray}

\begin{figure}[htp]
\begin{center}
\includegraphics[scale=0.55]{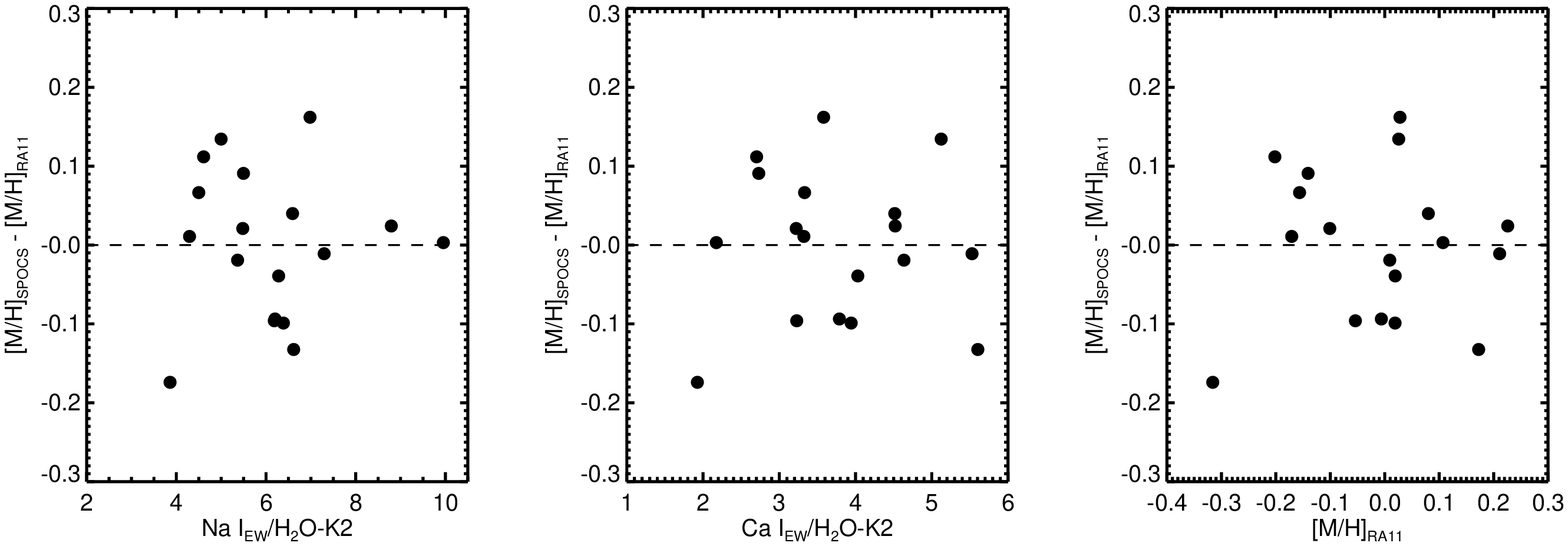}
\caption{[[M/H] residuals versus the independent variables of the [M/H] fit in Equation \ref{overmet} (\nai$/$H$_2$O-K2, left; \cai$/$H$_2$O-K2, middle), and as a function of the predicted metallicity (right).}
\label{resovermet}
\end{center}
\end{figure}

The K band [M/H] fit has a $RMSp$([M/H])=0.010 and an adjusted R square $R$$^2_{ap}$([M/H])= 0.67. For the [M/H] model in Equation \ref{overmet}, the $RMSE_{V}$ is equal to 0.111 dex ($\sim$ 70$\%$ confidence interval or 1-$\sigma$), and the 95$\%$ confidence interval is equal to $\pm$ 0.237. None of the previous photometric or spectroscopic works on M dwarf metallicities have estimated [M/H] values, therefore, a comparison to other methods cannot be performed. The [M/H] residuals versus the dependent variables of the fit in Equation \ref{overmet} and the SPOCS [M/H] values are shown in Figure \ref{resovermet}. The residuals show just random  scatter: , as with the [Fe/H] residuals, no structure is evident in the residuals as function of the normalized Na and Ca line strengths or the computed [M/H] values. 

\begin{figure}[htp]
\begin{center}
\includegraphics[scale=0.8]{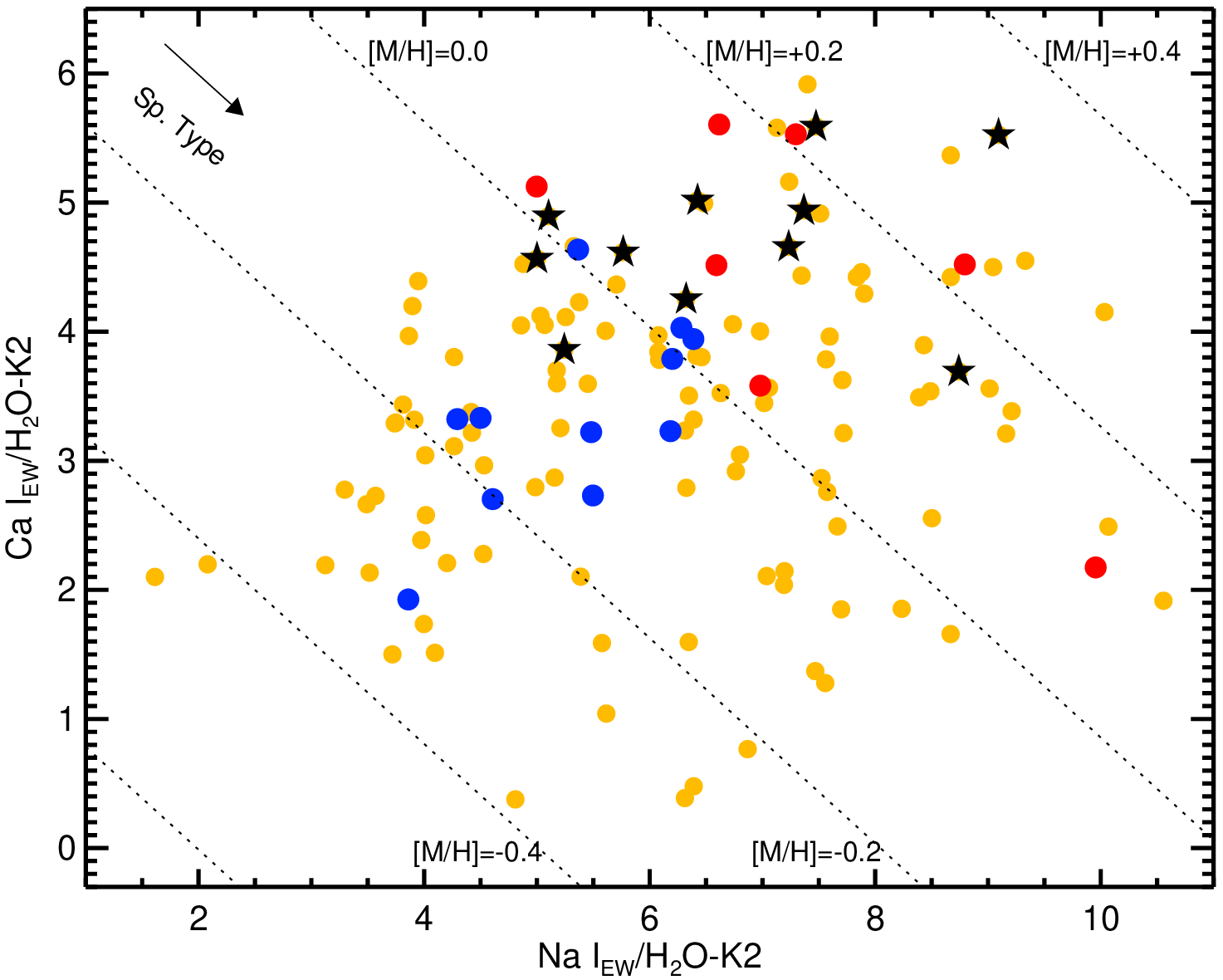}
\caption{The EW of the \cai vs the EW of the \nai  both weighted by the H$_2$O-K2 index, for the M dwarfs in our sample. The nomenclature is the same as Figure \ref{scatteriron}. The dashed line is a iso-metallicity contour for [M/H] $=$0.0. The dotted lines are iso-metallicity contours for [M/H] values +0.4, +0.2, -0.2, and -0.4 dex from the top right corner to the bottom left corner, respectively. The isometallicity contours were calculated from Equation \ref{overmet}. The M dwarf planet hosts all have [M/H]$>$0.0 dex, except Gl 581 and Gl649, whose [M/H] are equal to -0.06 dex and -0.02 dex. The most metal-rich star in the sample HIP 79431 has a [M/H]=+0.33 dex and the least metal-rich V$*$ V1513 Cyg has [M/H]=-0.45 dex.}
\label{scatterover}
\end{center}
\end{figure}

Figure \ref{scatterover} shows the same Ca/H$_2$O-K2 versus Na/H$_2$O-K2 plane as Figure \ref{scatteriron}, but the dashed and dotted lines represent iso-overall-metallicity contours calculated from Equation \ref{overmet}. The M dwarf planet hosts all have [M/H]$>$0.0 dex, with the exception of Gl 581 and Gl 649 (the same stars which were subsolar in [Fe/H]) with [M/H] values equal to -0.06 dex and -0.02 dex, respectively. The most metal-rich star in the sample HIP 79431 has a [M/H]=+0.33 dex and the least metal-rich V$*$ V1513 Cyg has [M/H]=-0.45 dex.  The predicted [M/H] values from Equation \ref{overmet} for the stars in our sample are shown in Table \ref{spectral_table}.  

\begin{figure}[htp]
\begin{center}
\includegraphics[scale=0.8]{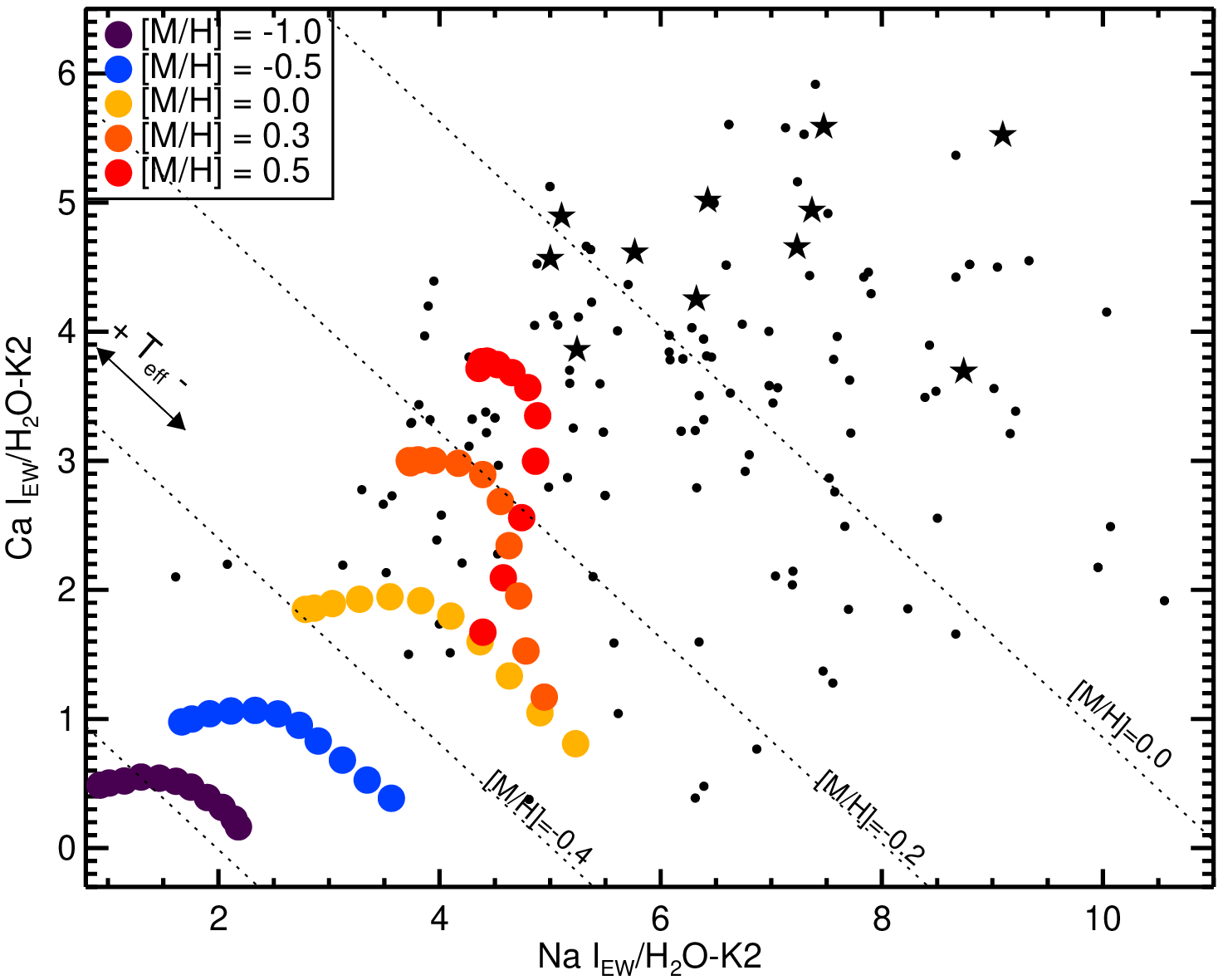}
\caption{EW of \cai vs EW of \nai, weighted by H$_2$O-K2 index, for the BT-Settl-2010 synthetic models by \citet{2010arXiv1011.5405A} with T$_{\mathrm{eff}}$$\leq$ 3000K. The dashed lines are isometallicity contours for [M/H] values, from the top right corner to the bottom left corner, of -0.2, -0.4, and -0.6. The isometallicity contours for [M/H] were calculated from Equation \ref{overmet}. There is a clear distinction between the different metallicity models in this plane. The BT-Settl-2010 synthetic models follow the same direction as the isometallicity contours but their weaker strengths of the \nai doublet and \cai triplet  do not allow a direct metallicity comparison with empirical data. Subsolar and solar models of temperatures between 3700K and 3000K appear as parallel lines in this plane,but the supersolar models intersect the solar model at 3000K, due to the odd behavior of the \nai doublet features of the supersolar models mentioned in the text}
\label{scatter_synthetic}
\end{center}
\end{figure}

As the BT-Settl-2010 theoretical model atmospheres are calibrated as a function of their overall metallicity, we are now able to directly compare the feature strengths measured from our model grid against the iso-metallicity contours we inferred from the empirical calibration sample above.  Figure \ref{scatter_synthetic} shows the EW of \cai vs the EW of \nai, both weighted by H$_{2}$O-K2 that we measured from the BT-Settl-2010 model spectra with temperatures equal to or higher than 3000K. This plane splits the synthetic models by metallicity, in a similar way as it also splits the calibration stars for the metallicity fits. The [M/H]=-1.0 dex models lie in the bottom-left corner, and the [M/H]=+0.5 dex models lie upwards and to the right of the [M/H]=-1.0 dex models, consistent with the direction of the offsets predicted by the isometallicity contours.
However, since the synthetic models predict weaker \nai doublet and \cai triplet features than are observed in the empirical spectra, the models do not directly align with the empirical iso-metallicity contours (depicted by dashed lines) obtained with the K band [M/H] fit of Equation \ref{overmet}. The [M/H]=-1.0 dex models are located above the [M/H]=-0.6 dex empirical contour, while the [M/H]=+0.5 dex model points are far below the [M/H]=0.0 dex empirical contour. The odd behavior of the \nai doublet features of the supersolar models mentioned above becomes evident as temperature decreases in Figure \ref{scatter_synthetic}. Subsolar and solar models of temperatures between 3700K and 3000K appear as parallel lines in this plane, but the supersolar models intersect the solar model at 3000K.

\subsection{The K band [Fe/H] Method Compared with Other [Fe/H] Techniques}\label{comparisonwphoto}

\subsubsection{Statistics of the K band [Fe/H] and Photometric Methods.}

We now seek to compare the quality of our [Fe/H] fit, and the resultant values it predicts for our calibration stars, to similar evaluations of the [Fe/H] predictions produced by the photometric calibrations.  \citet[][hereafter N11]{2011arXiv1110.2694N} performed a similar exercise, comparing the $RMSp$, $R$$^2_{ap}$, and the dispersion around the mean ($rms$) values of the photometric metallicity scales using a sample of 23 M dwarfs with solar-type companions of known metallicity, and refined the \cite{2010A&A...519A.105S} calibration based on their results. 

Unfortunately, for the same reason that N11 did not include the RA10 metallicity predictions in their analysis, because our calibrator samples do not overlap well, it is difficult to directly compare the statistics of our fit to those computed by N11. N11 analyzed 23 M dwarfs with reliable metallicity estimates, V magnitudes, and trigonometric parallaxes; of these, only 7 are members of our sample of M dwarfs with moderate resolution NIR spectra, providing minimal leverage for evaluating the accuracy of the various techniques with a common set of calibrators. This demonstrates an important point: the statistics calculated above, by definition, describe how well a given formalism can reproduce measurements of a particular calibration set.  The meaning of those statistics, therefore, is inextricably linked to the particular calibration set from which they were computed: comparing those statistics amongst various metallicity calibrations requires that the statistics be computed with respect to a specific, and shared, set of calibrators.  
 
To make matters worse, the \textit{veracity} of the calibration set is as important, if not more so, as the \textit{number} of calibrators.  For example, while the B05 functional fit has overall better statistics (as we show below), some of the stars in its calibration sample have \textit{poor} photometry values. The JA09, SL10, and N11 calibrations improved the B05 calibration by including only stars with high precision photometry/parallax measurements, excluding several of the stars in B05, and including stellar population and kinematic considerations. Moreover, we find differences up to 0.3 dex in the [Fe/H] values assigned for the M dwarfs in our calibration sample and the values used to derived the photometric scales. For example, B05, SL10 and N11 assign a [Fe/H] value of -0.15 dex to Gl 250 B, while the SPOCS [Fe/H] value we adopt for this star in our own calibration sample is +0.14 dex\footnote{We note that as this paper went to press, \citet{2011arXiv1112.0141O} reported a third metallicity for Gl 250B of -0.05 dex, based on an analysis of high-resolution ($R\sim$50,000) high signal-to-noise J band spectrum of Gl 250 B itself.}.  As the metallicity calibrations were derived with reference to these [Fe/H] values, the choice of which set of [Fe/H] values to adopt as a benchmark will necessarily bias the comparison towards the calibrations derived from those [Fe/H] values.  As a result, a truly fair comparison of the photometric and spectroscopic techniques will only be possible once the calibration sample has been significantly enlarged, and the veracity of the [Fe/H] values adopted for the calibration sample has been well established.  

Given the limitations outlined above, we conclude that the fairest comparison of the spectroscopic and photometric techniques that can currently be made is to qualitatively compare the internal quality of each fit, as benchmarked by its native calibration sample, as well as to compare the size and character of the calibration sample from which each fit was derived.  Table \ref{rms_r2a} lists the $rms$, $RMSp$ and $R$$^2_{ap}$ values for each calibration technique, along with the number of predictors $p$ and sample size $N$ for each model. The uncertainties were estimated using the bootstrapping method described in N11. 

The number of calibrators and the dispersion around the mean value evaluate the quality of the calibration sample that each technique is based on.  Large samples (large $N$) that cover a large range of values (large dispersion) are preferred. According to Table \ref{rms_r2a}, the B50 calibration sample is the largest ($N$$=$48), and also the most diverse ($rms$=0.483), followed by our calibration (RA11), SL10 and N11, respectively. The JA09 calibration sample has a very low $rms$ value since its composed only by high [Fe/H] stars.

The $RMSp$ and $R$$^2_{ap}$ statistics evaluate the internal quality of each functional fit; they describe how well the values predicted by each fit match the measured for the calibrator stars that were used to construct the fit. Small $RMSp$ values (small residuals) and $R$$^2_{ap}$ values closer to 1 are preferred (see Appendix \ref{statistics_appendix}). The JA09 fit has the smallest $RMSp$, but its $R$$^2_{ap}$ has a large uncertainty. N11 noted that the $R$$^2_{ap}$ is a noisy statistical estimator for small samples, however, and the JA09 fit only has 6 stars in its calibration sample. The B05 fit has the largest $R$$^2_{ap}$. 

In summary, the B05 calibration has the better statistics when its native calibration sample is used. However, the B05 calibration underestimates M dwarf metallicities, probably a result of using \textit{old} visual magnitudes, instead of Johnson V-band magnitudes, and the lack of high [Fe/H] M dwarfs in their calibration sample. The JA09 fit has a very small calibration sample composed only of high metallicity M dwarfs, and therefore, not a reliable fit for [Fe/H] prediction for M dwarfs. We find that the RA11, SL10 and N11 fits have calibration samples of similar size, similar $rms$ and $RMSp$ values; our calibration has the highest $R$$^2_{ap}$ values of the three. However, as noted earlier, none of these statistics can evaluate the veracity of the input values of the calibration sample, which determines the accuracy of the predicted values by the functional fit, as the recalibration of the B05 fit demonstrates.  Constructed an enlarged calibration sample, and demonstrating the veracity of the adopted [Fe/H] estimates, are key steps to enable a full and detailed comparison of the reliability of these techniques.

\subsubsection{K band [Fe/H] Estimates Compared with Other [Fe/H] Techniques}

Supersolar metallicities for most of the M dwarf planet hosts are predicted by the K band [Fe/H] fit as well as the photometric calibrations by \citet{2009ApJ...699..933J} and \citet{2010A&A...519A.105S}. While these three methods agree on the systematic metal-richness of planet hosts, the [Fe/H] values they predict do differ for each of the planet hosts (Table \ref{planettable2}).

Photometric [Fe/H] methods need accurate V magnitudes and distances, to calculate absolute K$_S$ magnitudes, and thereby infer the metallicities of M dwarfs from the M$_V$ vs V-K$_S$ color-magnitude diagram. Forty eight of the M dwarfs in our sample have accurate Johnson-V magnitudes and parallaxes by HIPPARCOS \citep{2007A&A...474..653V}, and K$_S$ magnitudes by 2MASS \citep{2006AJ....131.1163S}.  By including the M dwarfs with Johnson V magnitudes and parallaxes in the Yale Trigonometric Parallaxes Catalogue \citep{2001yCat.1238....0V}, we can assemble a sample of ninety stars with the necessary data to compare the results from the new K band [Fe/H] fit and photometric [Fe/H] relations.

Table \ref{comptable} lists the ninety M dwarfs with Johnson-V magnitudes and parallaxes, with their [Fe/H] values obtained from the photometric calibrations and the K band [Fe/H] fit in Equation \ref{ironmet}.  Since the \citet{2011arXiv1110.2694N} photometric calibration is a marginal refinement of the \citet{2010A&A...519A.105S} calibration, we only included the later in this comparison. Eight-six of the stars in Table \ref{comptable} are also included in the Palomar-Michigan State University Nearby Star Spectroscopic Survey \citep[PMSU;][]{1997yCat.3198....0R}, and have measurements of the TiO and CaH feature strengths from their optical spectra. \citet{2009PASP..121..117W} created a spectroscopic optical technique based on these indices to estimate [Fe/H] metallicities for early M dwarfs. The [Fe/H] values predicted by the \citet{2009PASP..121..117W} technique from the PMSU indices are also listed in Table \ref{comptable}. The value of [Fe/H] is listed for a star in Table \ref{comptable} only if the parameters to calculate it satisfy the conditions of each technique:

\begin{itemize}

\item \citet[][hereafter B05]{2005A&A...442..635B}: The [Fe/H] photometric relation by B05 is valid for stars with 4 $\leq$ M$_K$ $\leq$ 7.5, 2.5 $\leq$ V-K$_s$ $\leq$ 6, and -1.5 $\leq$ [Fe/H] $\leq$ +0.2.

\item \citet[][hereafter JA09]{2009ApJ...699..933J}:  The [Fe/H] photometric relation by JA09 is valid for stars with 3.9 $\leq$ V-K$_s$ $\leq$ 6.6 and [Fe/H] $\geq$ -0.05 dex. However, we decreased the [Fe/H] lower limit to -0.12 dex, which corresponds to the extrapolation done for the star Gl 832 in JA09.

\item \citet[][hereafter SL10]{2010A&A...519A.105S}: The [Fe/H] photometric relation by SL10 does not explicitly mention any conditions. However, we only estimated [Fe/H] values for that stars with 3 $\leq$ V-K$_s$ $\leq$ 7, the color range in the color-magnitude diagrams in SL10.

\item \citet[][hereafter W09]{2009PASP..121..117W}:  The [Fe/H] relation based on TiO and CaH molecular absorption by W09 depends on the $\zeta_{TiO/CaH}$ parameter by \citet{2007ApJ...669.1235L}. The W09 calibration is valid for stars with -1.5 $\leq$ [Fe/H]$\leq$ +0.05 and 3500K $\leq$ T$_{\mathrm{eff}}$ $\leq$ 4000K. We used the stars with spectral types earlier than M3. The $\zeta_{TiO/CaH}$ parameter categorizes the main sequence M stars into 4 classes ordered by decreasing metallicity: dwarfs (dM), subdwarfs (sdM), extreme subdwarfs (esdM), and ultrasubdwarfs (usdM). The metallicity class for each star is also listed in Table \ref{comptable}.

\end{itemize}

\begin{figure}[htp]
\begin{center}
\includegraphics[scale=0.6]{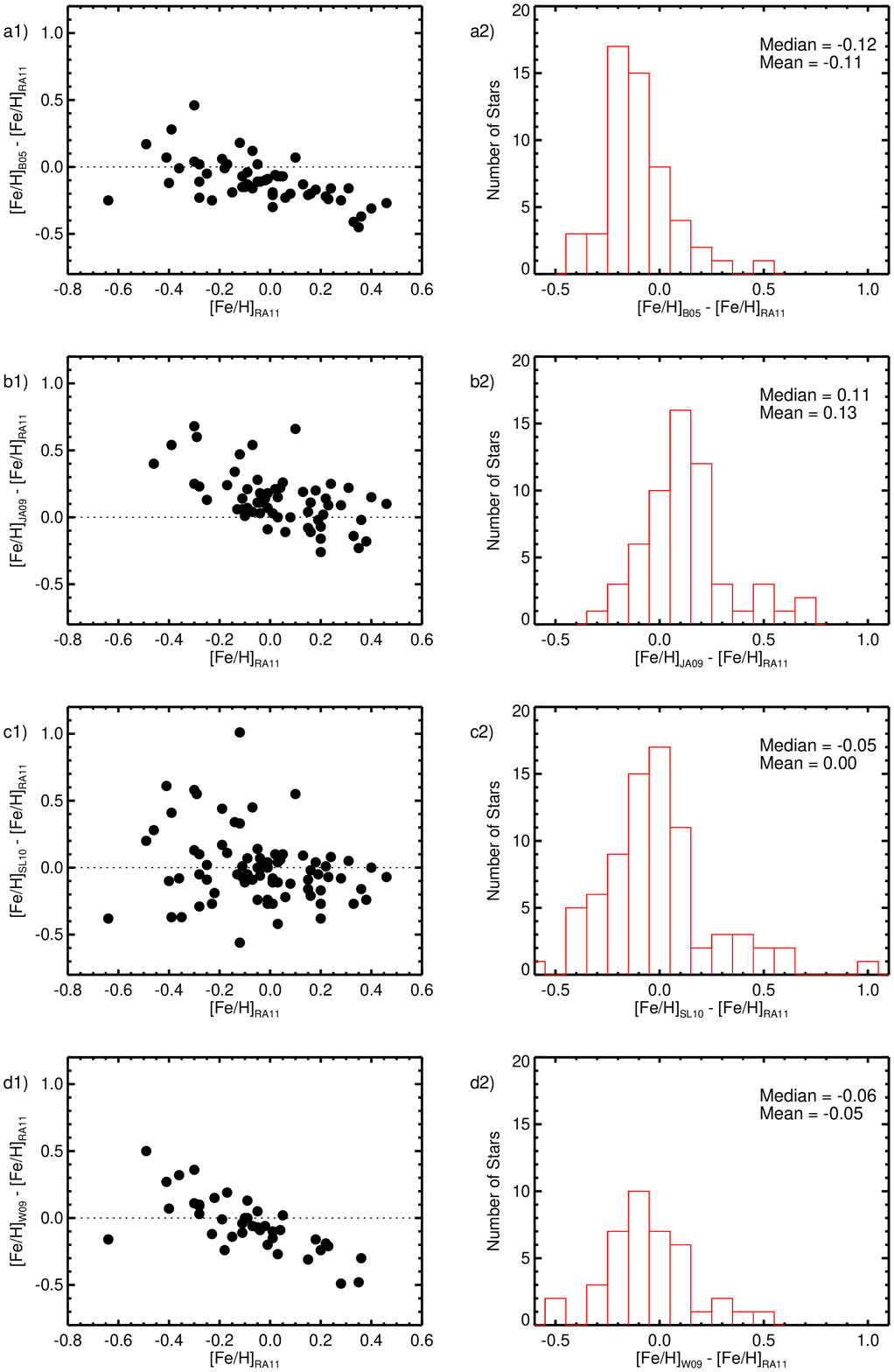}
\caption{Differences between the [Fe/H] values predicted for M dwarf stars in our sample from the K band [Fe/H] fit in Equation \ref{ironmet} and a1) the [Fe/H] from B05, b1) the [Fe/H] from JA09, c1) the [Fe/H] from SL10, d1) the [Fe/H] from W09. The B05 and W09 calibrations underestimate [Fe/H] values of the supersolar K band [Fe/H] stars. The JA09 calibration overestimates the [Fe/H] values of the subsolar K band [Fe/H] stars. The scatter of the residuals increases for the subsolar K band [Fe/H] stars when their [Fe/H] are compared with the SL10 [Fe/H] values.}
\label{res_compiron}
\end{center}
\end{figure}

Figure \ref{res_compiron} shows the difference between the [Fe/H] values obtained by Equation \ref{ironmet}, and the metallicity methods listed before. 

The B05 and W09 relations assign lower [Fe/H] values to the stars with solar and supersolar metallicities according to the K band [Fe/H] calibration. The stars with [Fe/H] differences within 0.15 dex have K band [Fe/H] metallicities between $\sim$ +0.1 dex and $\sim$ -0.5 dex, and correspond to 50$\%$ and 58$\%$ of the stars in the B05 and W07 samples, respectively. 

The JA09 relation assigns higher [Fe/H] values to the metal-poor stars in the K band [Fe/H] calibration. The stars with 0.15 dex [Fe/H] differences (i.e. $\sim$1$\sigma$ of the K band calibration) have K band [Fe/H] metallicities between -0.3 dex to +0.4 dex, and represent the 55$\%$ of the stars with JA09 metallicities. 

The SL10 relation [Fe/H] values agree within 0.15 dex with the K band [Fe/H] calibration for a 56$\%$ of the stars with SL10 metallicities. However, the dispersion for metal-poor stars is larger than for the stars with high K band [Fe/H] values, when compared to the SL10 technique. The SL10 technique ``underestimates" the values of some of the solar and supersolar K band [Fe/H] stars, but for low K band [Fe/H] values, it can either underestimate or overestimate that value by up to 0.6 dex.  GJ 1116 A (M6) has a K band metallicity of [Fe/H]$_{RA11}$ = -0.12 dex, but the SL10 method predicts a value of [Fe/H]$_{SL10}$ = +0.89 dex. The SL10 [Fe/H] value for GJ 1116 A is very unlikely and it may be a result of bad V photometry or a bad parallax measurement.

All the plots in Figure \ref{res_compiron} show a similar slope, except the  [Fe/H]$_{SL10}$ - [Fe/H]$_{RA11}$ vs. [Fe/H]$_{RA11}$ plot. The W09 relation was calibrated using the $\zeta_{TiO/CaH}$ parameter and metallicities inferred from \ion{Fe}{1} and \ion{Ti}{1} abundances of K-dwarfs and early M dwarfs from their high resolution spectra \citep{2005MNRAS.356..963W}. B05 acknowledge in their conclusion that their [Fe/H] values are consistent (where the paramenter spaces overlap) with the [Fe/H] values from \citet{2005MNRAS.356..963W}. Since B05 and W09 results agree, it is not surprising that both plots look quite similar. The JA09 calibration depends on the same observables as the B05 calibration, though supersolar metallicity stars were used as [Fe/H] calibrators for the fit. This could explain why the JA09 plot has the same slope as the B05 and W09, but with an offset towards higher metallicities. The observables in B05 were also used in the SL10 calibration, however, the kinematics of the FGK-calibrators and theoretical models were considered in the SL10 calibration. For the supersolar metallicity stars, the SL10 shows just an offset towards lower metallicities when compared to the RA11 calibration. For the metal-poor stars, the scatter is larger but there is not a distinct slope between SL10 and RA11.

\subsection{Inferring potential optical metallicity indicators}\label{optical_metallicity_indicators}

While M dwarfs are bright in the NIR, historically the optical regime has received more scrutiny as a means of studying M dwarfs. The molecules CaH, TiO, CaOH, and VO  cover the spectral range between 6300-9000 $\AA$. TiO absorption bands are known to be metallicity dependent  \citep[e.g.][]{2006PASP..118..218W}. Titanium oxide bands grow stronger with increasing metallicity as well as decreasing temperature. Hydride absorption bands also increase in strength with decreasing temperature, but, since the number density of neutral hydrogen increases as metallicity decreases, they grow stronger with decreasing abundance, too. Therefore, hydride absorption bands dominate the spectra of metal-poor stars, and the ratio between TiO and CaH has been used as a discriminant between disk and halo M stars \citep[e.g. the subdwarf classification by][]{1997AJ....113..806G}.

One hundred and nine M dwarfs studied in our sample are part of the Palomar/Michigan State University Nearby Star Spectroscopic Survey \citep[PMSU,][]{1995AJ....110.1838R,1996AJ....112.2799H}. The PMSU survey obtained optical spectroscopy of most of the M stars in The Third Catalogue of Nearby Stars \citep{1991STIA...9233932G}, allowing spectral types to be estimated from TiO band-strengths, chromospherically active stars to be identified by H$_\alpha$ emission, and metal-poor stars to be identified from the CaH strength relative to TiO strength. The overlap between our sample and the PMSU catalog allow us to test if a M dwarf's metallicity influences the spectral type obtained by the KHM technique. 

\begin{figure}
\begin{center}
\includegraphics[scale=0.6]{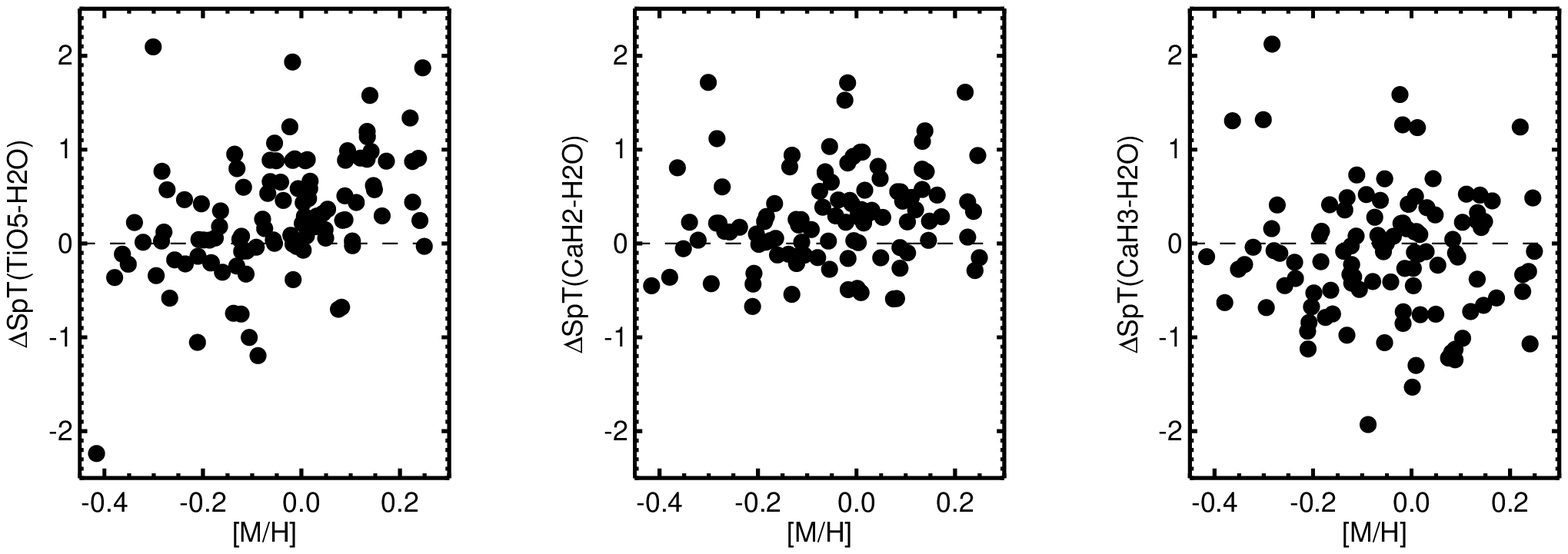}
\caption{Difference between the spectral types derived using the H$_2$O-K2 index and the TiO5 (left), CaH2 (middle), and CaH3 (right) indices versus [M/H]. The TiO5 index predicts later subtypes than the H$_2$O-K2 index for the solar and metal-rich stars, and earlier subtypes to the metal-poor stars. The CaH2 and CaH3 indices show less metallicity sensitivity. The CaH2 index predicts slightly later subtypes than the H$_2$O-K2 index, while the CaH3 index shows an opposite offset, but no bias with metallicity is visible for any of these two CaH indices.}
\label{scattplotsptype}
\end{center}
\end{figure}

\citet{1995AJ....110.1838R} defined nine narrow band spectroscopic indices (5 TiO indices, 3 CaH indices, and 1 CaOH index) to measure molecular features in the $~$6200-7200 $\AA$ wavelength range. Assuming that TiO band-strengths are primarily temperature dependent, \citet{1995AJ....110.1838R} measured the full strength of the TiO band at $\sim$7050 $\AA$ (the TiO5 index) for a large number of stars with KHM spectral types to calibrate TiO as a spectral type indicator. \citet{1997AJ....113..806G} used the \citet{1995AJ....110.1838R} index definitions to calibrate CaH2 and CaH3 indices as spectral type indicators, too. We calculated spectral types for the stars in our sample that are also in PMSU using the \citet{1995AJ....110.1838R} and \citet{1997AJ....113..806G} fit relations. The spectral types derived from the TiO5 and CaH features agree for most of the sample. The difference in spectral type predicted by the optical indices and H$_{2}$O-K2 index as function of overall metallicity is shown in Figure \ref{scattplotsptype}. The TiO5 index predicts later subtypes than the H$_{2}$O-K2 index for the solar and metal-rich stars, and earlier subtypes for the metal-poor stars. The CaH2 index predicts later subtypes than the H$_{2}$O-K2 index for most of the stars. The CaH3 index tends to give earlier subtypes than the H$_{2}$O-K2 index. However, considering their uncertainties, the spectral types derived from the TiO and CaH indices agree with the H$_{2}$O-K2 index based spectral types.

\begin{figure}[htp]
\begin{center}
\includegraphics[scale=0.45]{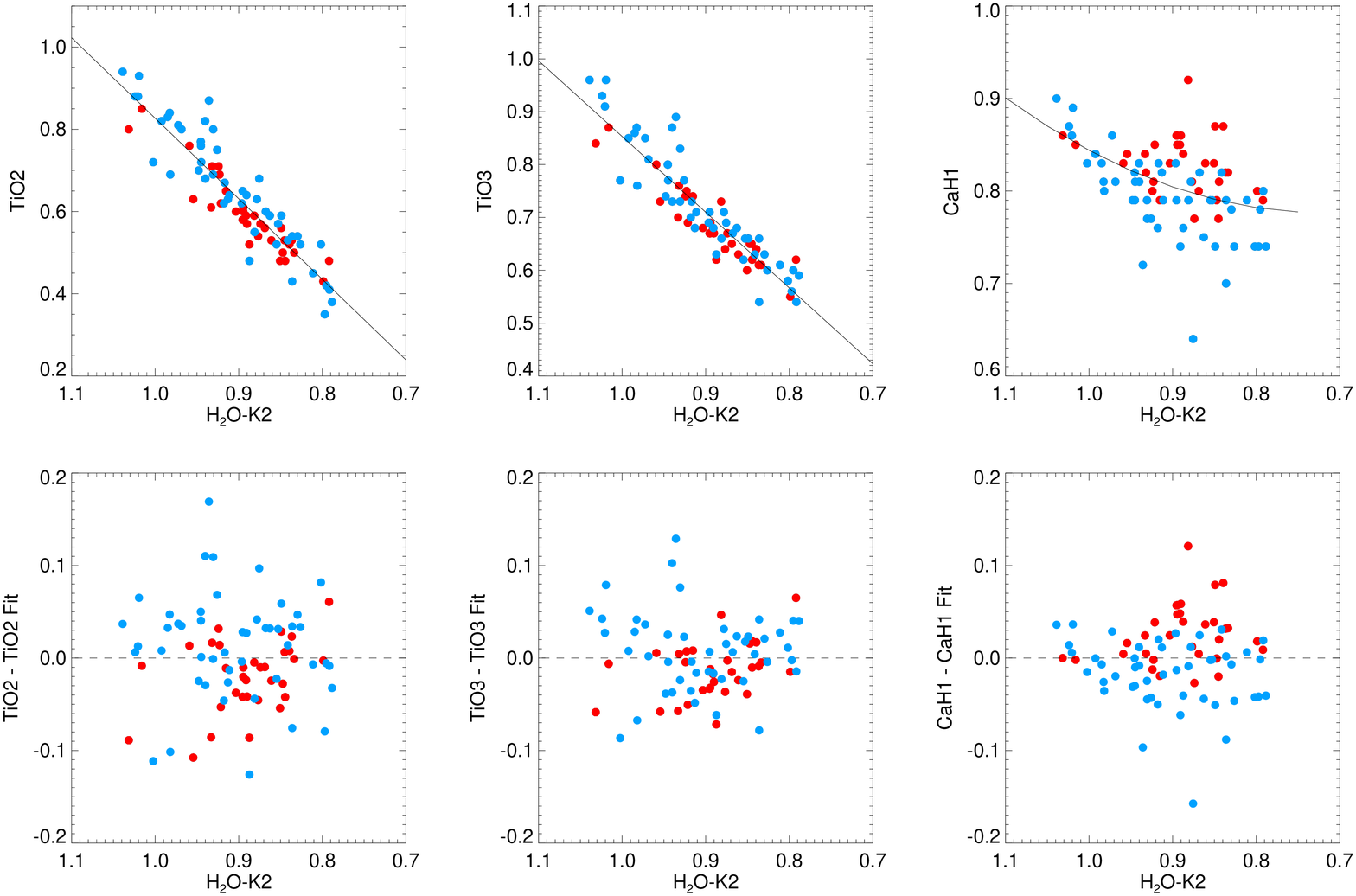}
\caption{The PMSU TiO2, TiO3, and CaH1 indices versus the H$_2$O-K2 index . M dwarfs with [M/H] $>$ 0.0 dex and [M/H]$<$ 0.0 dex are color-coded with red and blue, respectively. The TiO indices correlate with the H$_2$O-K2 index, with larger scatter towards early type M dwarfs. The CaH1 index seems to be more sensitive to metallicity than any of the other indices.}
\label{metallicitypmsu}
\end{center}
\end{figure}

The values of TiO2, TiO3, and CaH1 indices versus the H$_{2}$O-K2 index of the stars in the PMSU survey are shown in Figure \ref{metallicitypmsu}.  There is a strong correlation between the water index and the TiO indices up to H$_{2}$O-K2 $\sim$ 0.8. For spectral types later than M6, VO decreases the strength of the TiO features at $\sim$7050 $\AA$, and therefore, reverses the relation. The CaH1 index saturates at TiO5 $\sim$0.49 and, hence, was not used as a spectral type indicator for the optical calibrations by  \citet{1995AJ....110.1838R} and \citet{1997AJ....113..806G}. \citet{1995AJ....110.1838R} noticed that even when CaH1 was weaker that the CaH3 index, it appeared to be more sensitive to ``(presumably) abundance variations amongst the earlier type M dwarfs". Figure \ref{metallicitypmsu} has M dwarfs with [M/H] $>$ 0.0 dex and [M/H]$<$ 0.0 dex color-coded with red and blue, respectively. The considerable dispersion in the CaH1 vs H$_{2}$O-K2 relation can be explained by the metallicities of the stars. The CaH1 index saturates at $\sim$M3, due to the increasing TiO absorption in that region of the optical spectrum. Since both molecules, CaH and TiO, increase in strength towards lower temperatures, a ratio between these two molecules could be what CaH1 is really measuring. If a given star has low metallicity, the TiO absorption is not going to be as significant as the CaH, for any given temperature. The CaH1 feature will remain strong for low metallicities, which will correspond to low CaH1 index values. But, if the star is metal-rich, the pseudo-continuum used to estimate the CaH1 index is going to be affected by the strong TiO absorption. The CaH1 feature will become less prominent,  corresponding to high CaH1 index values. There is a similar split between the metal-rich and metal-poor M dwarfs for the TiO2 and TiO3 relations, but it is less significant. The CaH1 feature could be used as a metallicity discriminator for early solar-metallicity M dwarfs in the optical.

\section{Independent Tests of the K band [Fe/H] Fit: M-M Binaries, UVW Space Motions and H$\alpha$ Activity}

\subsection{Do M-M Binaries Return Similar Metallicities?}

\begin{figure}[htp]
\begin{center}
\includegraphics[scale=0.8]{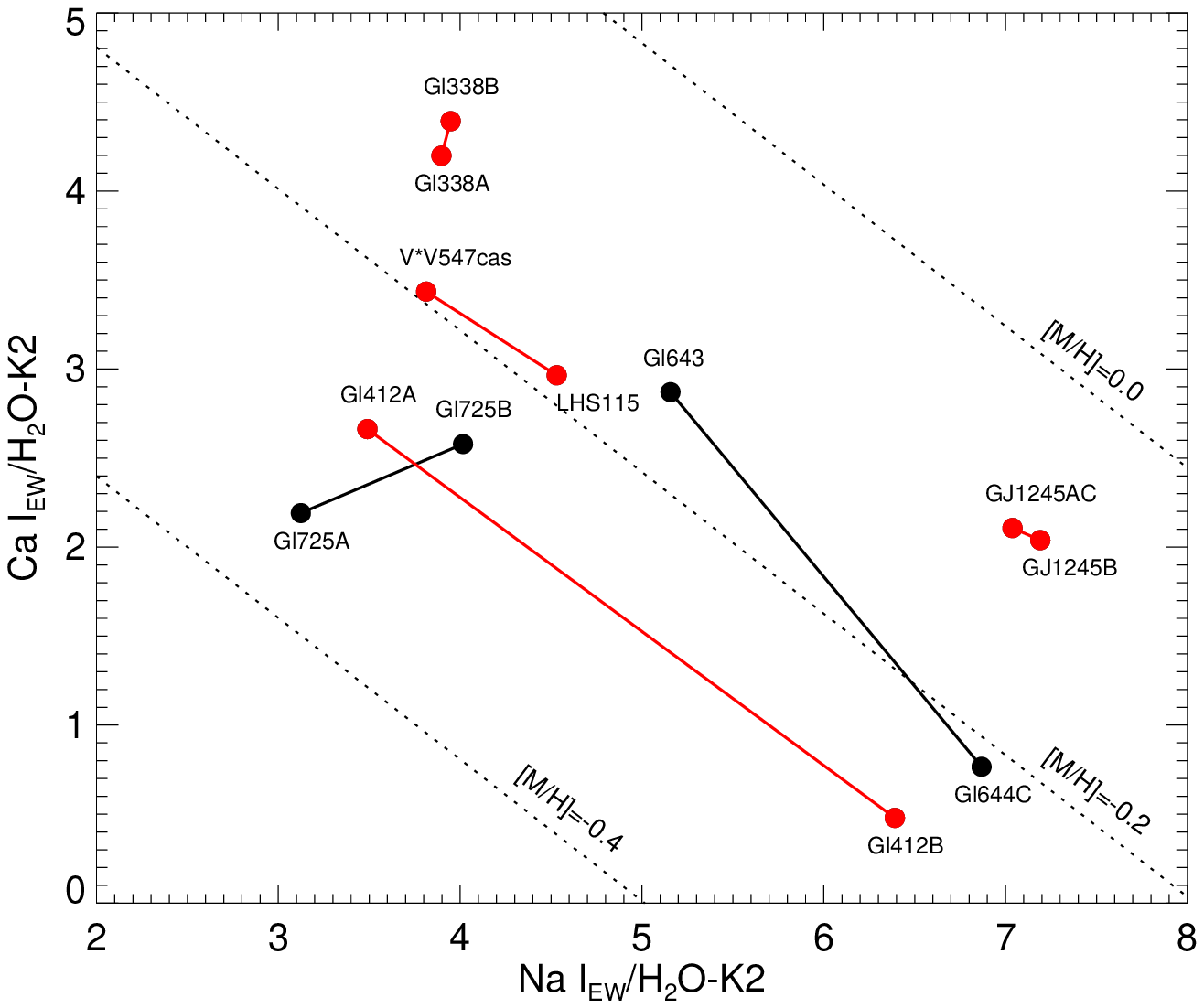}
\caption{EW of \cai vs EW of \nai, weighted by H$_2$O-K2 index, for the M-M binaries in Table \ref{tab_binaries}. The binary components are linked with lines. The dotted lines are isometallicity contours for [M/H] calculated from Equation \ref{overmet}. The red dots/lines indicate the systems where the K band metallicity estimates for the individual components agree to within 0.02 dex.}
\label{scatterbinaries}
\end{center}
\end{figure}

The K band [Fe/H] calibration and the photometric methods rely on the assumption that stars in binary systems share the same metallicity, as a result from being formed from the same interstellar material and at the same time. This assumption, which allows an M dwarf secondary's metallicity to be inferred from measurements of its FGK primary has been validated for binaries with multiple solar type components \citep{2004A&A...420..683D,2006A&A...454..581D}, and implies that any metallicity estimation technique should return consistent values for two M dwarfs in a wide multiple system.  As a sanity check, therefore, we inspect the [Fe/H] values predicted by each technique for the individual components of double M dwarf wide binary systems in our sample.  Table \ref{tab_binaries} shows the K band and photometric metallicities of 5 M-M binary or multiple systems in our sample, along with their NIR spectral types; Gl 643 and Gl 644C are also included as a pair, since both are proper motion companions to the Gl 644AB pair/triplet according to \citet{2000A&A...364..665S}. Photometric estimates cannot be made for all the stars in the systems since their photometry and/or parallaxes do not satisfy the conditions listed in Section \ref{comparisonwphoto} for each calibration. We also show these systems in Figure \ref{scatterbinaries}, where we connect the individual components of a multiple system with dotted lines to examine how well each system agrees with the slope of our derived iso-metallicity contours. 

The K band [Fe/H] and [M/H] estimates for individual components of (non-calibrator) binary systems agree remarkably well; four of the six systems have [Fe/H] and [M/H] metallicity differences of less than 0.02 dex, and none of the six have individual metallicity estimates that disagree by more than 0.13 dex, comparable to the 1$\sigma$ accuracy of the underlying calibration itself.  Indeed, the largest discrepancy for the K band calibration is for the Gl 725AB pair, whose 0.09 dex and 0.13 dex [M/H] and [Fe/H] differences are similar to the dispersion ($RMSE$) of each fit in Equations \ref{overmet} and \ref{ironmet}, respectively.  

The metallicity estimates predicted by the photometric techniques, however, do not reproduce this agreement: the photometric techniques predict metallicities for these individual binary components that typically differ by 0.1-0.2 dex, comparable to the single largest metallicity difference predicted by the K band technique. The Gl 412 system is the binary in our sample with the greatest spectral type difference, consisting of M1 and M6 components.  Despite the significant range in temperatures between these two stars, the metallicities predicted by the K band calibration only differ by 0.01 dex, suggesting our calibration is not strongly biased with respect to T$_{eff}$. The GJ1245 system provides another a good case study: both components have the same spectral-type in the optical and in K band, and their K band spectra share a very similar morphology, so it is perhaps not surprising that the metallicity estimates predicted by the K band calibration for the individual components agree quite well ($\Delta$ RA11 [Fe/H] = 0.01). The photometric [Fe/H] estimates for each component, however, disagree much more ($\Delta$ [Fe/H] = 0.25-0.3), reflecting underlying differences in the source photometry: while the distance measurements for GJ 1245 AC and GJ 1245 B agree well ($\delta$ d = 0.2 pc), their photometry does not, with GJ 1245 AC appearing $\sim$0.55 mag brighter in both V and K$_s$, placing it significantly higher in the M$_V$ vs. V-K$_s$ diagram and thus implying a more metal-rich composition via the photometric technique. \citet{1988ApJ...333..943M} resolved each component of the GJ 1245 triple system with speckle interferometry and obtained K band photometry for each of its components. \citet{1988ApJ...333..943M}'s photometric measurements revealed almost identical magnitudes for the A and B components, $\sim$1.1 mag brighter than C. However, the combined K magnitude for the GJ 1245 AC system was $\sim$ 0.5 mag brighter than GJ 1245 B, but the optical spectra for AC and B were almost identical. Therefore, the faint GJ 1245 C component could explain the photometric [Fe/H] discrepancy: it makes the star GJ 1245 A to appear brighter without affecting its optical and K band spectra. This implies that an unknown unresolved binary could appear to be more metal-rich under the photometric calibration if it is assumed to be a single star.  

\subsection{Do K band Metallicities Reproduce the Metallicity/Kinematic Correlations Expected for Galactic Sub-populations?}

It has been known for the better part of a century that there are clear correlations between the spatial, kinematic, and chemical properties of stars in the Milky Way.  These correlations allow the Milky Way to be decomposed into distinct stellar populations and structural components \citep[e.g., halo vs. disk;][]{Oconnell1958}, although some debate remains concerning the exact number and properties of sub-components \citep[e.g., thin vs. thick disk;][]{Ivezic2008}.  \citet{2010IAUS..265..300B} recently showed that when 899 F- and G- dwarf stars were separated by their kinematic properties, a considerable number of thick disk stars exhibit metallicities consistent with the metallicity trend of the thin disk, and vice versa. When the solar-type stars were separated by age rather than kinematics, however, the number of outliers in each metallicity trend was reduced. 

Age estimates for M dwarf stars are difficult to obtain, however, making kinematic properties the best available means for associating low-mass stars with various Galactic components.  \citet{2007AJ....134.2418B} investigated metallicity differences between M dwarfs associated with the thin and thick disk populations, bisected kinematically using the method described by \citet{2003A&A...410..527B}. \citet{2007AJ....134.2418B} measured the (CaH2+CaH3)/TiO5 ratio, previously shown by \citet{2003AJ....125.1598L} to roughly discriminate between solar-metallicity ([M/H]$\sim$0), subdwarf ([M/H]$\sim$-1.2) and extreme-subdwarf ([M/H]$\sim$-2.0) M dwarfs, for 6577 M dwarfs with Sloan Digital Sky Survey (SDSS) spectra.  \citet{2007AJ....134.2418B} found that the observed (CaH2+CaH3)/TiO5 distributions for the kinematically divided thin and thick disk stars did not differ greatly, implying metallicity differences less than 1 dex, but suggested that the kinematically selected thick disk stars were consistent with an older population due to the lower fraction of stars with \ha  emission, a tracer of chromospheric activity. 

\begin{figure}[htp]
\begin{center}
\includegraphics[scale=0.45]{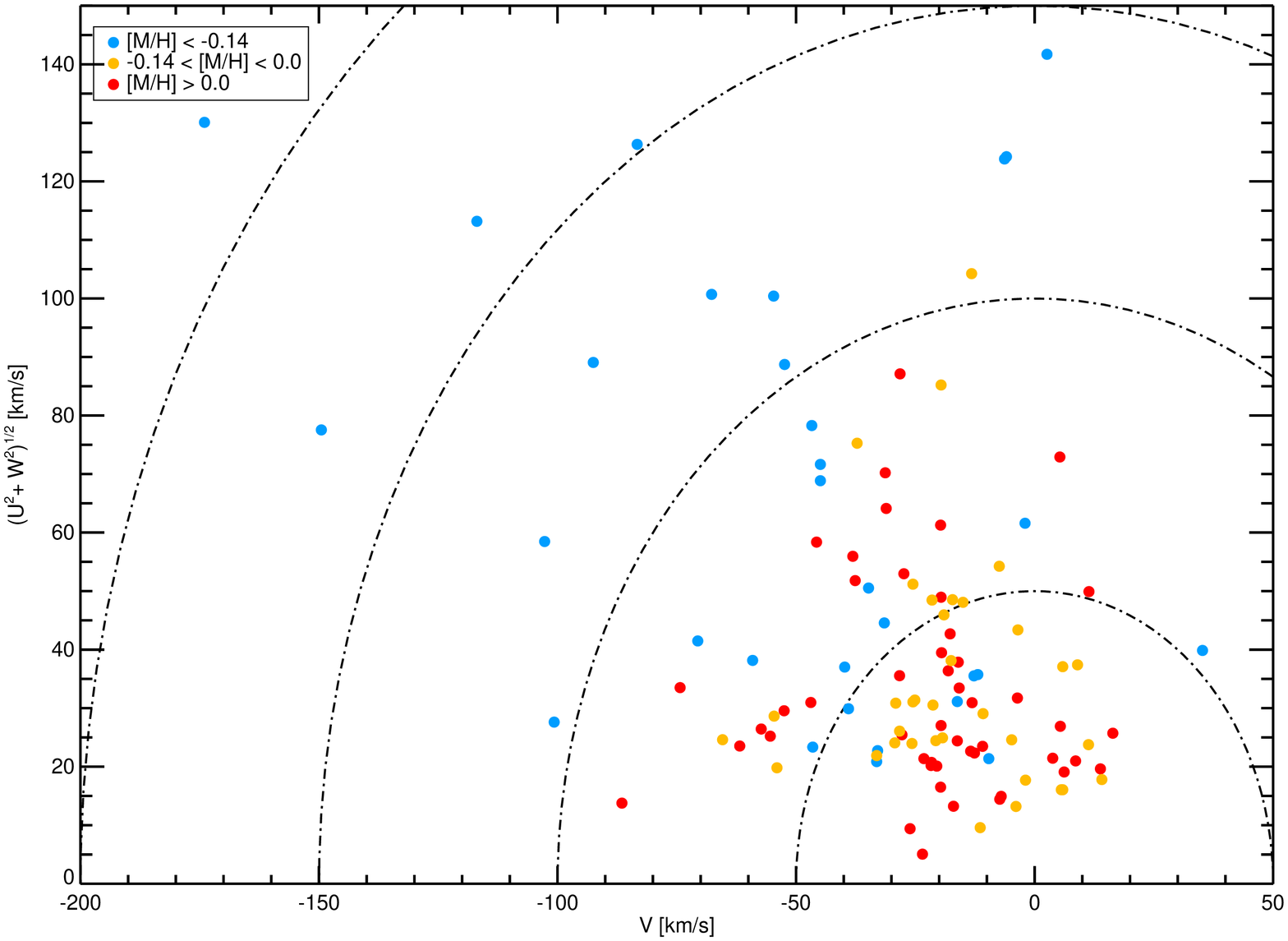}
\caption{Toomre diagram for the stars in our sample with PMSU kinematics, color-coded by metallicity. The metal-poor stars ([Fe/H]$<$-0.14 dex), depicted by blue dots, cover the whole velocity space, while the metal-rich stars ([Fe/H]$>$0.0 dex) exhibit V$_{tot}$ velocities no larger than $\sim$100 km/s.}
\label{toomre_met}
\end{center}
\end{figure}

\begin{figure}[htp]
\begin{center}
\includegraphics[scale=0.25]{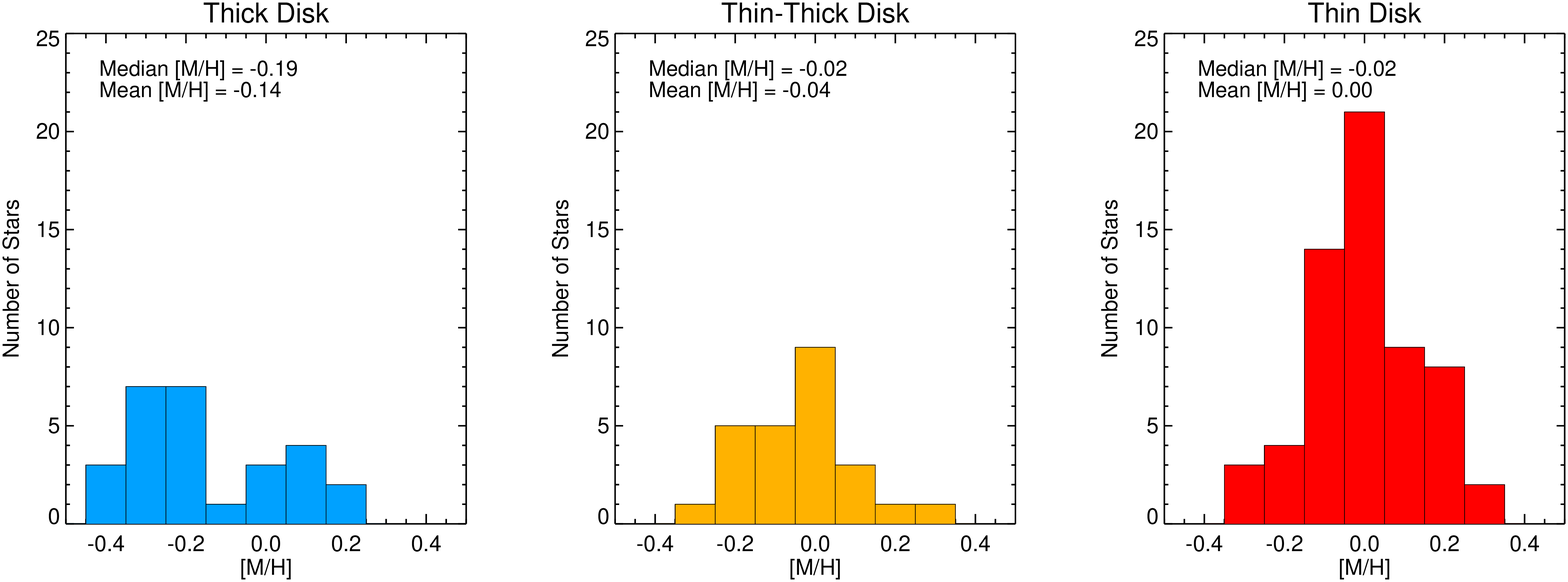}
\caption{K band metallicity distributions for the stars in our sample with PMSU kinematics.  The boundaries of these kinematic sub-groups are: thin disk, V$_{tot}$$<$50 km/s; thin-thick, 50 km/s $<$V$_{tot}$$<$70 km/s; thick disk, 70 km/s $<$V$_{tot}$$<$200 km/s. The metallicity distributions of these kinematically selected subgroups are consistent with the trends seen for solar-type stars: stars with large space motions are preferentially metal-poor, while stars with small space motions are preferentially metal-rich.}
\label{distributiontoomre}
\end{center}
\end{figure}

One hundred fourteen M dwarfs studied in this work have UVW space motions from the Palomar-Michigan State University Nearby Star Spectroscopic Survey \citep[PMSU;][]{1997yCat.3198....0R}, enabling us to test if the K band metallicity calibration can replicate the metallicity trends identified for higher-mass members of the kinematic sub-populations of the Milky Way.  One way to visualize the different kinematic populations of the Milky Way is through the so-called Toomre diagram\footnote{\citet{1987AJ.....93...74S} were the first to show this diagram in their paper on the collapse of the Galaxy, but acknowledged the diagram to A. Toomre [1980]}. The Toomre diagram is a representation of the stars' combined vertical and radial kinetic energies, (U$^{2}$ + W$^2$)$^{1/2}$, as a function of the stars' rotational energy, V. Constant values of the total space velocity for the stars, V$_{tot}$= (U$^{2}$ + W$^2$ +V$^2$)$^{1/2}$, can be represented by circles centered at [0,0], as is shown in steps of 50 km/s in Figure \ref{toomre_met}. We base our kinematic sub-divisions on the boundaries defined by \citet{2010IAUS..265..300B}, where thin disk stars possess V$_{tot}$$<$50 km/s and thick disk stars possess $\sim$70 km/s $<$V$_{tot}$$<$200 km/s; we additionally classify stars with intermediate velocities (50 km/s $<$V$_{tot}$$<$70 km/s) as thin-thick stars.  We show in Figure \ref{distributiontoomre} the distributions of K band overall metallicity estimates for members of each of these kinematics groups, with the mean and median [M/H] indicated for each distribution.  Consistent with results seen for higher mass stars, we find the thin disk population is noticably enriched ($\Delta$ [M/H] $\sim$ 0.2 dex) relative to the thick disk population, and a two-sided K-S test identifies that a $<$0.003\% probability that the two samples are consistent with the same parent distribution. 

To further explore the differences between these populations, we show in Figure \ref{toomre_met} the Toomre diagram for the stars with kinematic measurements in the PMSU survey, color-coded by metallicity, with boundaries for the metallicity sub-groups chosen from the mean values of the metallicity distributions of Figure \ref{distributiontoomre}.  Visualizing the sample in this way, it is apparent that the metal-poor stars span a much larger range of velocities than the metal-rich stars: more than half the stars with [M/H] $<$ -0.14 have V$_{tot} >$ 70 km/s, while every star with [M/H] $>$ -0.15 lies within the V$_{tot}$$\sim$100 km/s contour.  These results reinforce the trends seen by \citet{2010IAUS..265..300B} for higher mass stars: \citet{2010IAUS..265..300B} found that older (t $>$ 8 Gyrs) stars are a kinematically hot sample that cover most of the velocity space in the Toomre diagram, just like the thick-disk M dwarfs in Figure \ref{toomre_met}.  \citet{2010IAUS..265..300B} also found kinematically hot stars within the younger (t $<$ 7 Gyrs) population, but the majority have V$_{tot}$$\leq$ 100 k/m, consistent with what we see in Figure \ref{toomre_met} for our (metallicity selected) thin-disk M dwarfs.

\subsection{Are K band Metallicities Consistent With an Assumed Age-Metallicity-Activity Relation?}

A final observable that can be used to probe the existence of an underlying age-metallicity relation is chromospheric activity.  Common tracers of chromospheric activity, such as emission in \ion{Ca}{2} H $\&$ K or H$_{\alpha}$, indicate that young stars demonstrate high levels of chromospheric activity, with the activity decaying as the stars age; this effect has been seen in solar type stars \citep{Skumanich1972} and well into the M dwarf regime \citep{Stauffer1986}. Chromospheric activity timescales of M dwarfs expand from $\sim$ 1Gyr for early-type/higher-mass stars, to $\sim$ 10 Gyr for the later-type/lower-mass stars \citep{Hawley1999,2008AJ....135..785W}.  

As noted in the previous section, \citet{2010IAUS..265..300B} found that the link between age and metallicity appears to be stronger for solar-type stars than the link between kinematics and metallicity.  While it is not possible to derive an exact age for an M dwarf based on its observed activity level, one can use the activity lifetimes noted above to statistically divide stars into older and younger samples.  \citet{2002AJ....123.3356G} measured the equivalent widths of \ha on high-resolution spectra of 676 M dwarfs, and identified distinct groups according to their \ha properties. Our sample includes eighty two stars analyzed by \citet{2002AJ....123.3356G}, providing an opportunity to use activity as a crude proxy for age, and examine the evidence for an age-metallicity relationship within the M dwarfs in our sample.  

\begin{figure}[htp]
\begin{center}
\includegraphics[scale=0.45]{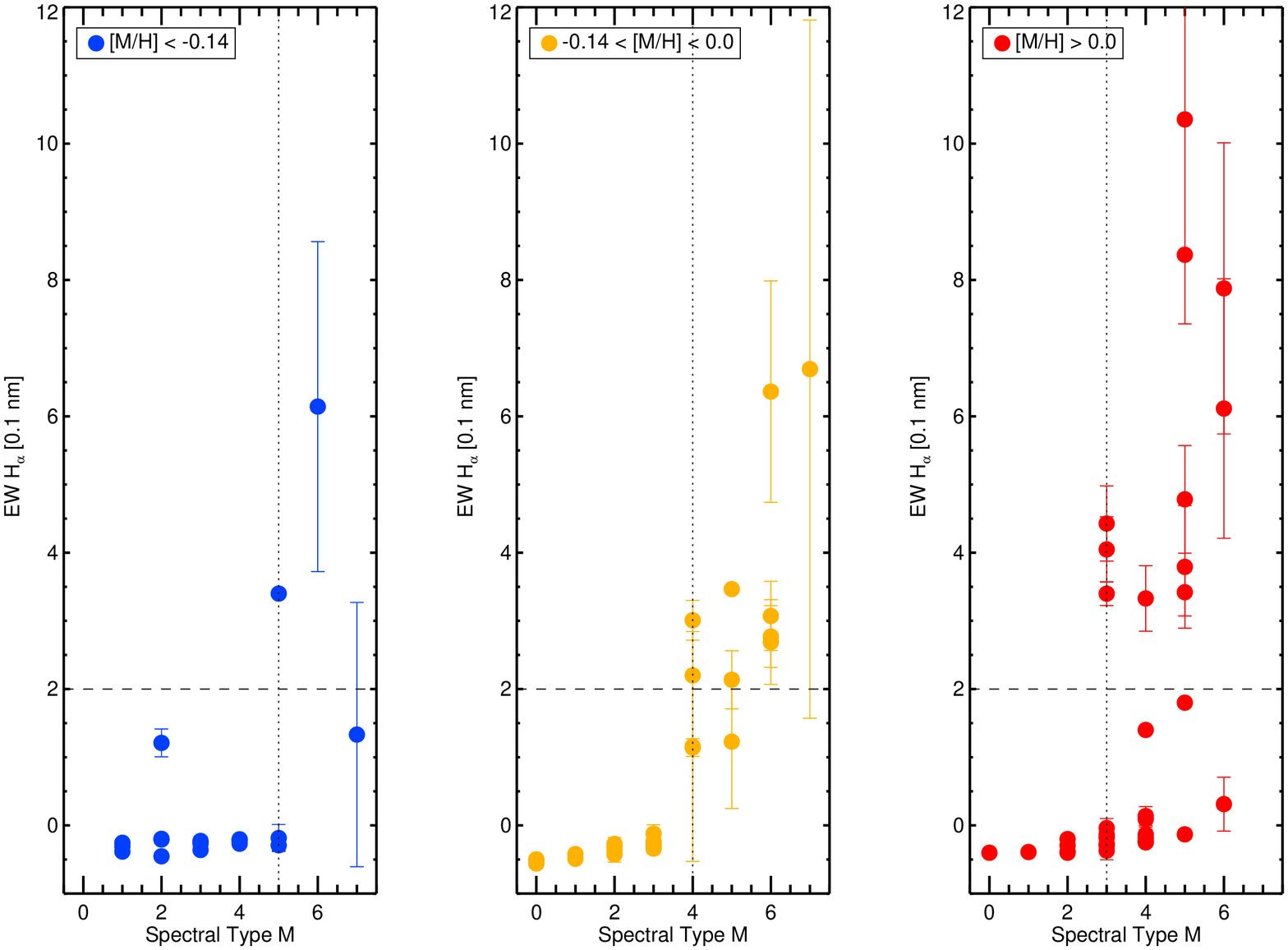} 
\end{center}
\caption{H$_{\alpha}$ strength as function of spectral type for the stars in our sample with H$_{\alpha}$ measurements by \citet{2002AJ....123.3356G}. The dotted-line represents a 2 \AA\ lower limit in \ha to define M dwarfs with significantly enhanced activity (i.e., more active than required to satisfy the 0.75 \AA\ limit adopted by \citet{2008AJ....135..785W}).  All late-type ($>$M5) stars are still active in all three metallicity categories, but activity persists to earlier types in the metal-rich samples, consistent with an underlying correlation between a star's age, activity, and metallicity. Chromospheric activity is noticeable for M3 stars with [M/H]$>$-0.01 dex, for M4 stars with -0.15$<$[M/H]$<$-0.01 dex, and for a M5 stars with  [M/H]$<$-0.15 dex. V$*$ V1513 Cyg is a metal-poor M2 dwarf that still exhibits chromospheric activity.}
\label{halpha_plot}
\end{figure}

 Figure \ref{halpha_plot} shows the \ha emission strength versus spectral type for these stars, divided according to the metallicity boundaries inferred from Figure \ref{distributiontoomre} and applied previously in Figure \ref{toomre_met}. Across all metallicity categories, most of the early-type (M0 to M2) stars show \ha absorption, implying either a weak/moderate activity or no activity at all, while the later-type stars show \ha emission, consistent with the longer activity lifetimes expected for the lowest-mass stars.  The metal-rich category seems to have a slightly higher average level of \ha emission when compared with the other categories, consistent with the metal-rich stars being a statistically younger sample than the more metal-poor samples. If we adopt a 2 \AA\ threshold for significant H$_{\alpha}$ activity, the earliest star in each category with elevated activity is clearly correlated with metallicity: the earliest active metal-poor star is an M5, while the earliest active intermediate and metal-rich stars are M4 and M3, respectively.  This correlation is precisely what would be expected if there were an underlying age-metallicity relation, such that the sample of metal-poor stars consists of statistically older stars, where activity is only seen for the lower-mass stars with relatively long activity lifetimes, while metal-rich stars are statistically young, with higher mass stars not yet having transitioned into the inactive stage.

\section{Particularly promising M dwarf planet hosts}

The nearly 600 confirmed extrasolar planets discovered to date reveal that planets are a natural and frequent by-product of star formation.  One of the most important observational constraints is that planet frequency rises steeply with host star abundance above solar metallicity in FGK-stars \citep[e.g. ][]{2004A&A...415.1153S,2005ApJ...622.1102F}. As mentioned in the introduction, the first photometric and spectroscopic metallicity measurements for M dwarf planet hosts indicated that they were slightly metal-poor \citep{2005A&A...442..635B, 2006ApJ...653L..65B}. However, recent metallicity calibrations \citep[][]{2009ApJ...699..933J,2010ApJ...720L.113R,2010A&A...519A.105S} have convincingly shown that the small sample of nearby M dwarf planet hosts have solar and super-solar metallicities.

\begin{figure}[htp]
\begin{center}
\includegraphics[scale=0.75]{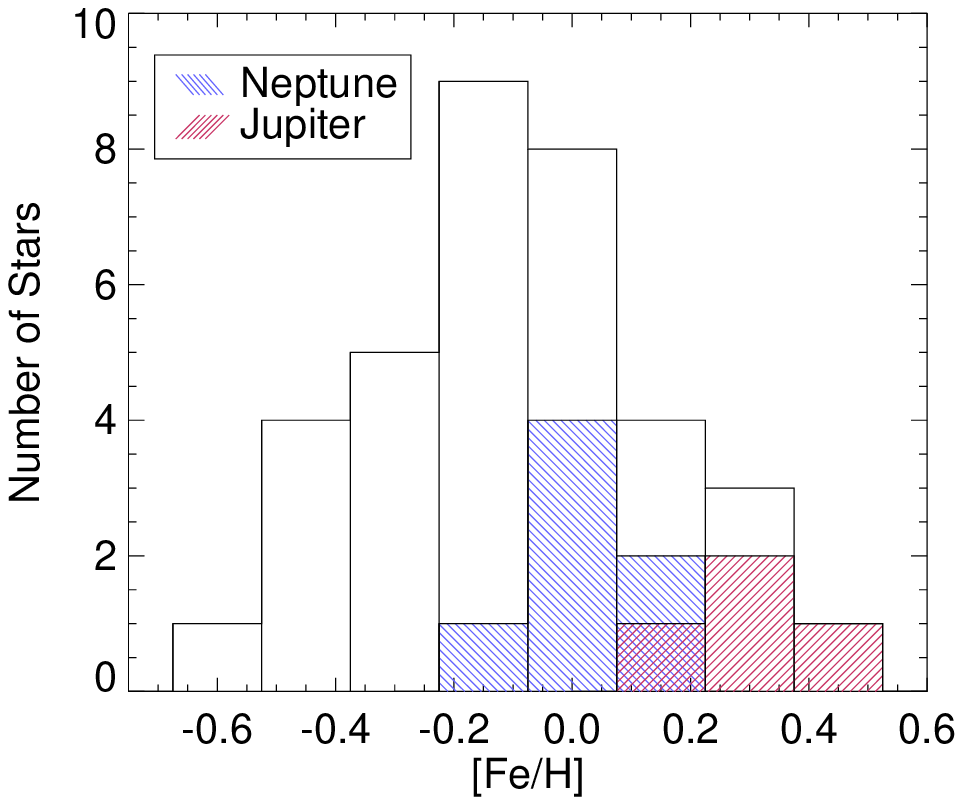}
\end{center}
\caption{[Fe/H] distribution of 35 M dwarfs whose RV measurements have rule out the presence of Jupiter-size planets within several AU by the CPS team, along with the [Fe/H] distribution of Jupiter hosts and Neptune hosts in our sample. }
\label{cpsdist}
\end{figure}

Table \ref{spectral_table} flags the M dwarf planet hosts in our sample. According to the metallicity calibration derived in Section 5.2, the eleven M dwarf planet hosts have overall metallicities higher than -0.06 dex, with the Jovian hosts being more metal-rich than the Neptune or super-Earth hosts. The sample of M dwarf planet hosts is small, but their K band metallicities suggest the existence of a planet-metallicity correlation for M dwarfs.  We compared the [Fe/H] distribution of the eleven planet hosts to the [Fe/H] distribution of 35 M dwarfs in our sample that are also members of the California Planet Survey (CPS) sample of low-mass stars (John Johnson, private communication). These  35 M dwarf have been subjected to an intensive RV monitoring campaign by the CPS team, and their RV measurements have ruled out the presence of Jupiter-size planets within several AU. Figure \ref{cpsdist} shows the metallicity distributions of the 35 CPS M dwarfs and of the 11 planet hosts in our sample. A two-sided K-S test identifies that there is a $<$1.6\% probability that the K band metallicities we calculate for these samples of M dwarfs with and without planets share the same parent distribution. If the planet sample is divided by Jupiter and Neptune/super-Earth hosts, then a two-sided K-S test gives a 0.8\% chance that the Jovian hosts and the 35 CPS stars are drawn from the same parent distribution. The probability that the Neptune and super-Earth hosts share the same [Fe/H] distribution as the 35 CPS stars without detected planets is much higher: 8.6\%. These results are in agreement with what is observed for FGK stars, where Neptunes and super-Earths, unlike Jupiters, do not seem to form preferentially around highly metal-rich ([Fe/H]$>$ 0.2 dex) stars \citep[][]{2008A&A...487..373S,2011arXiv1109.2497M}. However, we acknowledge that the small size of these samples prevents any conclusive result.

According to the [Fe/H] distribution of the planet hosts in our sample, M dwarfs with solar and super-solar metallicities should be the preferred targets to look for planets around cool stars. The most metal-rich stars ([M/H]$>$+0.2 dex) in this sample are the Jovian hosts HIP 79431 and Gl 849, and the M dwarfs Gl 285 (M5), G 203-47 (M4), Gl 169.1 A (M3), LHS 3799 (M5), and Gl 205 (M0). \citet{2009ApJ...701.1922B} used radio astrometry to search for planets around a sample of active M stars within 10 pc and their analyzed sample included two of the metal-rich stars listed above (Gl 285 and LHS 3799). \citet{2009ApJ...701.1922B} were able to rule out the existence of 3-6 M$_J$ planets within 1 AU at the 99$\%$ confidence level. \citet{2009ApJ...705...89L} searched for an infrared excess around Gl 205, and found none, excluding the presence of debris at distances beyond the snow-line.

\begin{figure}[htp]
\begin{center}
\includegraphics[scale=0.55]{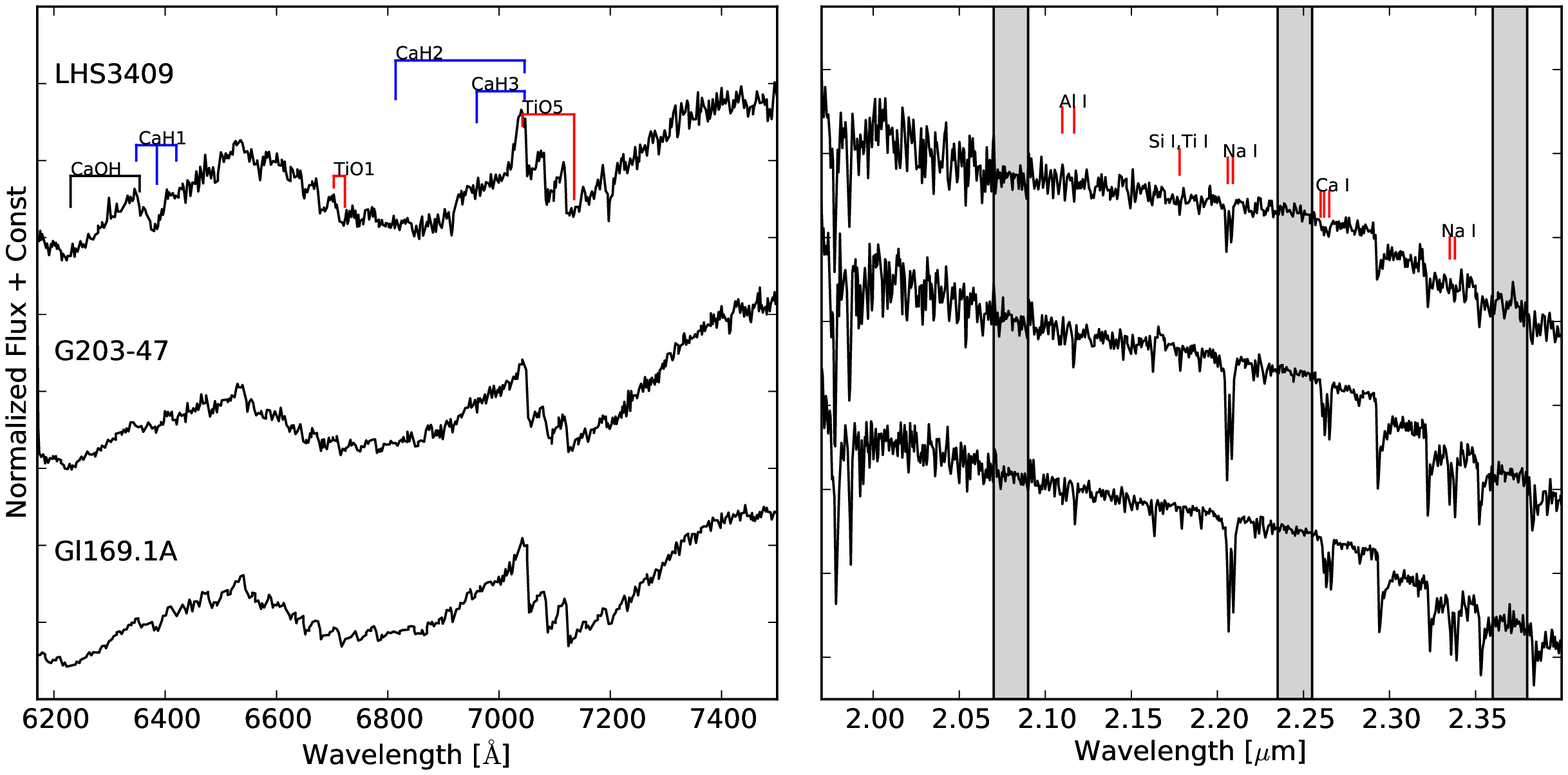}
\end{center}
\caption{Optical PMSU spectra and K band spectra of LHS 3409 (top), G 203-47 (middle), and Gl 169.1A (bottom).  The PMSU TiO and CaH indices are shown in red and blue, respectively.  The most prominent atomic features in the M dwarf K band spectra are also indicated.  The regions used to calculate the water index , H$_2$O-K2, are shown in gray.  G 203-47 and Gl 169.1A belong to binary systems with white-dwarf companions. LHS 3409, a metal-poor star ([M/H]=-0.36) of similar spectral type, is shown for comparison. The K band absorption features are weaker in the spectrum of LHS 3409, but the CaH absorption in its optical spectrum is stronger than in the 2 metal-rich stars, which indicates that G 203-47 and Gl 169.1A have supersolar metallicities.}
\label{wdfig}
\end{figure}

Interestingly, the metal-rich stars G 203-47 and Gl 169.1 A do not have planetary companions, but they do possess  white-dwarf (WD) companions. Gl 169.1 B is a relatively young and featureless white dwarf with an age estimate of $\sim$ 2.8 Gyr based on a model with a core composed of pure carbon \citep{1998ApJ...497..294L}. Gl 203-47 B does not have an age estimate, however \citet{1999A&A...344..897D} predicted that its primary (Gl 203-47) would have a peculiar composition since the system was likely in a contact configuration during the post main-sequence evolution of the current day WD component. The semi-major axis of the present orbit of the G 203-47 system is $\sim$ 0.05 AU \citep{1999A&A...344..897D}. Therefore, G 203-47 could have accreted nucleosynthesis products dredged up to the surface of Gl 203-47B's AGB progenitor. Figure \ref{wdfig} shows the optical PMSU spectra and K band spectra of G 203-47 and Gl 169.1 A, along with the spectra of LHS 3409, a star with similar spectral type but subsolar metallicity. The strong \ion{Na}{1} and \ion{Ca}{1} features in K band, and the weak CaH absorption in their optical spectra, indicates that these stars have indeed supersolar metallicities. 

Most of exoplanet searches have been performed at visible wavelengths. M dwarfs are intrinsically faint in the optical, so the largest stellar component in the solar neighborhood haven't yet been searched thoroughly for planets, except for the brightest, closest early M dwarfs. However, planet searches are now being performed with near-infrared radial velocity instruments \citep[e.g.][]{2010ApJ...713..410B,2011ApJ...735...78C,2011arXiv1103.0004M} and with far red filters for transits \citep{2009IAUS..253...37I}. For these searches, the visible faintness of M dwarfs is not a problem, but other intrinsic properties of M dwarfs should be considered to optimize detectability. 

Therefore, the most ``desirable" M dwarf targets for planet searches should be:

\begin{itemize}

\item Bright: The optical faintness of M dwarfs is not a problem for the near-infrared searches, but a large number of M dwarfs per night is desirable. Brighter M star are favorable for achieving a fixed signal-to-noise ration and for ease of follow-up for any planets that are detected.

\item Slow rotators: Stars with high rotation rates have radial velocity measurements with lower precision since stellar rotation widens the spectral features. \citet{2001A&A...374..733B} found that the radial velocity uncertainties increase by a factor of $\sim$4 when the rotation velocity increases from $\sim$1 km/s to 10 km/s. Stars with \vsini $>$ 10 km/s should be avoided to provide optimal RV sensitivity.

\item Inactive: The activity of M dwarfs can also cause false positives (e.g. starspot-crossing events; \citet{2011ApJ...730...82C}) and can cause radial velocity and photometric jitter \citep{2005wright,2011AAS...21734311B}. Depending on the starspot coverage, the planet-to-star radius ratio can change up to few percent \citep{2011ApJ...730...82C}. Stars with H$_{\alpha}$ emission should be avoided to avoid false positives and jitter.

\item Metal-rich: The M dwarf planet host shown in Table \ref{planettable2} all have metallicities [M/H]$\ge$ -0.05 dex. Therefore, it is probable that solar and supersolar metallicity M dwarfs are more likely to host planets.

\end{itemize}

Table \ref{desirabletargets} lists the stars that satisfy the conditions mentioned above for activity, rotation, and metallicity in our M dwarf sample. The \ha values are from \cite{2002AJ....123.3356G}. The \vsini  values were taken from \citet[][Tables 1 and 3]{2009ApJ...704..975J} and \citet{2010AJ....139..504B}. Some of the stars in Table \ref{desirabletargets} have detected stellar or planetary companions already (five of the eleven M dwarf planet hosts have their \ha and \vsini  listed in the papers mentioned above).  It will be interesting to see how many planets are discovered around the targets in Table \ref{desirabletargets} in the near future, and what  ``flavor" they might be.

\section{Conclusions}

We present K band spectra for a sample of 133 M dwarfs, including 18 M dwarfs with reliable metallicity estimates (as inferred from an FGK type companion), 11 planet hosts, more than 2/3 of the Northern 8 pc sample, and additional M dwarfs from the Lepine Shara Catalog. From these spectra, we measured EWs of the \cai and \nai lines, as well as an index quantifying the absorption due to H$_{2}$O opacity, features we have previously shown to be of use for predicting M dwarf temperatures and metallicities.

\begin{enumerate}

\item{From the subset of our stars which possess primary or secondary KHM spectral types, we calibrated the H$_2$O-K2 index as a spectral type indicator. We also estimated effective temperatures for the stars in the sample by interpolating their H$_2$O-K2 indices onto a solar metallicity [H$_2$O-K2, T$_{eff}$]  model grid.}

\item{We revise the functional form adopted to predict the [Fe/H] metallicities of M dwarfs based on measurements of their \nai, \cai, and H$_{2}$O features.  We perform a linear regression upon 18 M dwarfs with reliable metallicity estimates from FGK companions to calibrate this updated relationship.  Statistical tests demonstrate that the [Fe/H] estimates produced by this relationship are accurate to $RMSE$=0.141 dex, and confirm that this relation accounts for more of the variance within our calibrant sample than any other existing technique for estimating M dwarf metallicities.}

\item{For the first time, we derive an expression for an M dwarf's overall metallicity [M/H]; quantitative comparisons to model atmospheres, which are benchmarked according to overall metallicity, provide a qualitative validation of our approach for estimating metallicities from \nai, \cai, and H$_{2}$O features, but also demonstrate an overall offset between the atomic line strengths predicted by models as compared to actual observations.}

\item{We examined previous optical molecular indices sensitive to stellar metallicity using out metallicity estimates. We identify that the CaH1 feature as a potentially valuable optical metallicity discriminator for solar-metallicity early type M dwarfs.}

\item{We perform several sanity checks of our metallicity estimates, confirming that our metallicity estimates reproduce expected correlations between metallicity and Galactic space motions and H$\alpha$ emission line strengths, and return statistically identical metallicities for members of M-M multiple systems.} 

\item{We recovered the results from previous metallicity studies that nearby M dwarf planet host exhibit solar to supersolar metallicities, where stars with Jovian-mass planets are more metal-rich than stars with Neptune-likes or super-Earths. A list of the best targets in the sample for planetary searches selected by metallicity, activity and rotation-rates is given.}

 \end{enumerate}

\acknowledgements We thank the staff and telescope operators of Palomar Observatory for their support. We thank Travis Barman, John Bochanski, Jeff Valenti, and Andrew West for helpful discussions about various topics. We thank John Johnson for providing the M dwarf targets in the CPS target list for the K-S test calculations. We thank the anonymous referee for her/his helpful comments that improved our manuscript.

K.R.C. acknowledges support for this work from the Hubble Fellowship Program, provided by NASA through Hubble Fellowship grant HST-HF-51253.01-A awarded by the STScI, which is operated by the AURA, Inc., for NASA, under contract NAS 5-26555. 

This research has made use of NASA's Astrophysics Data System
Bibliographic Services, the SIMBAD database, operated at CDS,
Strasbourg, France, the NASA/IPAC Extragalactic Database, operated by
the Jet Propulsion Laboratory, California Institute of Technology,
under contract with the National Aeronautics and Space Administration,
and the VizieR database of astronomical catalogs
\citep{Ochsenbein2000}. 

This publication makes use of data products from the Two Micron All Sky Survey, which is a joint project of the University of Massachusetts and the Infrared Processing and Analysis Center/California Institute of Technology, funded by the National Aeronautics and Space Administration and the National Science Foundation. 

\facility{Palomar: 200 inch (TSPEC)}

\appendix
\appendix

\section{Statistical tests and reliable uncertainties for functional fits \label{statistics_appendix}}

A well-fitting regression model results in predicted values close to the observed data values. The following statistics are commonly used to evaluate a model fit: 

\noindent\textbf{Residual Mean Square (RMSp)}

The RMSp  is defined as

\begin{eqnarray}
\mathrm{RMSp} &=&\sum _{i=1}^{n} \frac{\mathrm{(\widehat{y}_{i} - y_{i})^{2}}}{n~-~p} \\
\end{eqnarray}

where n is the number of data points (in this case the number of calibrators), p is the number of predictors in the model (including the constant term), $y_{i}$ is the value of the response variable i and $\widehat{y}_{i}$ is the prediction value given by the regression model for $y_{i}$. The RMSp is the variance of the residuals. In general, an RMSp value closer to 0 indicates a model fit that is more useful for prediction.

\noindent\textbf{Root Mean Square Error (RMSE)}

The RMSE is the square root of the variance of the residuals

\begin{eqnarray}
\mathrm{RMSE} &=& \sqrt{\mathrm{RMSp}}\\
\end{eqnarray}

The RMSE indicates how accurately the regression model predicts the response. The RMSE is an absolute measure of it and has the same units as the response variable. Lower values of RMSE indicate better fit.

\noindent\textbf{Adjusted Square of the Multiple Correlation Coefficient ($R$$^2_{ap}$)}

The $R$$^2_{ap}$ represents the proportion of variability in a data set that is accounted for by a regression model. The $R$$^2_{ap}$ is defined as

\begin{eqnarray}
\mathrm{R^2_{ap}} &=&~1~-~\frac{(n~-~1)\sum _{i=1}^{n} \mathrm{(\widehat{y}_{i} - y_{i})^{2}}}{(n~-~p)\sum _{i=1}^{n} \mathrm{(y_{i} - \overline{y})^{2}}} \\
&=&~1~-~\frac{\mathrm{RMSp}}{\mathrm{MST}}
\end{eqnarray}

where $\overline{y}$ is the overall mean of the observations $y_{i}$, and $MST$ is the total mean of squares. The $R$$^2_{ap}$ compares the unbiased variance of the residuals ($RMSp$) and the unbiased variances of the observations ($MST$), therefore, can be interpreted as the proportion of total variance that is explained by the model. A value of $R$$^2_{ap}$=1 indicates that the regression model explains all of the variance in the sample, while $R$$^2_{ap}$=0 indicates that the regression models explains none of the variance. $R$$^2_{ap}$ should always be used with models with more than one predictor variable. 

\noindent\textbf{Root Mean Square Error of Validation (RMSE$_{V}$)}

A better estimate of the uncertainty in the K band [Fe/H] fit can be obtained by performing the Predicted Residual Sum of Squares procedure \citep[PRESS;][]{weisberg2005applied}. The PRESS statistic is equivalent to "leave-one-out" method, where, in this case, the regression procedure is repeated leaving one of the n calibrators out and performing the regression with the other n-1 calibrators, for each one of the n calibrators. The PRESS statistics is defined as  

\begin{eqnarray}
\mathrm{PRESS} &=& \sum _{i=1}^{n} \mathrm{\widehat{e}^{2}_{i}}\\
&=&\sum _{i=1}^{n} \mathrm{(\widehat{y}_{(-c_{i})i} - y_{i})^{2}} \nonumber
\end{eqnarray}

where $\widehat{e}_{i}$ is the residual for the calibrator $c_{i}$, computed as the difference between the observed value of the predictand $y_{i}$ and the prediction $\widehat{y}_{(-c_{i})i}$ from a regression model calibrated using the calibration sample but without the calibrator $c_{i}$. Then, it is possible to use as a measure of uncertainty a statistic based on this validation data

\begin{eqnarray}
\mathrm{RMSE_{V}} &=& \sqrt{\frac{\mathrm{PRESS}}{n_{\footnotesize{V}}}}
\end{eqnarray}

Equation (2.6) is known as the Root-Mean-Squared Error of validation (RMSE$_{V}$), where n$_{\footnotesize{V}}$ is the number of calibrators. The RMSE$_{V}$ value is a sensible estimate of average prediction error according to \citet{weisberg2005applied}. The RMSE$_{V}$ value can be used then to obtain confidence intervals at a desired significance level around the predictions  

\begin{eqnarray}
\mathrm{\widehat{y}_{i}} \pm \mathrm{C}(\emph{t}_{\alpha/2, n-p})~ \mathrm{RMSE_{V}}\nonumber
\end{eqnarray}

where the multiplier of RMSE$_{V}$ corresponds to the desired probability point on the cumulative distribution function. The correct cumulative distribution function for this specific fit is the two-sided $\emph{t}$-student distribution with n-p=15 degrees of freedom, where n=18 is the sample size for calibration and p=3 is the number of regressors in the model fit, including the constant term. The two-sided $\emph{t}$-student distribution is preferred instead of a normal distribution due to the small number of calibrators (n$=$ n$_{V}$ $=$18). For approximate 70$\%$, 95$\%$, and 99$\%$  confidence intervals, the values for C($\emph{t}_{\alpha/2, n-p=15}$) are equal to 1.07, 2.13, and 2.95, respectively. 

\setlength{\baselineskip}{0.6\baselineskip}
\bibliography{./bibliography}
\setlength{\baselineskip}{1.667\baselineskip}

\clearpage

\figsetstart
\figsetnum{1}
\figsettitle{K band spectra of 133 nearby M-dwarfs}

\figsetgrpstart
\figsetgrpnum{1.1}
\figsetgrptitle{K band spectra of LHS 3576, Gl 338 A, and Gl 338 B. The main spectral features are indicated.}
\figsetplot{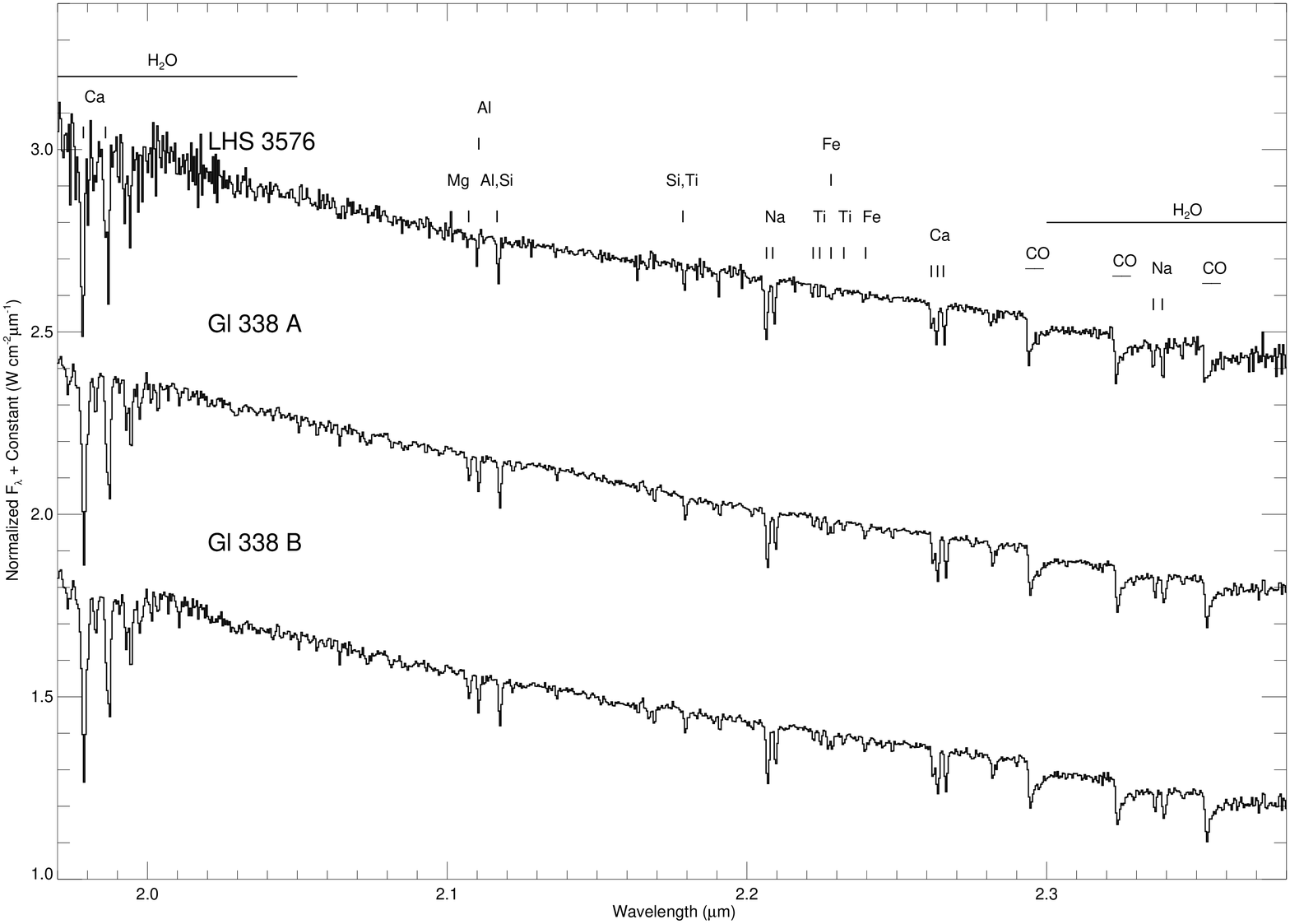}
\figsetgrpnote{TripleSpec 200-inch K band spectra of 133 nearby M-dwarfs. }
\figsetgrpend

\figsetend

\begin{figure}
\figurenum{1}
\plotone{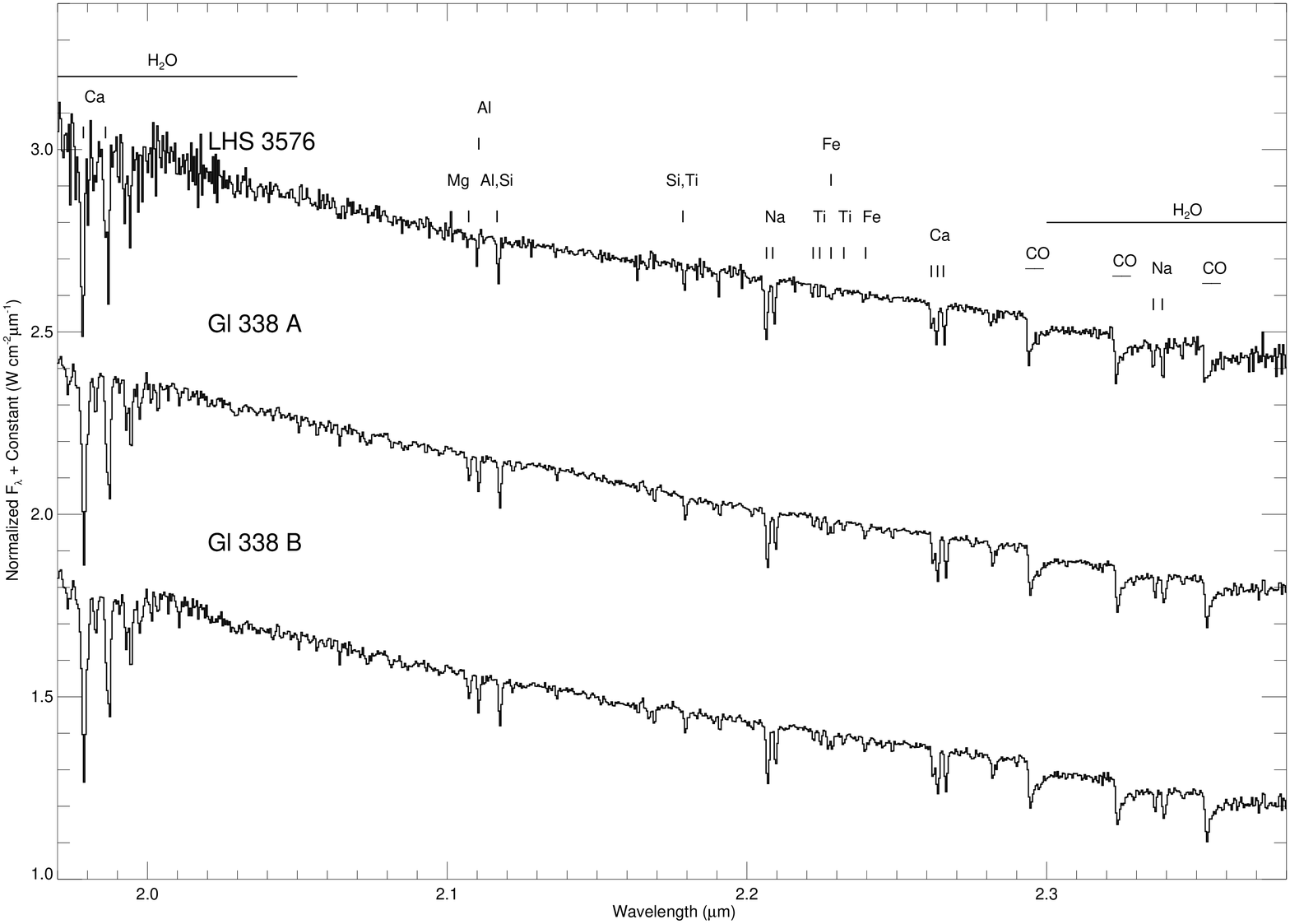}
\caption{TripleSpec 200-inch K band spectra of 133 nearby M-dwarfs. }
\end{figure}

\begin{figure}
\begin{center}
\includegraphics[scale=0.6,angle=90]{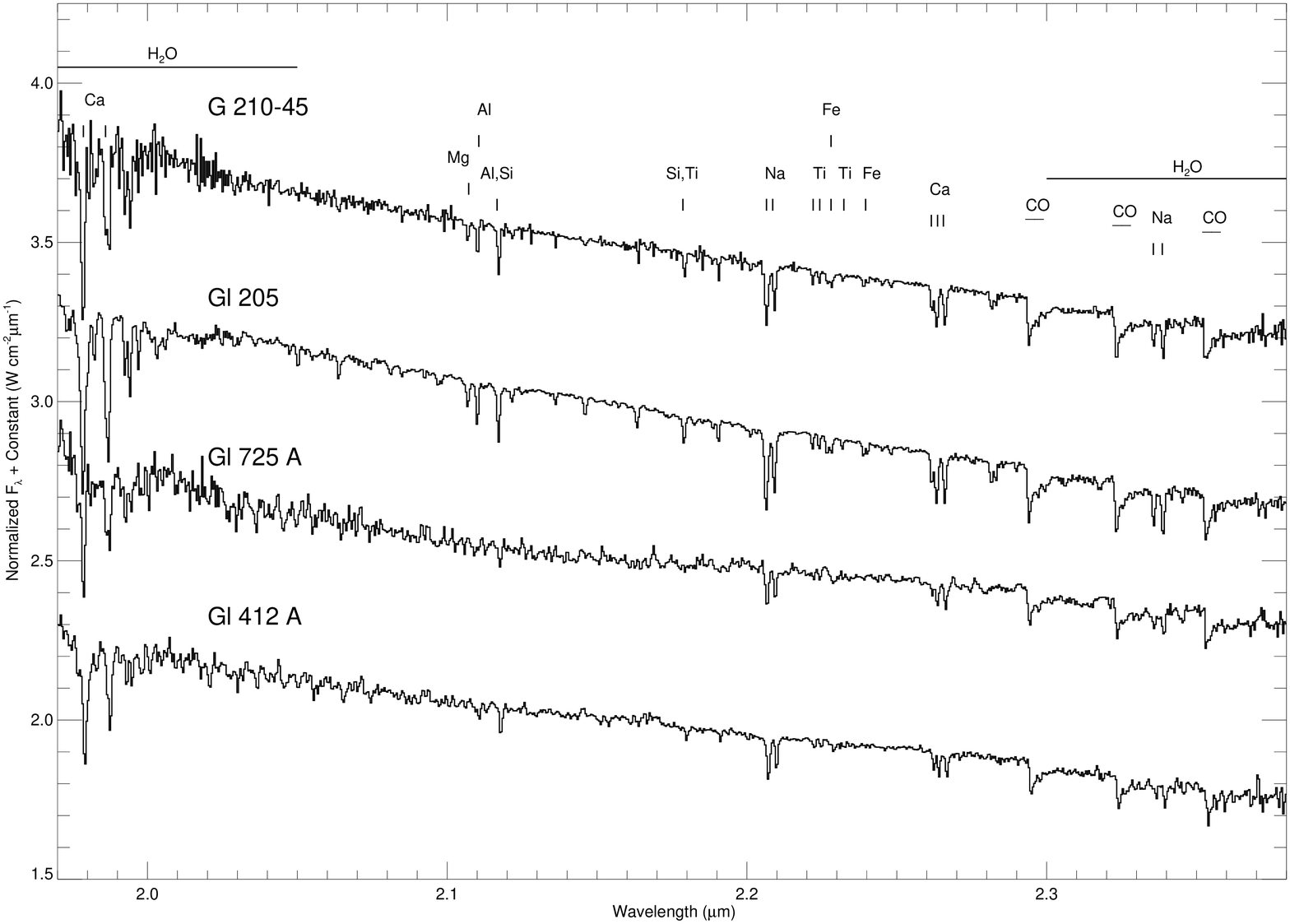}
\caption{\footnotesize{K band spectra of G 210-45, Gl 205, Gl 725 A, and Gl 412 A. The main spectral features are indicated.}}
\end{center}
\end{figure}

\begin{figure}
\begin{center}
\includegraphics[scale=0.6,angle=90]{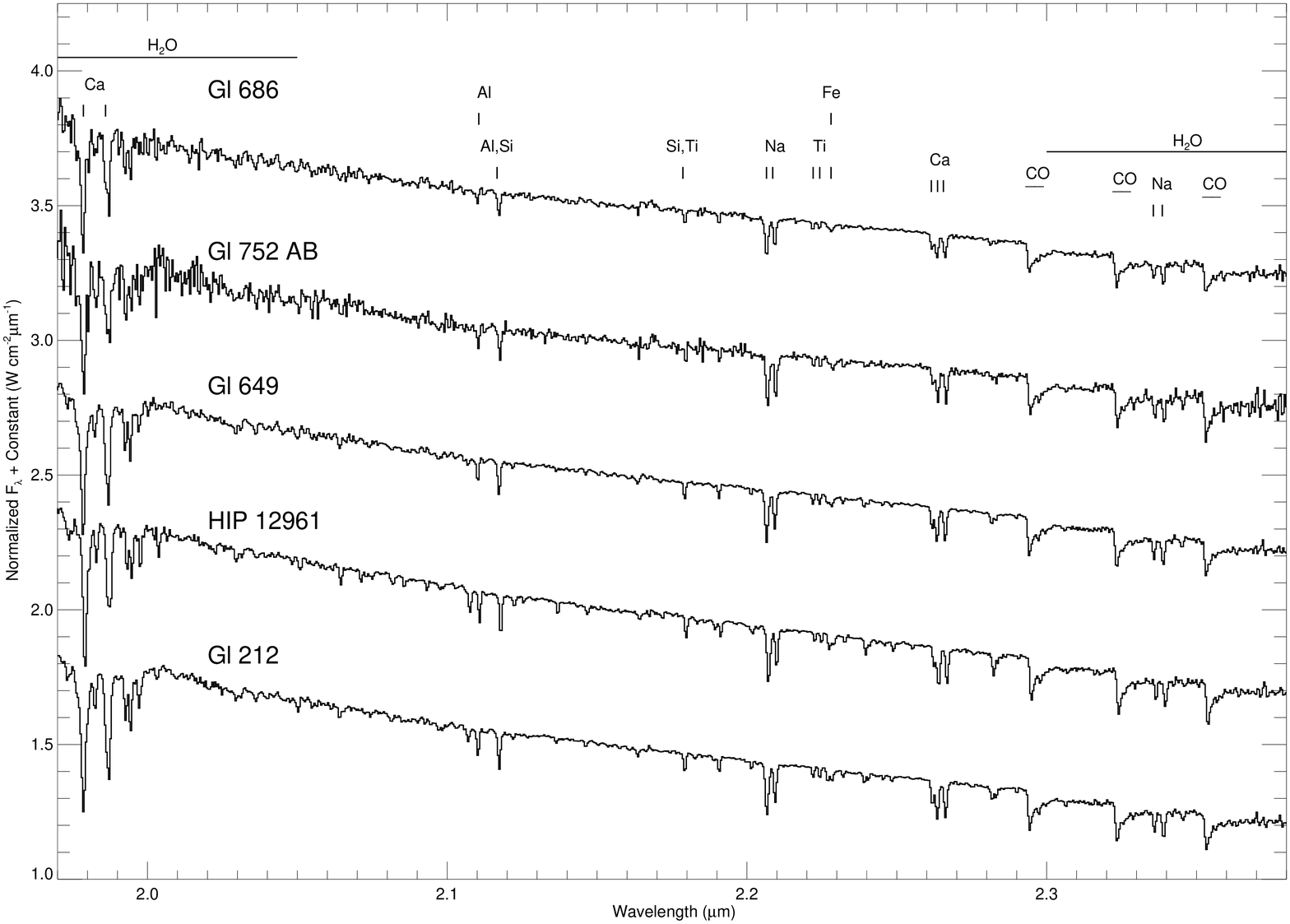}
\caption{\footnotesize{K band spectra of Gl 686, Gl 752 AB, Gl 649, HIP 12961, and Gl 212. The main spectral features are indicated.}}
\end{center}
\end{figure}

\begin{figure}
\begin{center}
\includegraphics[scale=0.6,angle=90]{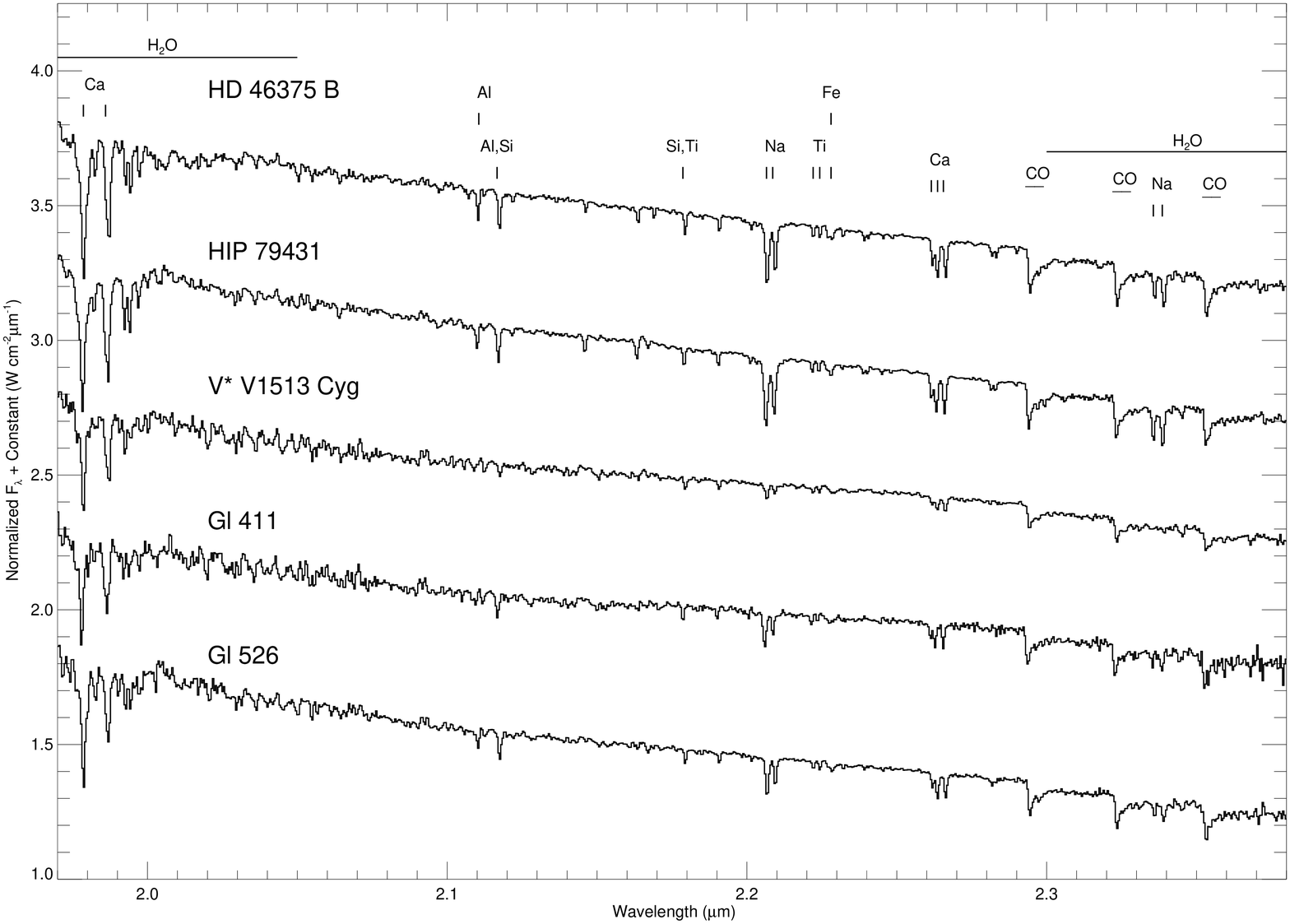}
\caption{\footnotesize{K band spectra of HD 46375 B, HIP 79431, V* V1513 Cyg, Gl 411, and Gl 526. The main spectral features are indicated.}}
\end{center}
\end{figure}

\begin{figure}
\begin{center}
\includegraphics[scale=0.6,angle=90]{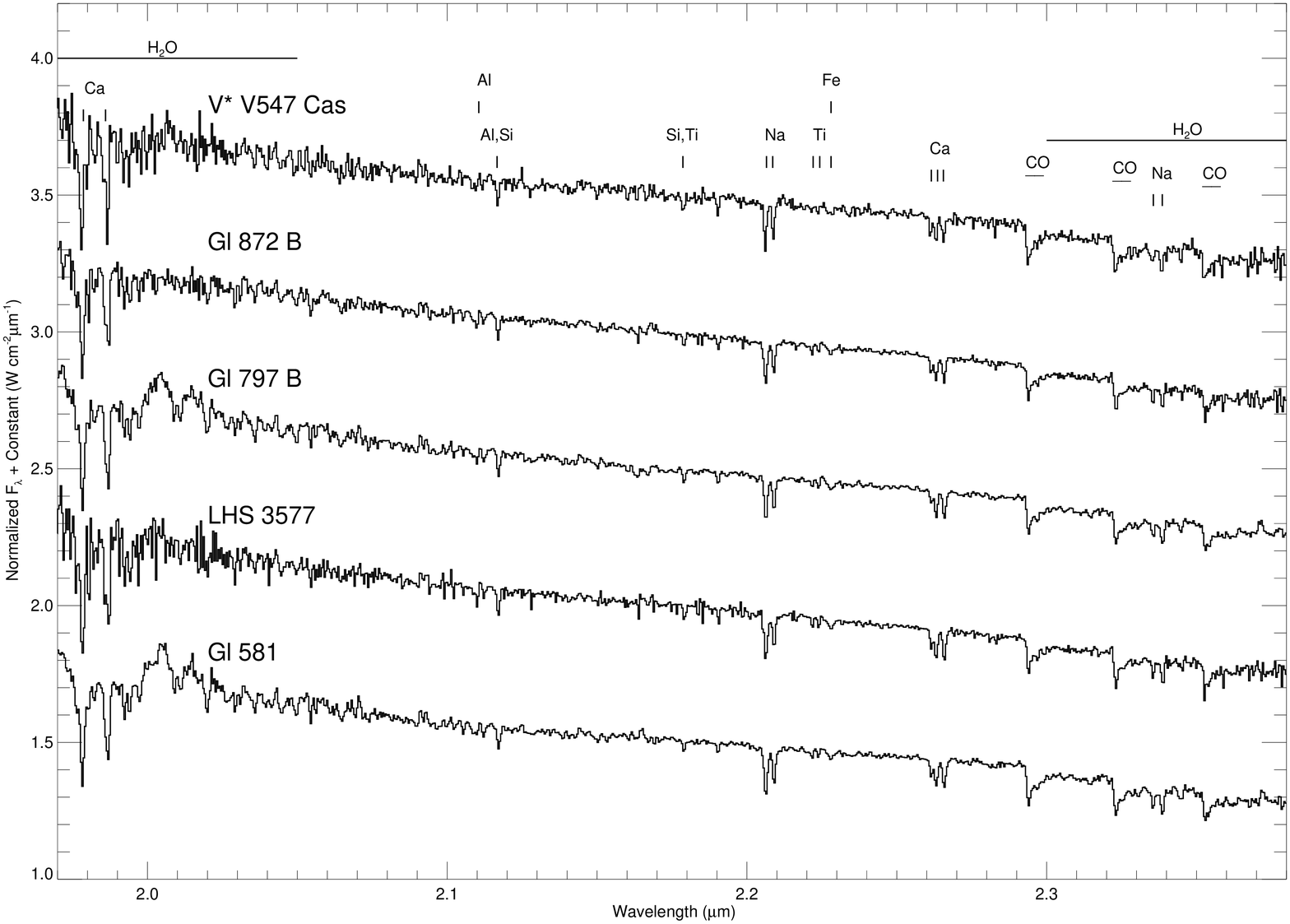}
\caption{\footnotesize{K band spectra of V* V547 Cas, Gl 872 B, Gl 797 B, LHS 3577, and Gl 581. The main spectral features are indicated.}}
\end{center}
\end{figure}

\begin{figure}
\begin{center}
\includegraphics[scale=0.6,angle=90]{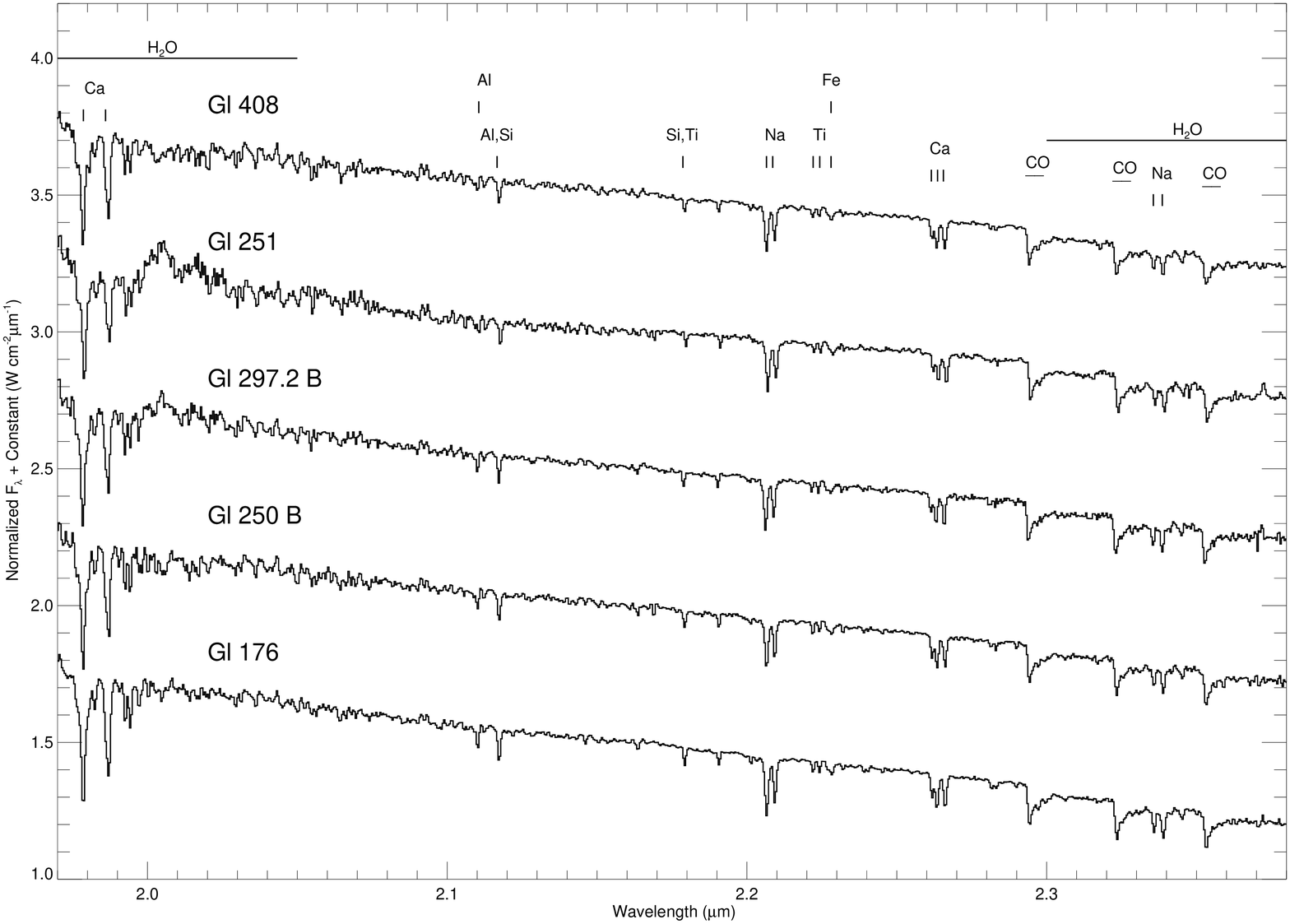}
\caption{\footnotesize{K band spectra of Gl 408, Gl 251, Gl 297.2 B, Gl 250 B, and Gl 176. The main spectral features are indicated.}}
\end{center}
\end{figure}

\begin{figure}
\begin{center}
\includegraphics[scale=0.6,angle=90]{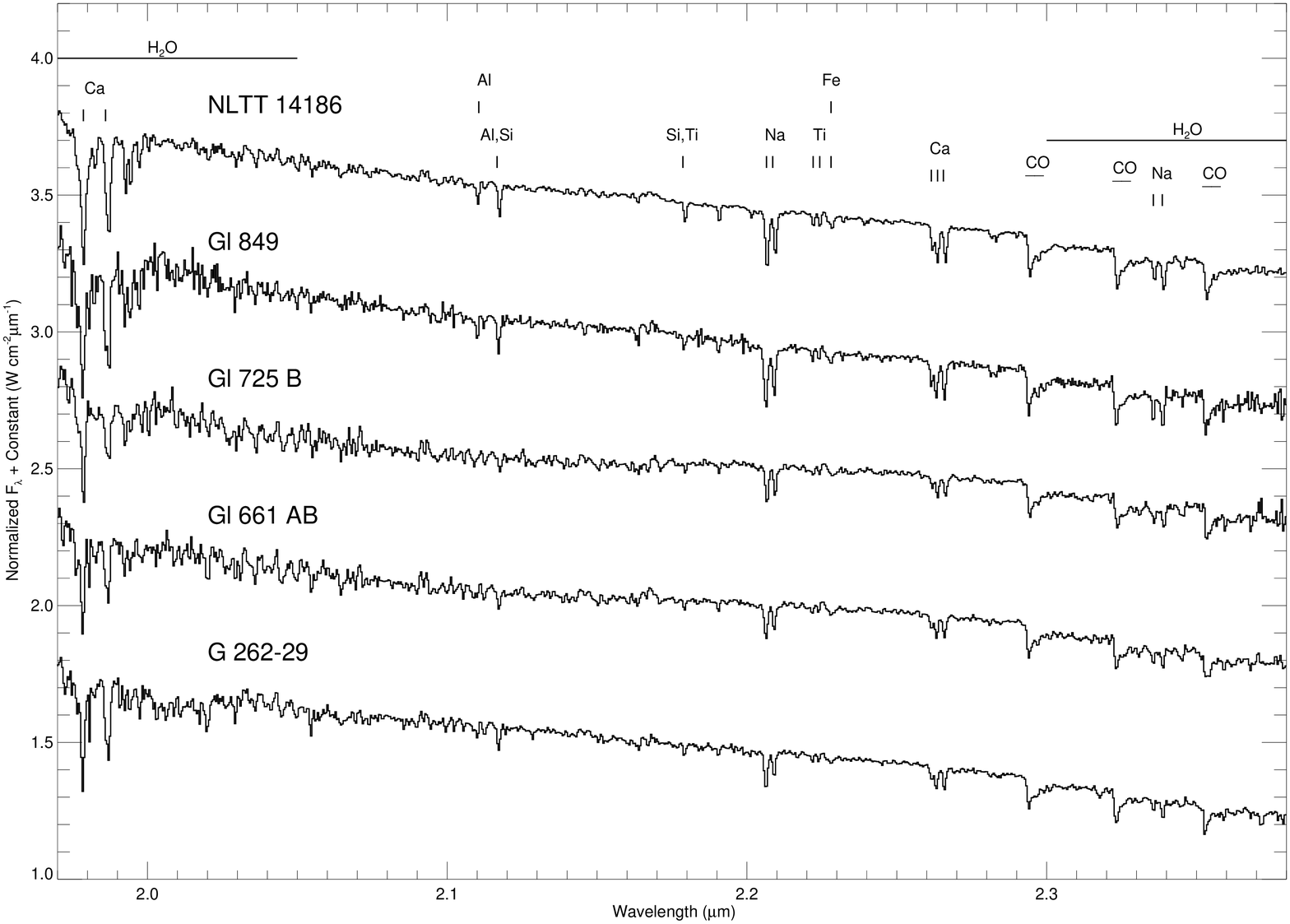}
\caption{\footnotesize{K band spectra of NLTT 14186, Gl 849, Gl 725 B, Gl 661 AB, and G 262-29. The main spectral features are indicated.}}
\end{center}
\end{figure}

\clearpage

\begin{figure}
\begin{center}
\includegraphics[scale=0.6,angle=90]{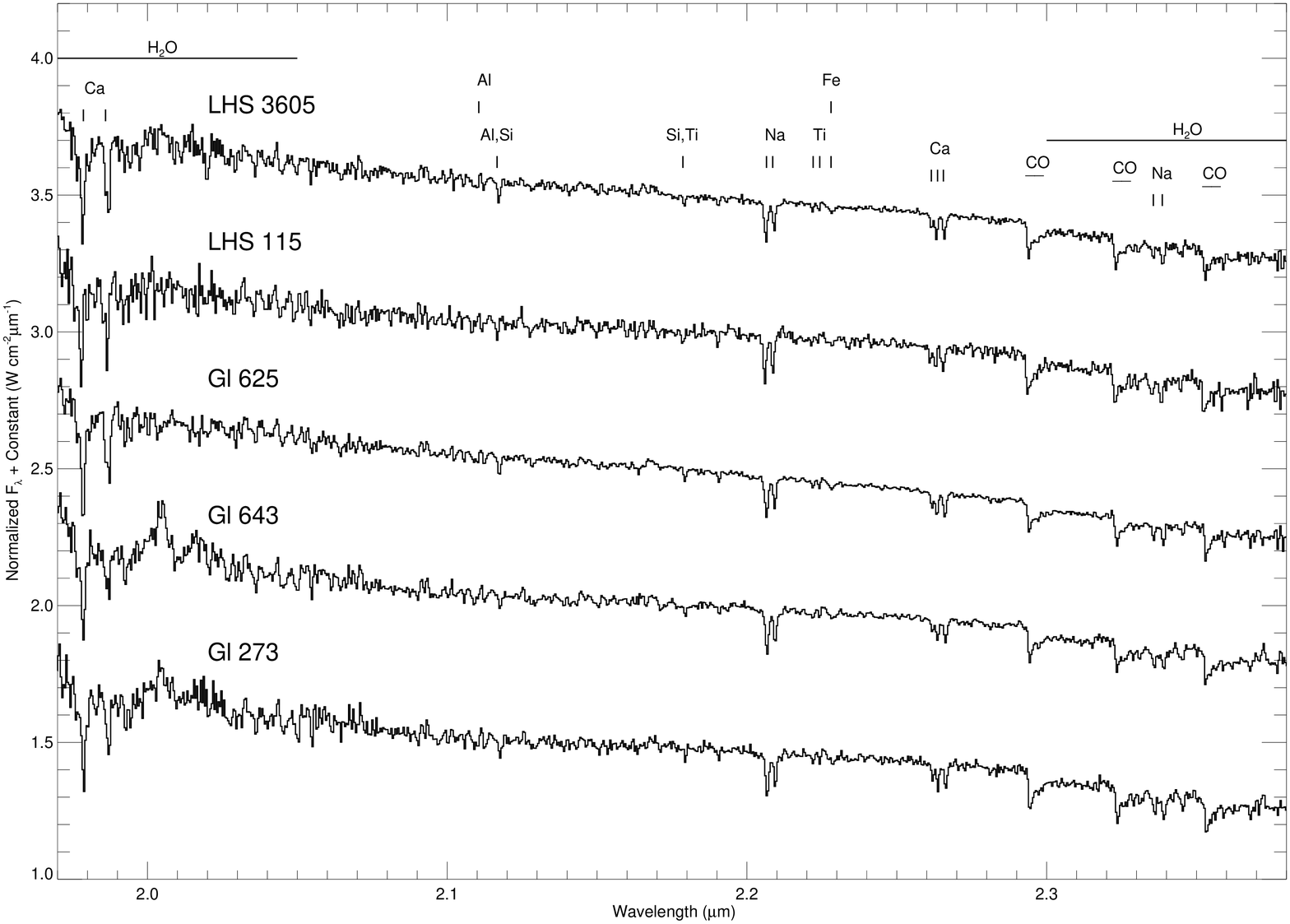}
\caption{\footnotesize{K band spectra of LHS 3605, LHS 115, Gl 625, Gl 643, and Gl 273. The main spectral features are indicated.}}
\end{center}
\end{figure}

\begin{figure}
\begin{center}
\includegraphics[scale=0.6,angle=90]{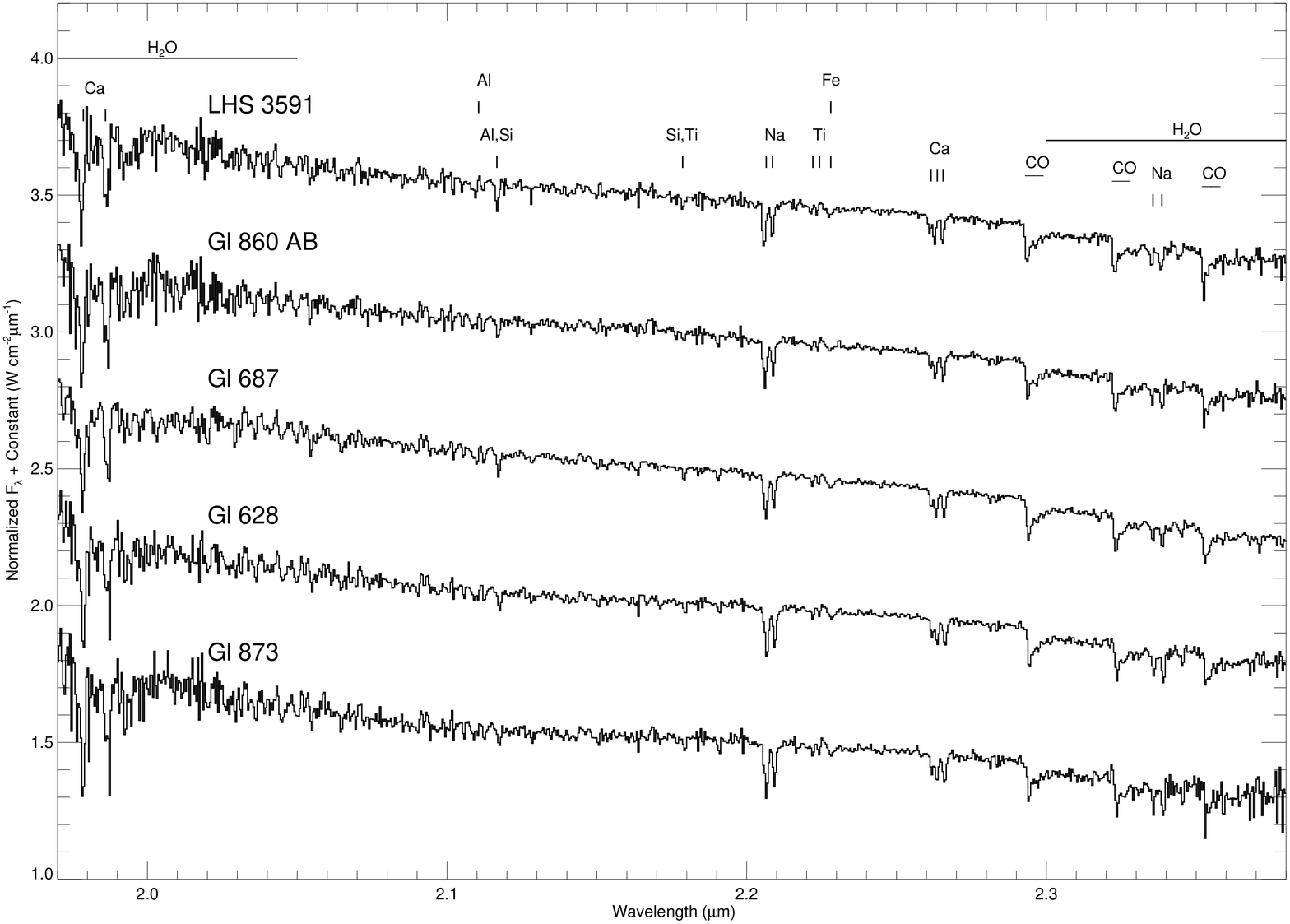}
\caption{\footnotesize{K band spectra of LHS 3591, Gl 860 AB, Gl 687, Gl 628, and Gl 873. The main spectral features are indicated.}}
\end{center}
\end{figure}

\begin{figure}
\begin{center}
\includegraphics[scale=0.6,angle=90]{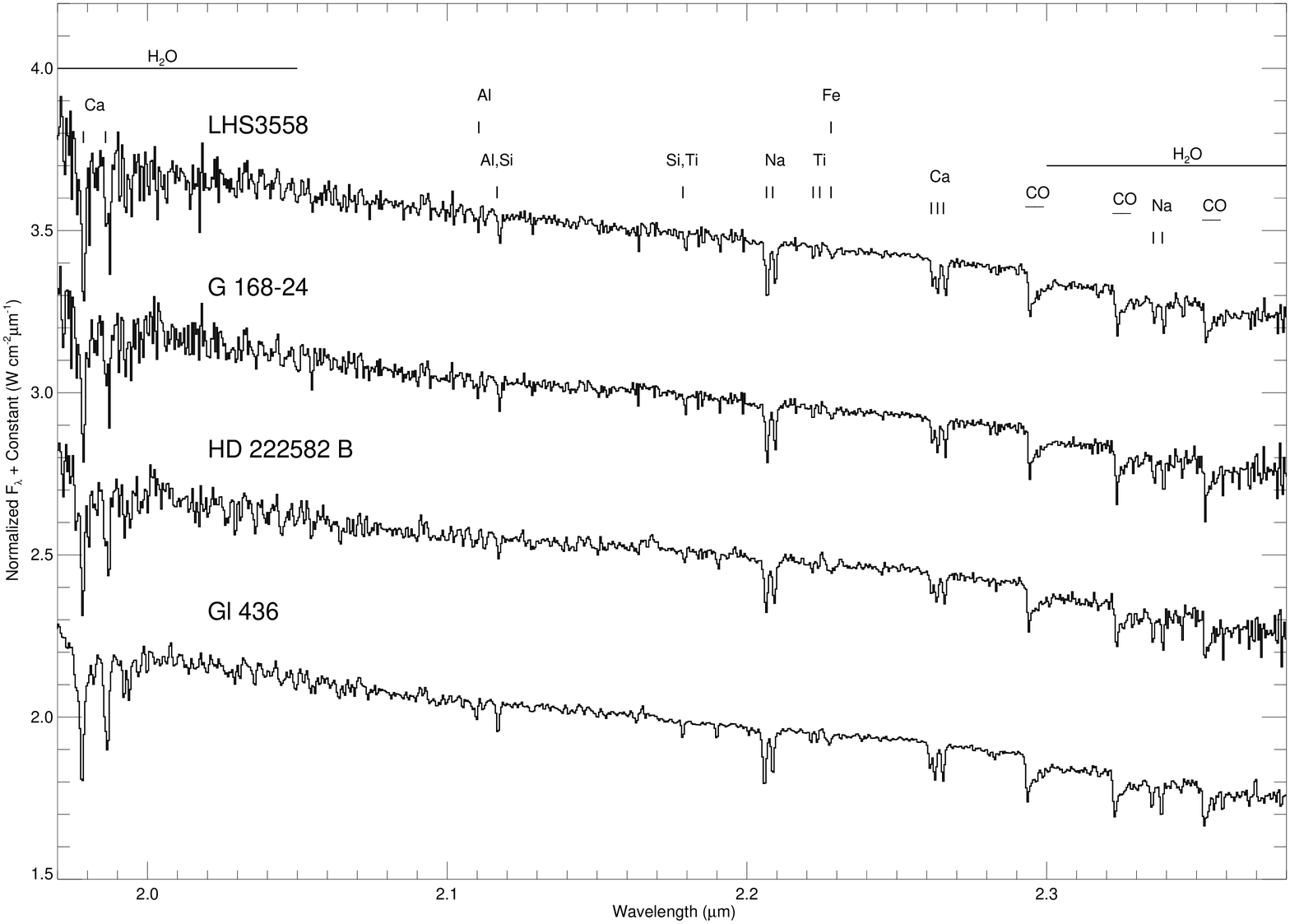}
\caption{\footnotesize{K band spectra of LHS 3558, G 168-24, HD 222582 B, and Gl 436. The main spectral features are indicated.}}
\end{center}
\end{figure}

\begin{figure}
\begin{center}
\includegraphics[scale=0.6,angle=90]{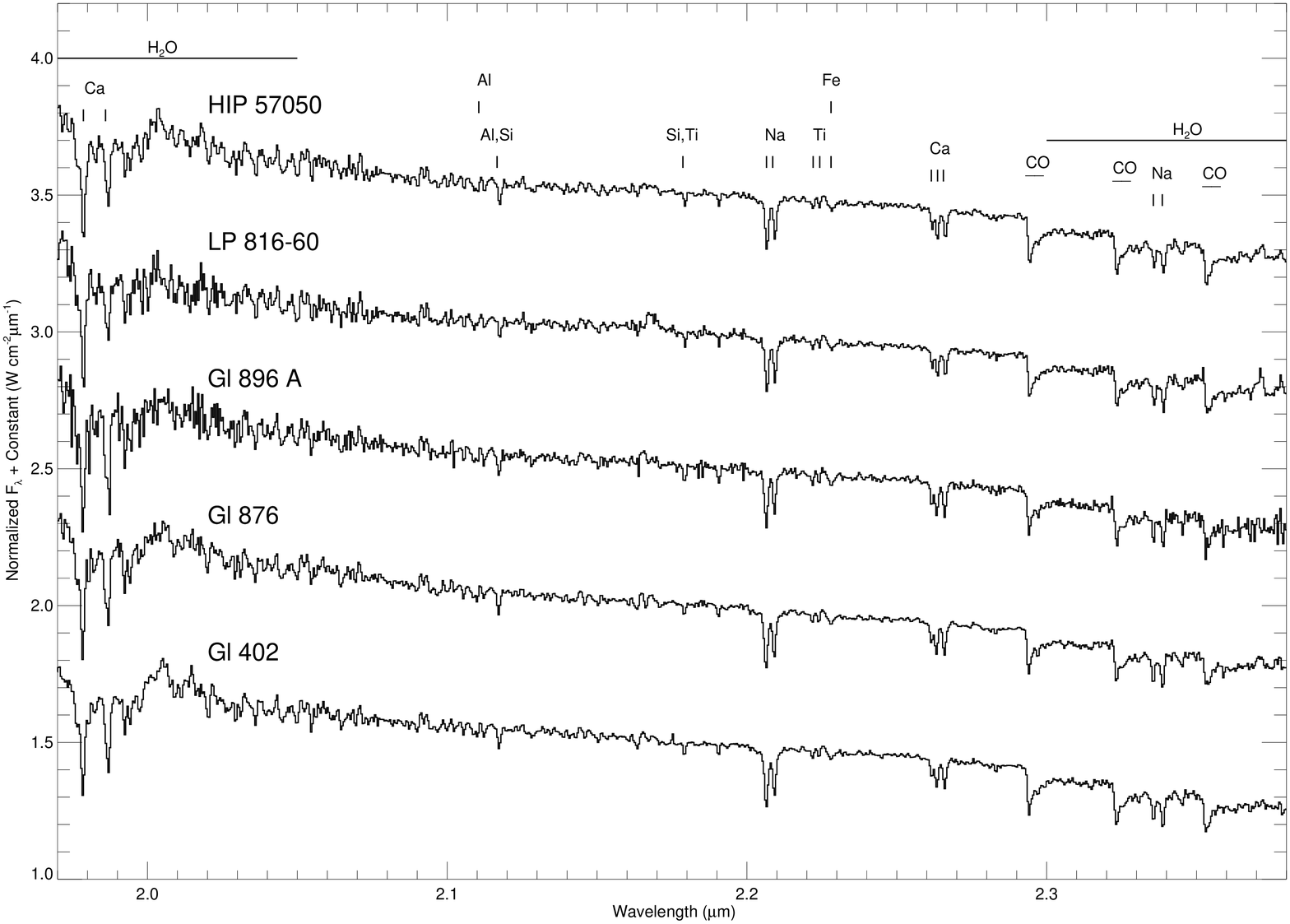}
\caption{\footnotesize{K band spectra of HIP 57050, LP 816-60, Gl 896 A, Gl 876, and Gl 402 . The main spectral features are indicated.}}
\end{center}
\end{figure}

\begin{figure}
\begin{center}
\includegraphics[scale=0.6,angle=90]{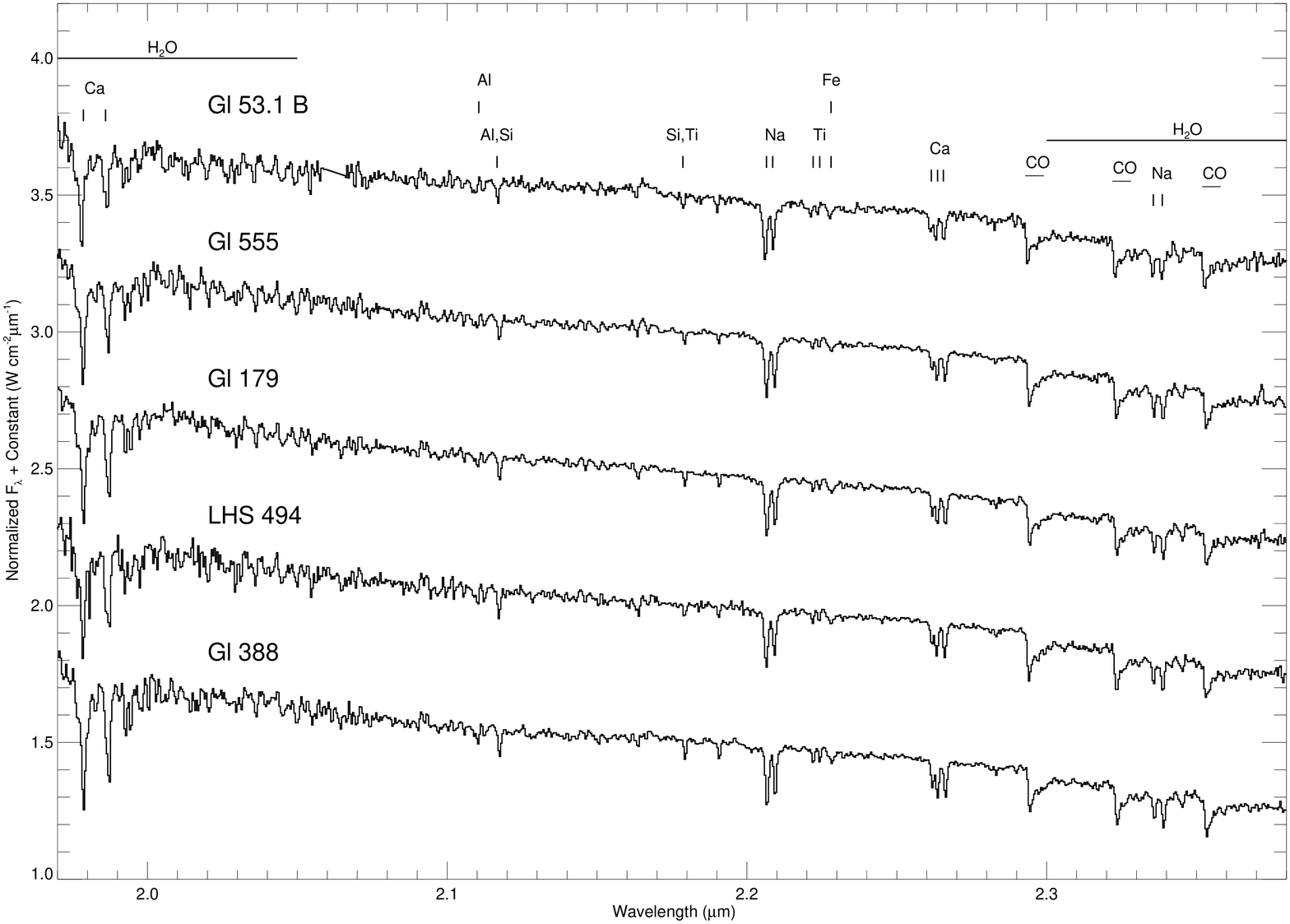}
\caption{\footnotesize{K band spectra of Gl 53.1 B, Gl 555, Gl 179, LHS 494, and Gl 388. The main spectral features are indicated.}}
\end{center}
\end{figure}

\begin{figure}
\begin{center}
\includegraphics[scale=0.6,angle=90]{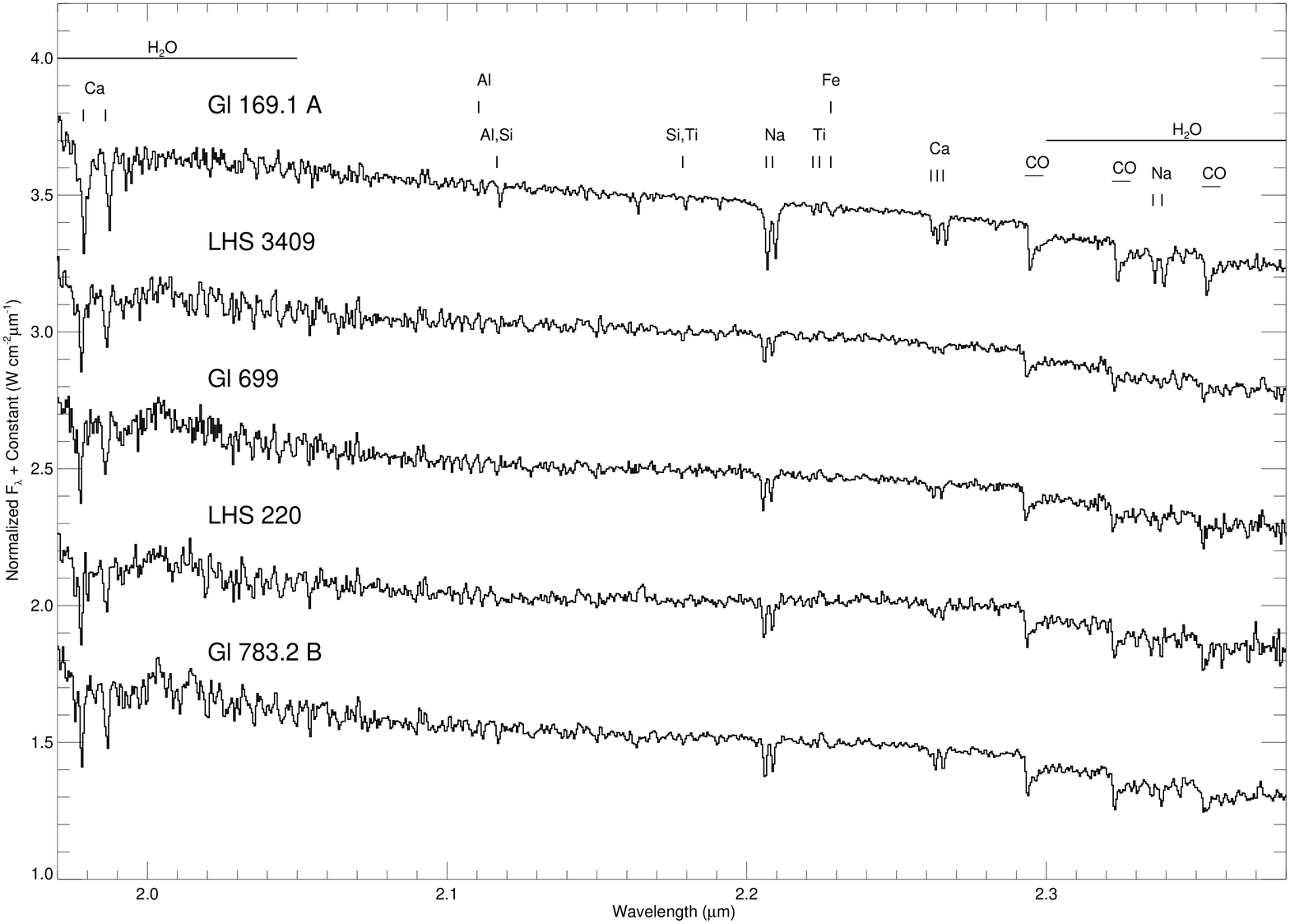}
\caption{\footnotesize{K band spectra of Gl 169.1 A, LHS 3409, Gl 699, LHS 220, and  Gl 783.2 B. The main spectral features are indicated.}}
\end{center}
\end{figure}

\begin{figure}
\begin{center}
\includegraphics[scale=0.6,angle=90]{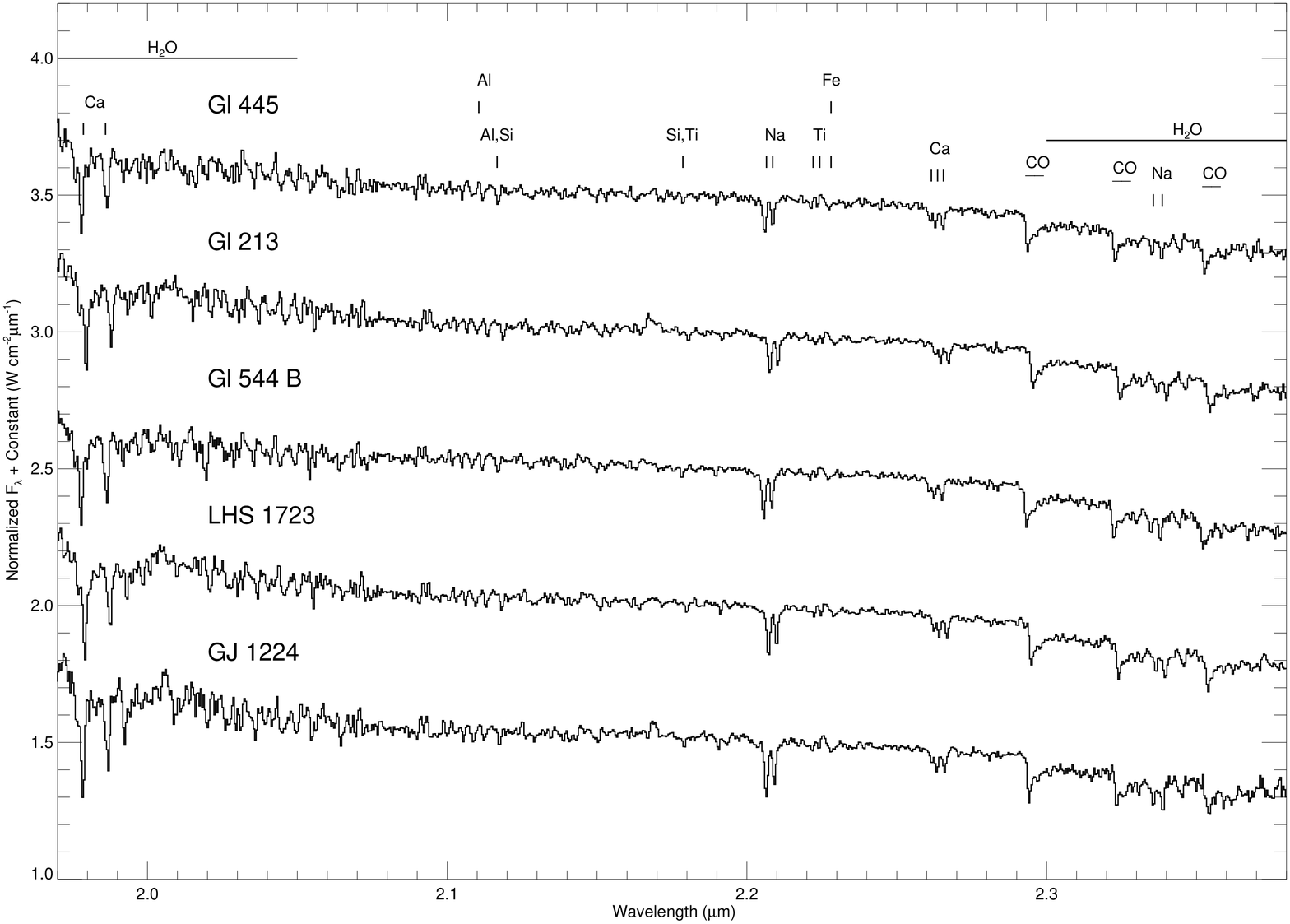}
\caption{\footnotesize{K band spectra of Gl 445, Gl 213, Gl 544 B, LHS 1723, and GJ 1224. The main spectral features are indicated.}}
\end{center}
\end{figure}

\clearpage

\begin{figure}
\begin{center}
\includegraphics[scale=0.6,angle=90]{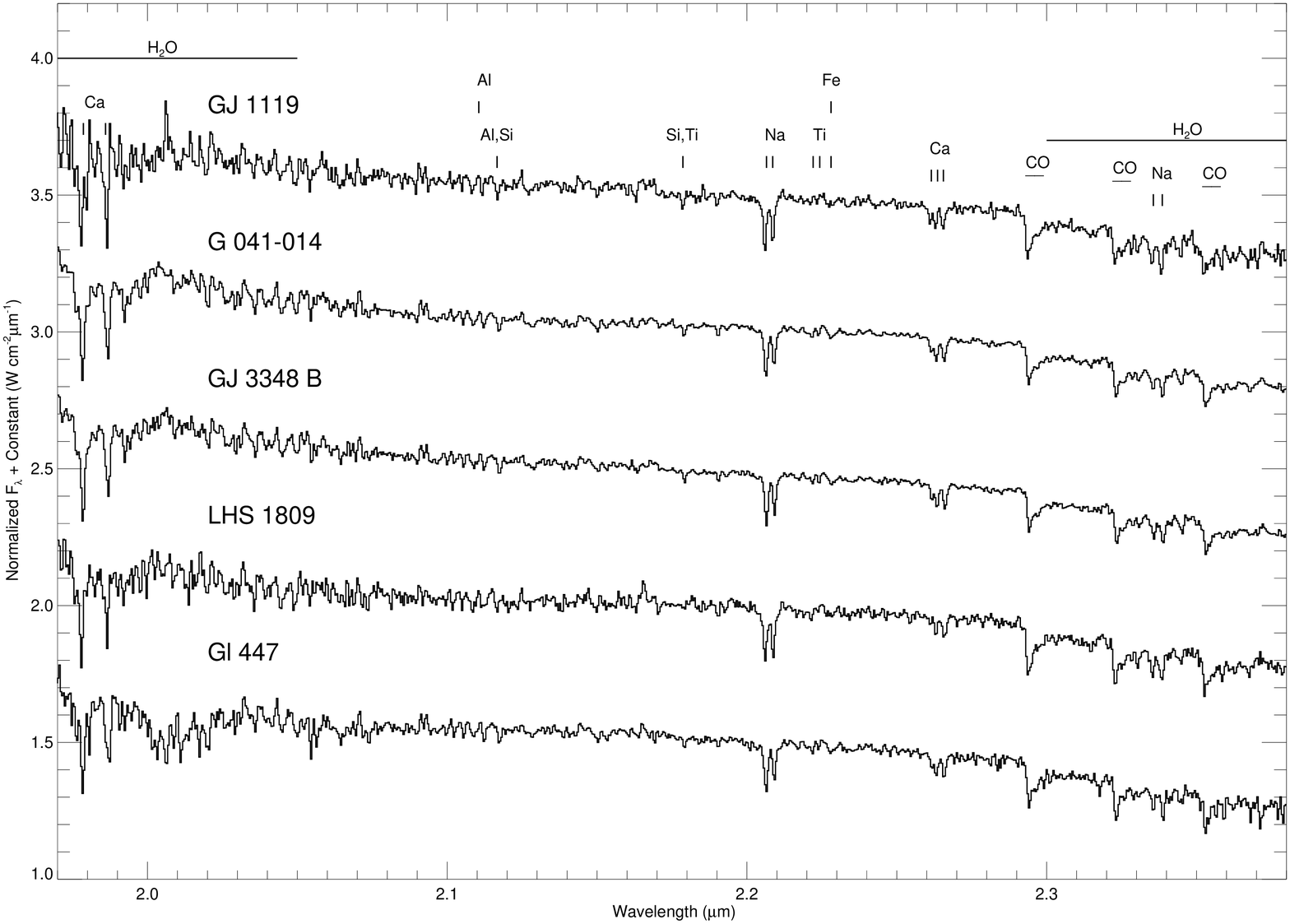}
\caption{\footnotesize{K band spectra of GJ1119, G 041-014, GJ 3348 B, LHS 1809, and Gl 447. The main spectral features are indicated.}}
\end{center}
\end{figure}

\begin{figure}
\begin{center}
\includegraphics[scale=0.6,angle=90]{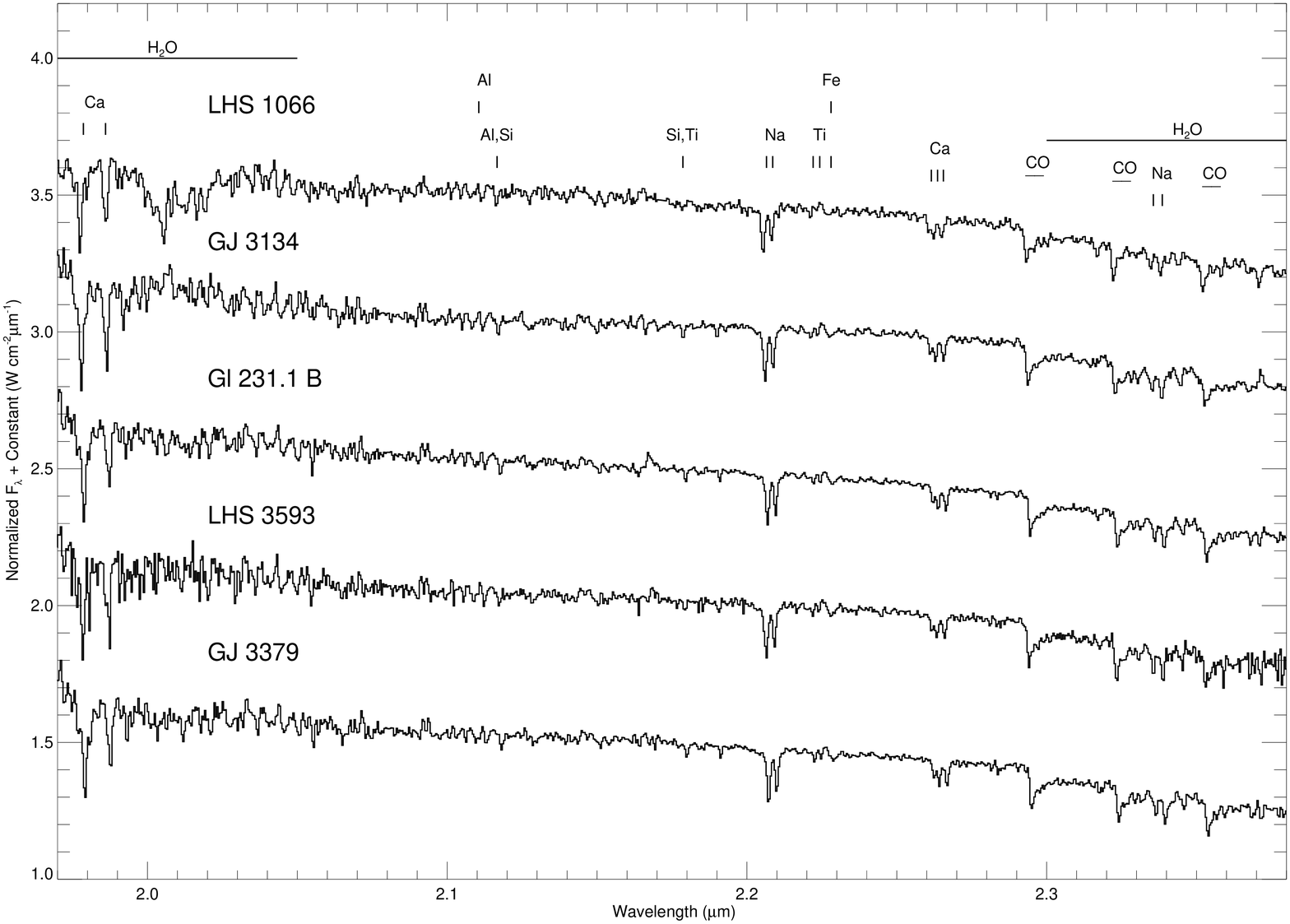}
\caption{\footnotesize{K band spectra of LHS 1066, GJ 3134, Gl 231.1 B, LHS 3593, and GJ 3379. The main spectral features are indicated.}}
\end{center}
\end{figure}


\begin{figure}
\begin{center}
\includegraphics[scale=0.6,angle=90]{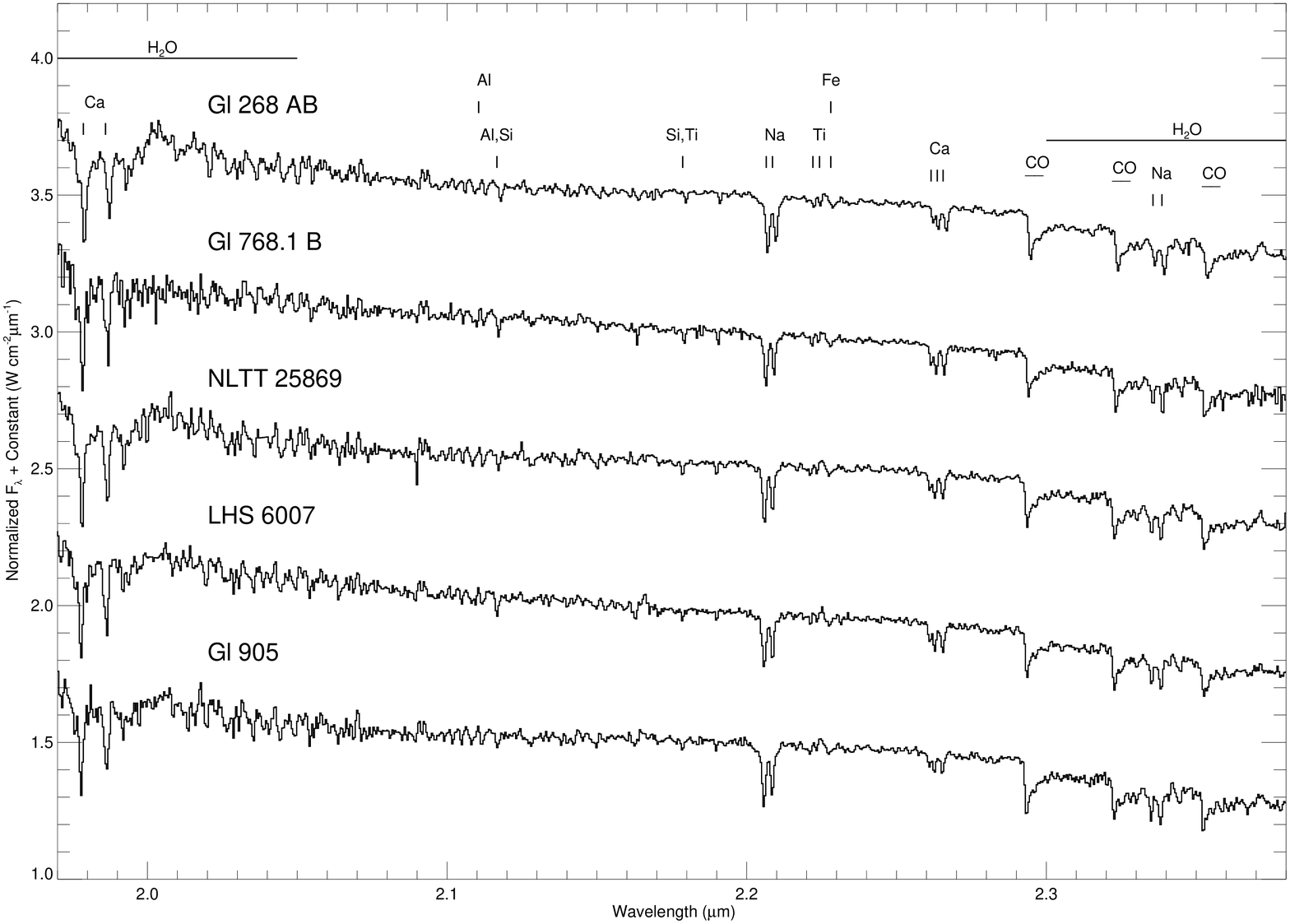}
\caption{\footnotesize{K band spectra of  Gl 268 AB, Gl 768.1 B, NLTT 25869, LHS 6007, and Gl 905. The main spectral features are indicated.}}
\end{center}
\end{figure}

\begin{figure}
\begin{center}
\includegraphics[scale=0.6,angle=90]{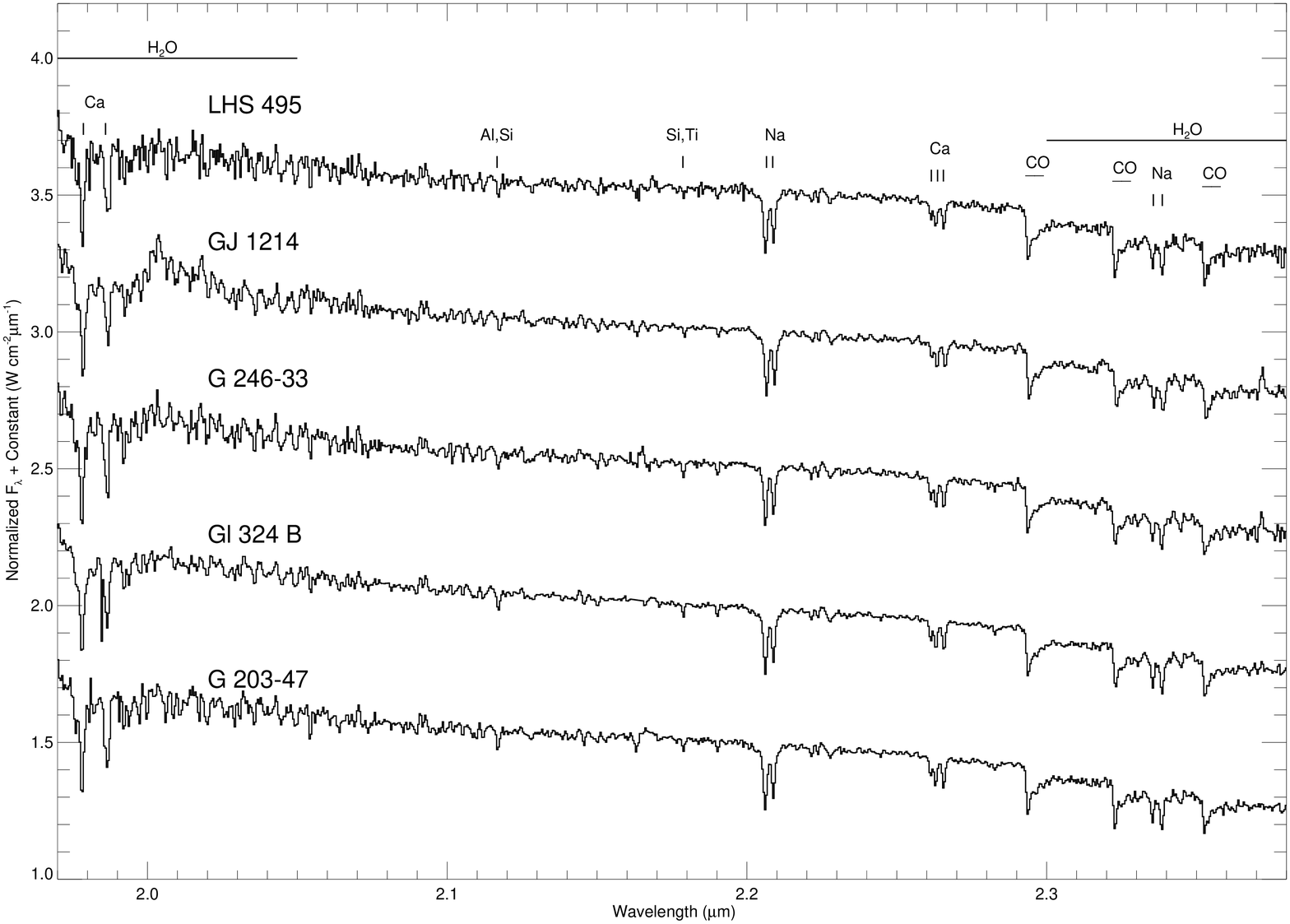}
\caption{\footnotesize{K band spectra of LHS 495, GJ 1214, G 246-33, Gl 324 B, and G 203-47. The main spectral features are indicated.}}
\end{center}
\end{figure}

\begin{figure}
\begin{center}
\includegraphics[scale=0.6,angle=90]{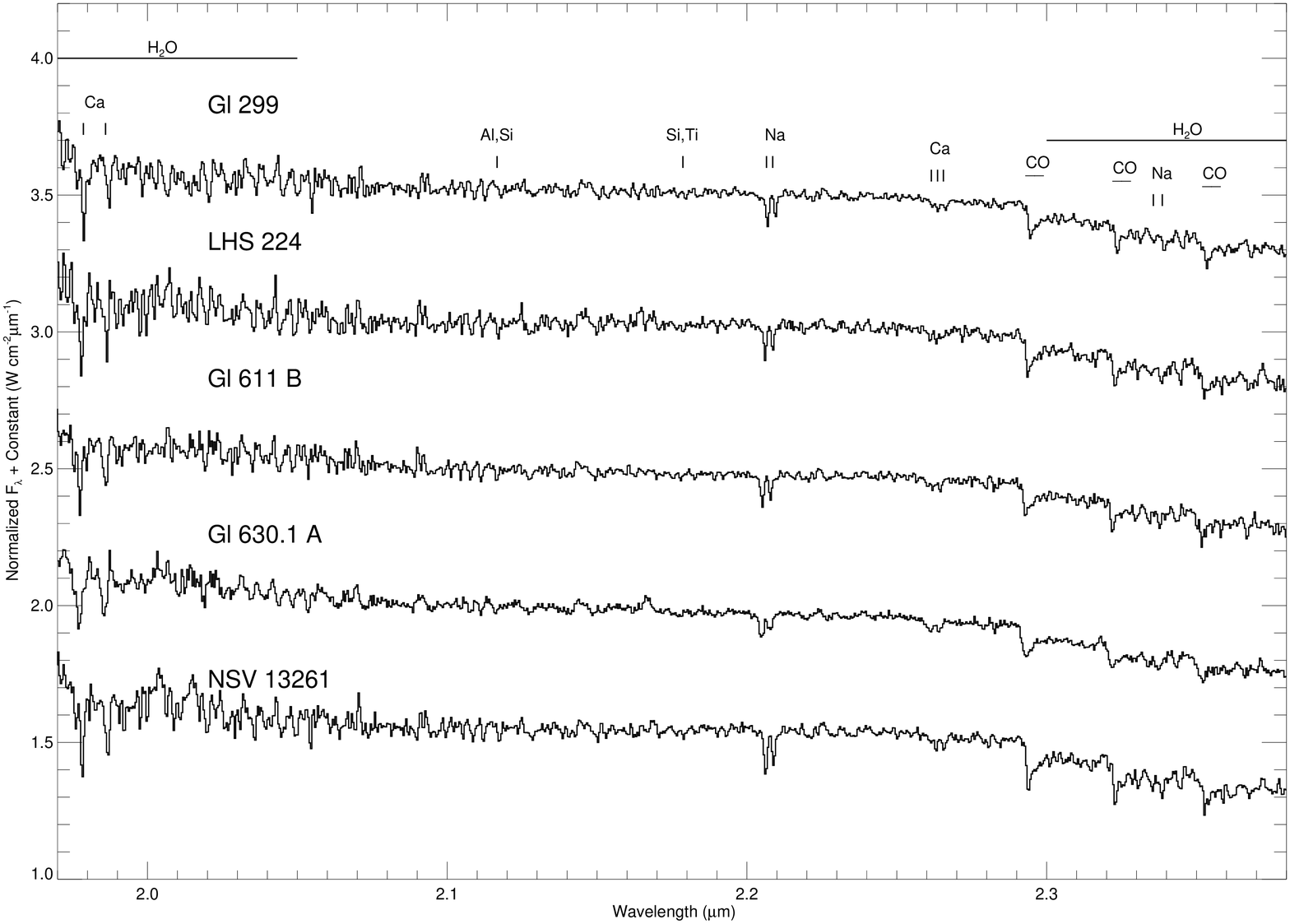}
\caption{\footnotesize{K band spectra of Gl 299, LHS 224, Gl 611 B, Gl 630.1 A, and NSV 13261. The main spectral features are indicated.}}
\end{center}
\end{figure}

\begin{figure}
\begin{center}
\includegraphics[scale=0.6,angle=90]{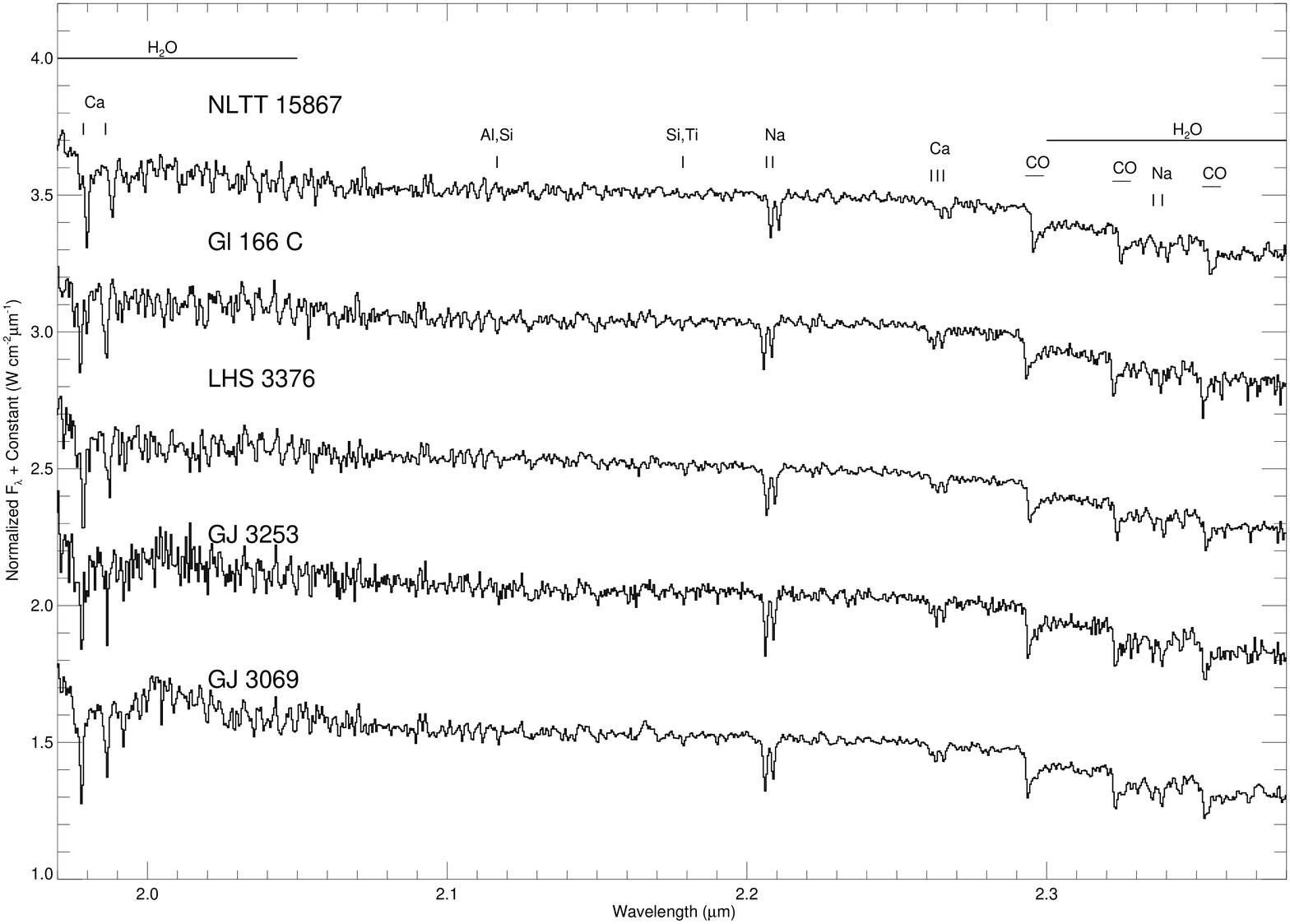}
\caption{\footnotesize{K band spectra of NLTT 15867, Gl 166 C, LHS 3376, GJ 3253, and GJ 3069. The main spectral features are indicated.}}
\end{center}
\end{figure}

\begin{figure}
\begin{center}
\includegraphics[scale=0.6,angle=90]{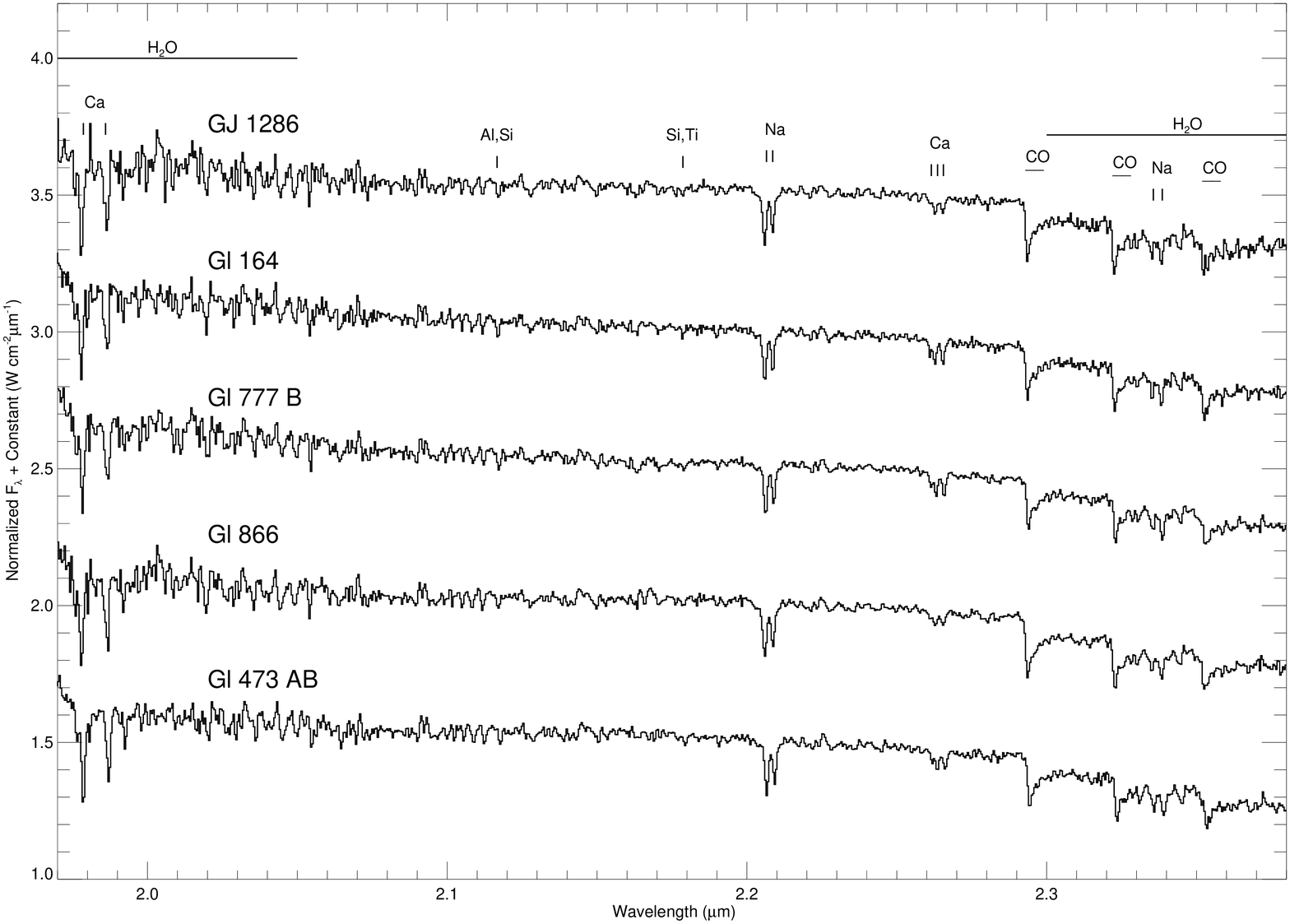}
\caption{\footnotesize{K band spectra of GJ 1286, Gl 164, Gl 777 B, Gl 866, and Gl 473 AB. The main spectral features are indicated.}}
\end{center}
\end{figure}

\clearpage

\begin{figure}
\begin{center}
\includegraphics[scale=0.6,angle=90]{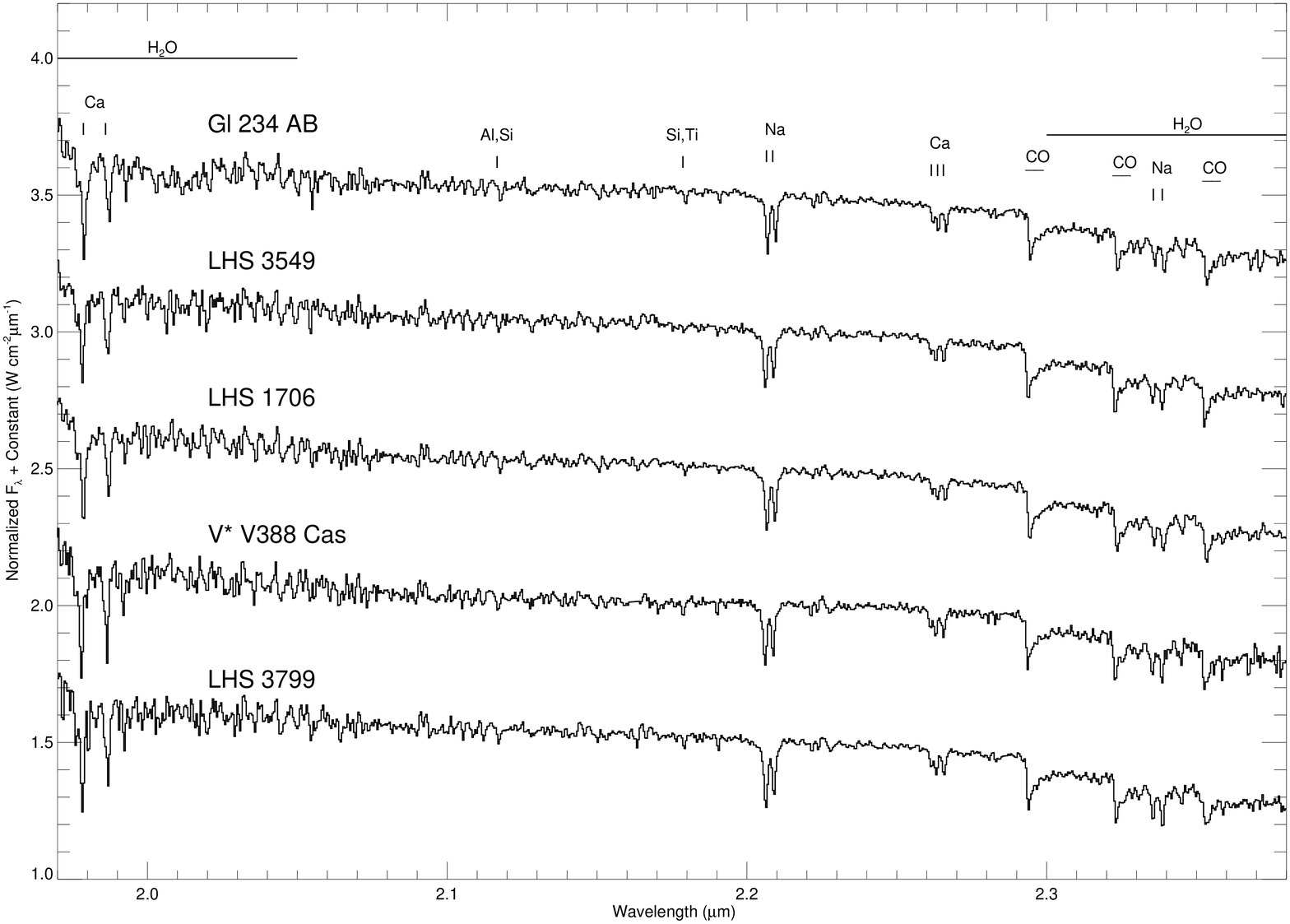}
\caption{\footnotesize{K band spectra of Gl 234 AB, LHS 3549, LHS 1706, V* V388 Cas, and LHS 3799. The main spectral features are indicated.}}
\end{center}
\end{figure}

\begin{figure}
\begin{center}
\includegraphics[scale=0.6,angle=90]{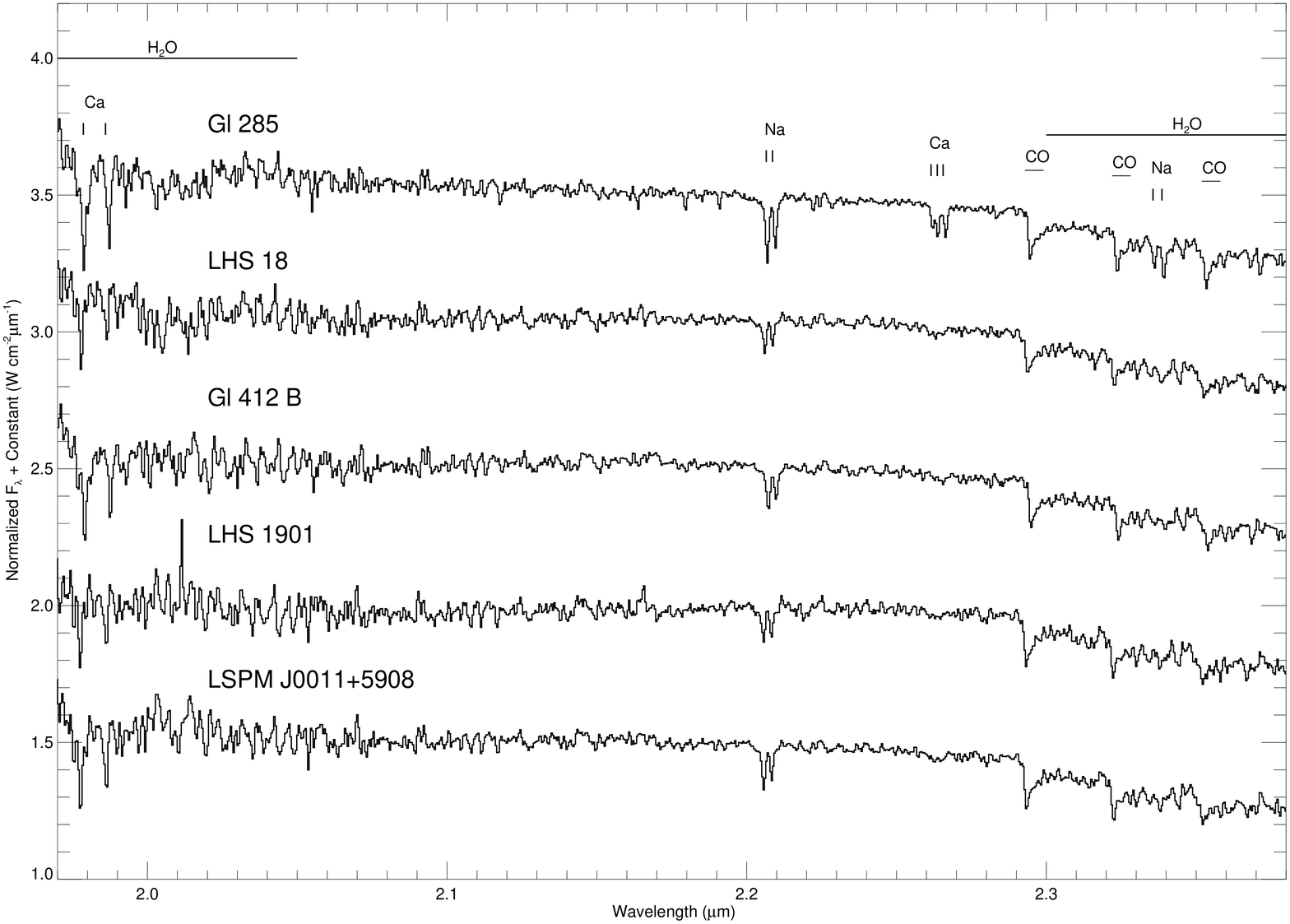}
\caption{\footnotesize{K band spectra of Gl 285, LHS 18, Gl 412 B, LHS 1901, and LSPM J0011+5908. The main spectral features are indicated.}}
\end{center}
\end{figure}

\begin{figure}
\begin{center}
\includegraphics[scale=0.6,angle=90]{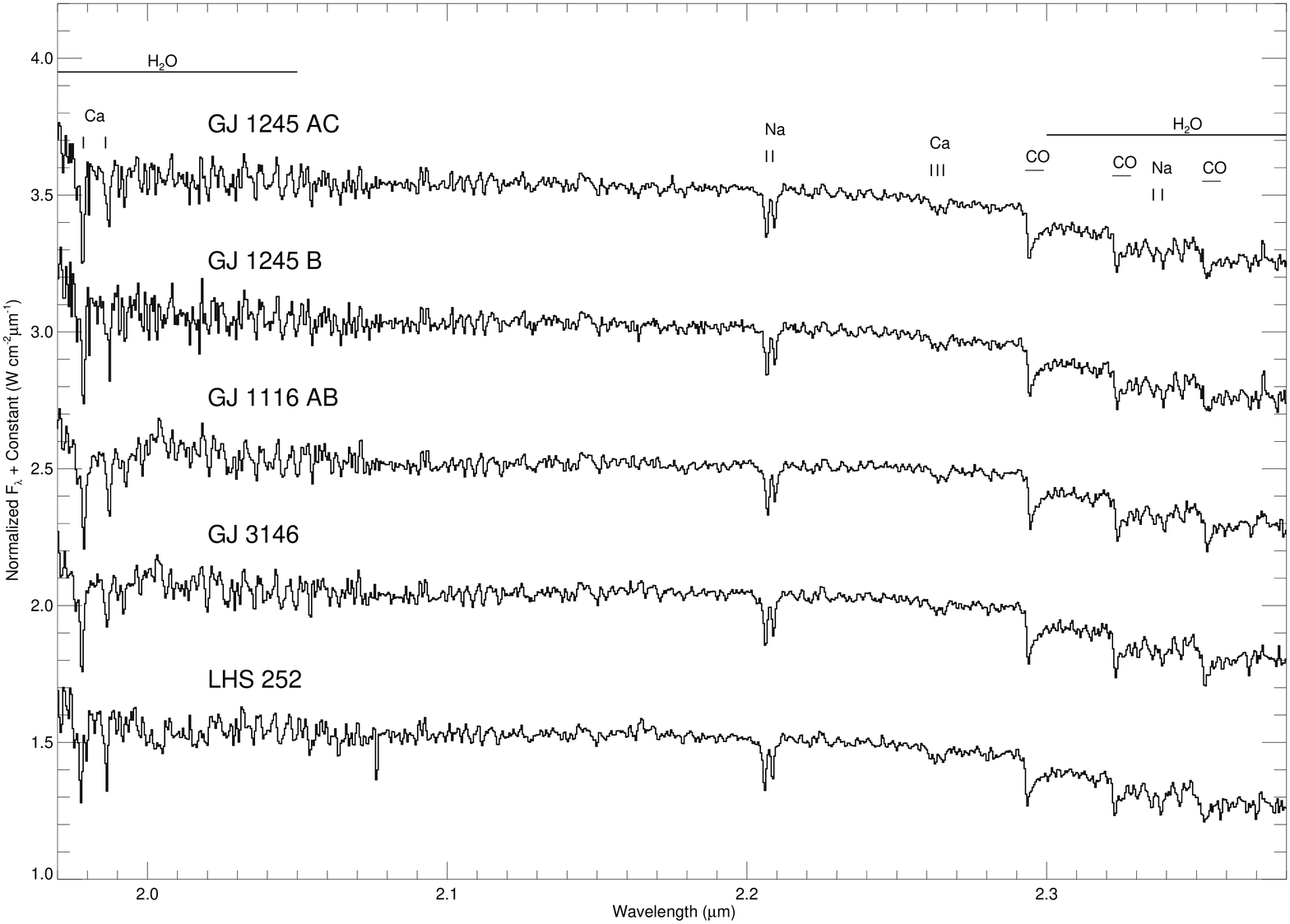}
\caption{\footnotesize{K band spectra of GJ 1245 AC, GJ 1245 B, Gl 1116 AB, GJ 3146, and LHS 252. The main spectral features are indicated.}}
\end{center}
\end{figure}

\begin{figure}
\begin{center}
\includegraphics[scale=0.6,angle=90]{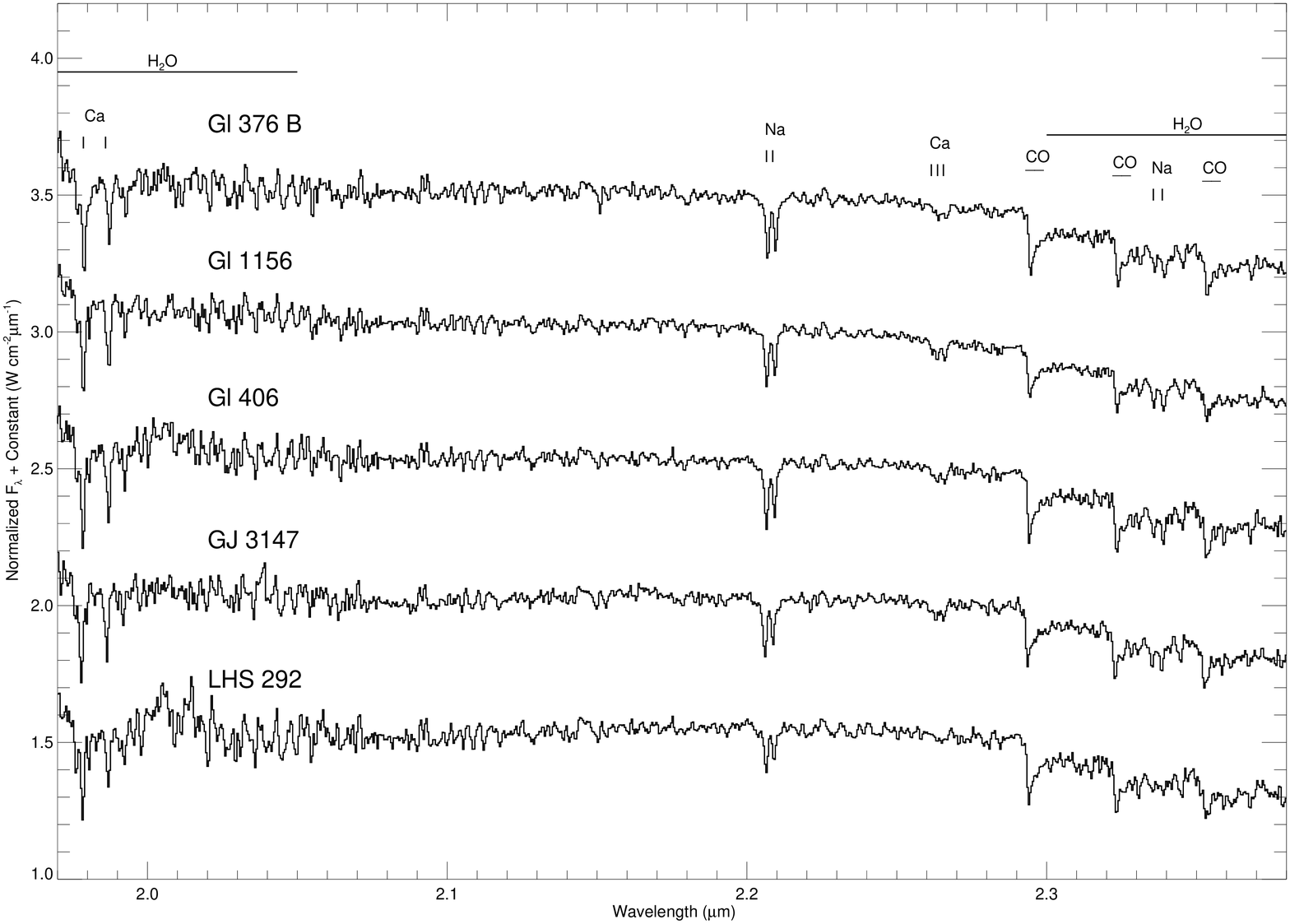}
\caption{\footnotesize{K band spectra of Gl 376 B, Gl 1156, Gl 406, GJ 3147, and LHS 292. The main spectral features are indicated.}}
\end{center}
\end{figure}

\begin{figure}
\begin{center}
\includegraphics[scale=0.6,angle=90]{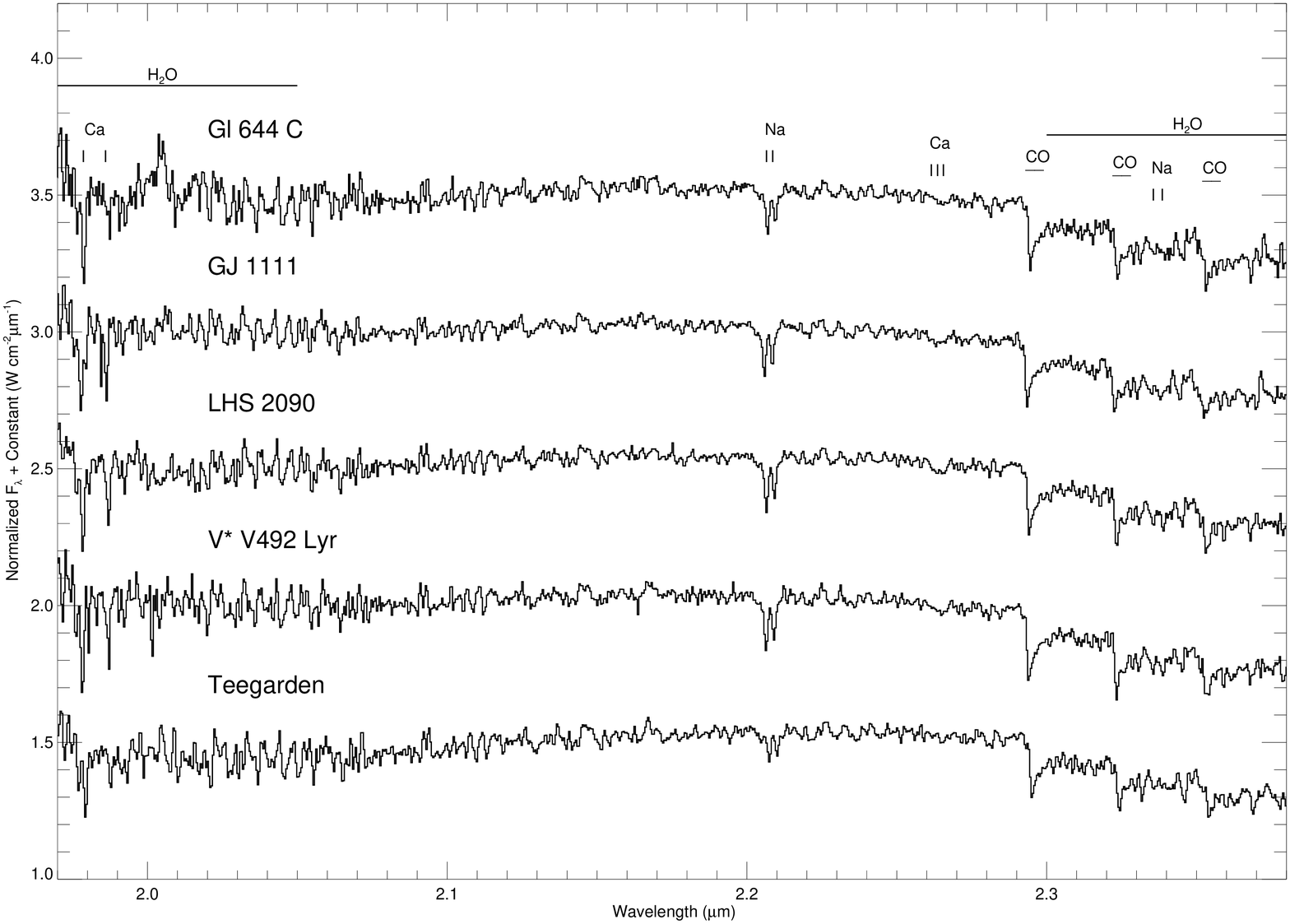}
\caption{\footnotesize{K band spectra of  Gl 644 C, GJ 1111, LHS 2090, V$*$ V492 Lyr, and Teegarden's star. The main spectral features are indicated.}}
\end{center}
\end{figure}

\clearpage

\clearpage

\begin{landscape}
\begin{deluxetable}{lccccccc}
\tablewidth{0pt}
\tabletypesize{\footnotesize}
\tablecaption{TripleSpec Nearby M dwarf Sample -- Source Properties}
\tablehead{
\colhead{Name} & 
\colhead{d (pc)} & 
\colhead{V (mag)} & 
\colhead{V Ref.} & 
\colhead{2MASS K}  & 
\colhead{Date} & 
\colhead{SNR} &
\colhead{Star} }
\startdata


LHS 3576	&	19.9	&	10.29	&	1	&	6.553	&	2009-07-20	&	278	&	LSPM	\\
Gl 338 A	&	6.2	&	7.64	&	1	&	3.99	&	2009-02-05	&	243	&	8pc	\\
Gl 338 B	&	6.3	&	7.7	&	1	&	4.14	&	2009-02-05	&	350	&	8pc	\\
G 210-45	&	23.4	&	11.21	&	1	&	7.251	&	2009-07-20	&	290	&	LSPM	\\
Gl 205	&	5.7	&	7.97	&	1	&	4.039	&	2009-09-30	&	652	&	8pc	\\
Gl 725 A	&	3.6	&	8.94	&	1	&	4.432	&	2009-07-19	&	866	&	8pc	\\
Gl 412 A	&	4.8	&	8.82	&	1	&	4.769	&	2009-02-05	&	337	&	8pc	\\
Gl 686	&	8.1	&	9.62	&	1	&	5.572	&	2009-07-19	&	702	&	LSPM	\\
Gl 752 AB	&	5.9	&	9.12	&	1	&	4.673	&	2009-07-19	&	416	&	8pc	\\
Gl 649	&	10.3	&	9.7	&	1	&	5.624	&	2010-05-22	&	862	&	planet$/$LSPM	\\
HIP 12961	&	23.8	&	10.25	&	1	&	6.736	&	2010-11-25	&	268	&	planet$/$LSPM	\\
Gl 212	&	12.5	&	9.78	&	1	&	5.759	&	2009-09-30	&	627	&	calibrator$/$LSPM	\\
HD 46375 B	&	26.4	&	11.8	&	3	&	7.843	&	2009-02-04	&	795	&	calibrator$/$LSPM	\\
HIP 79431	&	14.9	&	11.34	&	1	&	6.589	&	2010-05-22	&	659	&	planet$/$LSPM	\\
V* V1513 Cyg	&	\nodata	&	11.98	&	5	&	8.113	&	2009-07-20	&	864	&	LSPM	\\
Gl 411	&	2.5	&	7.49	&	1	&	3.254	&	2009-02-05	&	444	&	8pc	\\
Gl 526	&	5.4	&	8.46	&	1	&	4.415	&	2010-05-22	&	1996	&	8pc	\\
V* V547 Cas	&	10.1	&	10.27	&	1	&	6.037	&	2007-11-17	&	385	&	LSPM	\\
Gl 872 B	&	16.3	&	11.7	&	3	&	7.3	&	2009-07-20	&	356	&	calibrator$/$LSPM	\\
Gl 797 B	&	20.1	&	11.88	&	8	&	7.416	&	2009-07-19	&	581	&	calibrator$/$LSPM	\\
LHS 3577	&	12.5	&	10.79	&	1	&	6.533	&	2009-07-20	&	311	&	LSPM	\\
Gl 581	&	6.3	&	10.57	&	1	&	5.837	&	2009-02-05	&	359	&	planet$/$8pc	\\
Gl 408	&	6.6	&	10.03	&	1	&	5.503	&	2009-12-24	&	2079	&	8pc	\\
Gl 251	&	5.5	&	9.89	&	1	&	5.275	&	2009-02-04	&	422	&	8pc	\\
Gl 297.2 B	&	12.5	&	11.8	&	3	&	7.418	&	2008-02-16	&	260	&	LSPM	\\
Gl 250 B	&	8.7	&	10.05	&	3	&	5.723	&	2009-02-04	&	681	&	calibrator$/$LSPM	\\
Gl 176	&	9.4	&	9.95	&	1	&	5.607	&	2009-09-30	&	667	&	planet$/$LSPM	\\
NLTT 14186	&	34.5	&	14.60	&	8	&	7.621	&	2009-09-30	&	630	&	calibrator$/$LSPM	\\
Gl 849	&	8.8	&	10.41	&	1	&	5.594	&	2009-07-20	&	755	&	planet$/$LSPM	\\
Gl 725 B	&	3.5	&	9.7	&	1	&	5	&	2009-07-19	&	832	&	8pc	\\
Gl 661 AB	&	6.1	&	9.44	&	3	&	4.83	&	2009-07-19	&	708	&	8pc	\\
G 262-29	&	32.8	&	11.7	&	2	&	7.61	&	2009-07-20	&	425	&	LSPM	\\
LHS 3605	&	13.7	&	11.98	&	1	&	7.64	&	2009-07-20	&	307	&	LSPM	\\
LHS 115	&	10.2	&	12.19	&	3	&	6.377	&	2007-11-17	&	326	&	LSPM	\\
Gl 625	&	6.6	&	10.13	&	1	&	5.833	&	2009-07-19	&	573	&	8pc	\\
Gl 643	&	6.5	&	11.73	&	1	&	5.756	&	2009-07-19	&	530	&	8pc	\\
Gl 273	&	3.8	&	9.84	&	1	&	4.857	&	2009-02-04	&	327	&	8pc	\\
LHS 3591	&	32.4	&	12.73	&	2	&	8.238	&	2009-07-20	&	288	&	LSPM	\\
Gl 860 AB	&	4	&	9.59	&	1	&	4.777	&	2009-07-20	&	304	&	8pc	\\
Gl 687	&	4.5	&	9.15	&	1	&	4.548	&	2010-05-24	&	984	&	8pc	\\
Gl 628	&	4.3	&	10.1	&	1	&	5.075	&	2010-05-24	&	808	&	8pc	\\
Gl 873	&	5	&	10.29	&	1	&	5.299	&	2009-07-20	&	392	&	8pc	\\
LHS 3558	&	8	&	10.54	&	1	&	5.933	&	2009-07-20	&	628	&	8pc	\\
G 168-24	&	16.3	&	12.51	&	2	&	7.873	&	2009-07-20	&	752	&	LSPM	\\
HD 222582 B	&	41.9	&	14.50	&	7	&	9.583	&	2009-07-20	&	376	&	calibrator$/$8pc	\\
Gl 436	&	10.2	&	10.67	&	1	&	6.073	&	2008-02-17	&	997	&	planet$/$LSPM	\\
HIP 57050	&	11	&	11.86	&	1	&	6.822	&	2010-05-22	&	537	&	planet$/$LSPM	\\
LP 816-60	&	5.5	&	11.41	&	1	&	6.199	&	2009-07-19	&	899	&	8pc	\\
Gl 896 A	&	6.2	&	10.05	&	1	&	5.326	&	2009-07-20	&	923	&	8pc	\\
Gl 876	&	4.7	&	10.16	&	1	&	5.01	&	2009-07-20	&	577	&	planet$/$8pc	\\
Gl 402	&	5.6	&	11.64	&	1	&	6.371	&	2009-12-24	&	1452	&	8pc	\\
Gl 53,1 B	&	20.7	&	13.6	&	3	&	8.673	&	2007-11-16	&	508	&	LSPM	\\
Gl 555	&	6.1	&	11.32	&	1	&	5.939	&	2010-05-22	&	716	&	8pc	\\
Gl 179	&	12.1	&	11.94	&	1	&	6.942	&	2010-11-25	&	278	&	planet$/$LSPM	\\
LHS 494	&	15.9	&	12.51	&	2	&	7.397	&	2009-07-20	&	491	&	LSPM	\\
Gl 388	&	4.9	&	9.4	&	2	&	4.593	&	2009-02-05	&	260	&	8pc	\\
Gl 169.1 A	&	5.5	&	11.06	&	2	&	5.717	&	2009-09-30	&	1508	&	8pc	\\
LHS 3409	&	20.3	&	15.11	&	2	&	10.229	&	2010-05-24	&	280	&	LSPM	\\
Gl 699	&	1.8	&	9.54	&	1	&	4.524	&	2010-05-24	&	758	&	8pc	\\
LHS 220	&	13.3	&	13.77	&	2	&	8.773	&	2007-11-17	&	302	&	LSPM	\\
Gl 783.2 B	&	20.4	&	13.94	&	3	&	8.883	&	2009-07-19	&	602	&	calibrator$/$LSPM	\\
Gl 445	&	5.4	&	10.8	&	1	&	5.954	&	2009-02-05	&	201	&	8pc	\\
Gl 213	&	5.8	&	11.56	&	1	&	6.389	&	2009-02-04	&	318	&	8pc	\\
Gl 544 B	&	19	&	14.5	&	3	&	9.592	&	2008-02-16	&	811	&	calibrator$/$LSPM	\\
LHS 1723	&	7.5	&	12.16	&	3	&	6.736	&	2009-09-30	&	582	&	8pc	\\
GJ 1224	&	7.5	&	13.64	&	2	&	7.827	&	2009-07-19	&	876	&	8pc	\\
GJ 1119	&	10.3	&	13.32	&	2	&	7.741	&	2007-11-16	&	972	&	LSPM	\\
G 041-014	&	5.6	&	10.89	&	3	&	5.688	&	2009-12-24	&	857	&	8pc	\\
GJ 3348 B	&	23.5	&	13.98	&	3	&	8.791	&	2009-09-30	&	416	&	calibrator$/$LSPM	\\
LHS 1809	&	9.3	&	14.45	&	2	&	8.435	&	2007-11-17	&	746	&	LSPM	\\
Gl 447	&	3.3	&	11.12	&	1	&	5.654	&	2009-02-05	&	304	&	8pc	\\
LHS 1066	&	16.6	&	14	&	3	&	9.11	&	2007-11-17	&	668	&	LSPM	\\
GJ 3134	&	24	&	14.31	&	3	&	8.995	&	2007-11-17	&	499	&	LSPM	\\
Gl 231.1 B	&	19.9	&	13.27	&	3	&	8.267	&	2009-02-04	&	908	&	calibrator$/$LSPM	\\
LHS 3593	&	13.9	&	13.98	&	2	&	8.481	&	2009-07-20	&	228	&	LSPM	\\
GJ 3379	&	5.4	&	11.33	&	2	&	6.042	&	2009-02-04	&	662	&	8pc	\\
Gl 268 AB	&	6.4	&	11.65	&	1	&	5.846	&	2009-02-04	&	443	&	8pc	\\
Gl 768.1 B	&	19.4	&	13.1	&	3	&	8.012	&	2010-05-22	&	531	&	calibrator$/$LSPM	\\
NLTT 25869	&	11	&	14.5	&	3	&	8.64	&	2008-02-16	&	551	&	LSPM	\\
LHS 6007	&	21.3	&	14.25	&	2	&	8.852	&	2007-11-17	&	437	&	LSPM	\\
Gl 905	&	3.2	&	12.27	&	2	&	5.929	&	2009-07-20	&	824	&	8pc	\\
LHS 495	&	9.8	&	13.41	&	2	&	7.749	&	2009-07-20	&	258	&	LSPM	\\
GJ 1214	&	13	&	14.67	&	2	&	8.78	&	2010-05-24	&	415	&	planet$/$LSPM	\\
G 246-33	&	14	&	14.63	&	3	&	8.656	&	2007-11-16	&	618	&	LSPM	\\
Gl 324 B	&	12.5	&	13.15	&	3	&	7.666	&	2008-02-16	&	814	&	calibrator$/$LSPM	\\
G 203-47	&	7.3	&	11.77	&	1	&	6.485	&	2009-07-19	&	659	&	8pc	\\
Gl 299	&	6.8	&	12.83	&	3	&	7.66	&	2009-02-04	&	858	&	8pc	\\
LHS 224	&	9.2	&	13.3	&	2	&	7.776	&	2007-11-17	&	465	&	LSPM	\\
Gl 611 B	&	13.8	&	14.2	&	2	&	9.159	&	2008-02-16	&	970	&	calibrator$/$LSPM	\\
Gl 630.1 A	&	14.5	&	12.9	&	2	&	7.796	&	2008-02-17	&	483	&	LSPM	\\
NSV 13261	&	15.9	&	14.67	&	2	&	8.753	&	2009-07-20	&	407	&	LSPM	\\
NLTT 15867	&	25	&	16.49	&	3	&	10.312	&	2009-02-04	&	334	&	calibrator$/$LSPM	\\
Gl 166 C	&	5	&	11.17	&	3	&	5.962	&	2009-09-30	&	591	&	calibrator$/$8pc	\\
LHS 3376	&	7.3	&	13.46	&	2	&	7.948	&	2009-07-19	&	718	&	8pc	\\
GJ 3253	&	14.3	&	13.79	&	2	&	8.053	&	2007-11-16	&	417	&	LSPM	\\
GJ 3069	&	14.4	&	15.12	&	3	&	8.864	&	2007-11-17	&	412	&	LSPM	\\
GJ 1286	&	7.2	&	14.69	&	2	&	8.183	&	2009-07-20	&	228	&	8pc	\\
Gl 164	&	11.9	&	13.5	&	2	&	7.915	&	2008-02-17	&	495	&	LSPM	\\
Gl 777 B	&	11.7	&	14.33	&	2	&	8.712	&	2009-07-19	&	522	&	calibrator$/$LSPM	\\
Gl 866	&	3.4	&	12.18	&	3	&	5.537	&	2009-07-20	&	400	&	8pc	\\
Gl 473 AB	&	4.3	&	12.44	&	3	&	6.042	&	2009-02-05	&	413	&	8pc	\\
Gl 234 AB	&	4.1	&	11.12	&	1	&	5.486	&	2009-02-04	&	691	&	8pc	\\
LHS 3549	&	9.3	&	14.04	&	2	&	8.095	&	2009-07-20	&	442	&	LSPM	\\
LHS 1706	&	14.1	&	15.23	&	2	&	8.977	&	2009-09-30	&	372	&	LSPM	\\
V* V388 Cas	&	\nodata	&	13.78	&	5	&	7.718	&	2007-11-17	&	410	&	LSPM	\\
LHS 3799	&	7.4	&	13.25	&	2	&	7.319	&	2009-07-20	&	388	&	8pc	\\
Gl 285	&	5.9	&	11.19	&	1	&	5.698	&	2009-02-04	&	1447	&	8pc	\\
LHS 18 	&	12	&	14.66	&	2	&	9.083	&	2007-11-17	&	475	&	LSPM	\\
Gl 412 B	&	4.8	&	14.45	&	3	&	7.839	&	2009-02-05	&	367	&	8pc	\\
LHS 1901	&	8	&	15.87	&	3	&	9.126	&	2007-11-17	&	538	&	8pc	\\
LSPM J0011\tablenotemark{a}	&	11.7	&	15.87	&	3	&	9.093	&	2007-11-17	&	666	&	LSPM	\\
GJ 1245 AC	&	4.5	&	13.41	&	2	&	6.854	&	2009-07-19	&	539	&	8pc	\\
GJ 1245 B	&	4.7	&	13.99	&	3	&	7.387	&	2009-07-19	&	440	&	8pc	\\
GJ 1116 AB	&	5.2	&	14.06	&	2	&	6.889	&	2009-02-05	&	1094	&	8pc	\\
GJ 3146	&	8.5	&	15.79	&	2	&	8.981	&	2007-11-17	&	485	&	LSPM	\\
LHS 252	&	10	&	15.05	&	2	&	8.668	&	2007-11-17	&	772	&	LSPM	\\
Gl 376 B	&	13.9	&	16.13	&	3	&	9.275	&	2009-12-24	&	372	&	calibrator$/$LSPM	\\
Gl 1156	&	6.5	&	13.79	&	2	&	7.57	&	2009-02-05	&	505	&	8pc	\\
Gl 406	&	2.4	&	13.54	&	3	&	6.084	&	2009-12-24	&	1860	&	8pc	\\
GJ 3147	&	10.4	&	15.99	&	2	&	9.011	&	2007-11-16	&	507	&	LSPM	\\
LHS 292	&	4.5	&	15.6	&	2	&	7.926	&	2009-12-24	&	870	&	8pc	\\
Gl 644 C	&	6.5	&	16.7	&	3	&	8.816	&	2009-07-19	&	542	&	8pc	\\
GJ 1111	&	3.6	&	14.81	&	2	&	7.26	&	2008-02-16	&	746	&	8pc	\\
LHS 2090	&	6	&	16.1	&	3	&	8.437	&	2009-12-24	&	914	&	8pc	\\
V* V492 Lyr	&	14.1	&	18.23	&	2	&	10.308	&	2009-07-20	&	370	&	LSPM	\\
TeegardenÕs	&	3.6	&	15.4	&	3	&	7.585	&	2009-02-04	&	1092	&	8pc	\\
2MASS J18\tablenotemark{b}	&	6.2	&	18.27	&	3	&	9.171	&	2009-07-19	&	383	&	8pc	\\
LHS 2065	&	8.5	&	18.8	&	2	&	9.942	&	2008-02-16	&	264	&	LSPM	\\
LHS 2924	&	10.8	&	19.74	&	2	&	10.744	&	2008-02-17	&	560	&	LSPM	\\
Gl 908\tablenotemark{c}	&	6.0	&	8.98	&	3	&	5.043	&	2009-07-20	&	253	&	8pc	\\
Gl 809\tablenotemark{c}	&	7.0	&	8.54	&	3	&	4.618	&	2009-07-19	&	680	&	8pc	\\
Gl 644 AB	\tablenotemark{c}&	6.5	&	9.80	&	3	&	4.403	&	2009-07-19	&	587	&	8pc	\\
Gl 829 AB	\tablenotemark{c}&	6.7	&	10.30	&	3	&	5.453	&	2009-07-19	&	208	&	8pc	\\
\enddata
\tablenotetext{a}{$L$SPM J0011+5908}
\tablenotetext{b}{2MASS J18353790+3259545}
\tablenotetext{c}{Stars with low quality K band spectra}
\tablenotetext{*}{Reference of distance and V magnitude: (1)
  HIPPARCOS \citep{2007A&A...474..653V}, (2) YALE \citep{2001yCat.1238....0V}, (3) PMSU \citep{1997yCat.3198....0R}, (4) \citet{2010MNRAS.403.1949K}, (5) \citet{1992ApJS...82..351L}, (6) \citet{1990A&AS...83..357B}, (7) \citet{2004ApJS..150..455G}, (8) \citet{1991STIA...9233932G}}
\label{SampleTable}
\end{deluxetable}
\end{landscape}

\begin{deluxetable}{lccl}
\tabletypesize{\scriptsize}
\tablecolumns{4}
\tablewidth{0pt}
\tablecaption{EW Bandpasses and Continuum Points}
\tablehead{
\colhead{Feature}   & \colhead{Wavelength}  & \colhead{Integration Limits} &   \colhead{Continuum Points} \\
\colhead{} & \colhead{[$\mu$m]} & \colhead{[$\mu$m]}   & \colhead{[$\mu$m]}}   
\startdata

Na I 	&2.206 	&2.2020-2.2120	&2.1965, 2.2125,2.2175\\
	& 2.209 & & \\
&&&\\
Ca I	&2.261 &2.2580-2.2690	&2.2510, 2.2580, 2.2705, 2.2760\\	
          &2.263 &	&	\\
	&2.265 &	&  \\

\hline 
\enddata
\label{ewtable}
\end{deluxetable}

\begin{landscape}
\centering 
\begin{deluxetable}{lccccccc}
\tablecolumns{8} 
\tablewidth{0pt}
\tabletypesize{\scriptsize}
\tablecaption{TripleSpec Nearby M dwarf Sample -- Spectral Measurements}
\tablehead{
\colhead{Name}  & \colhead{EW(Na $\mathrm{I}$)} &\colhead{EW(Ca $\mathrm{I}$)} &\colhead{H$_2$O-K2} &\colhead{Sp.T.} &\colhead{T$_{eff}$}  &\colhead{[M/H]} &\colhead{[Fe/H]} }
\startdata

LHS 3576		&	4.355	$\pm$	0.107	&	3.883	$\pm$	0.095	&	1.021	$\pm$	0.004	&	0	&	3883	$\pm$	24&		-0.13	$\pm$	0.12	&	-0.19	$\pm$	0.17	\\
Gl 338 A		&	4.045	$\pm$	0.082	&	4.358	$\pm$	0.081	&	1.038	$\pm$	0.003	&	0	&	4031	$\pm$	56&		-0.13	$\pm$	0.12	&	-0.18	$\pm$	0.17	\\
Gl 338 B		&	4.023	$\pm$	0.119	&	4.475	$\pm$	0.107	&	1.019	$\pm$	0.004	&	0	&	3869	$\pm$	15&		-0.11	$\pm$	0.12	&	-0.15	$\pm$	0.17	\\
G 210-45		&	4.998	$\pm$	0.104	&	4.633	$\pm$	0.097	&	1.024	$\pm$	0.004	&	0	&	3910	$\pm$	37&		-0.03	$\pm$	0.12	&	-0.05	$\pm$	0.17	\\
Gl 205		&	7.636	$\pm$	0.043	&	6.105	$\pm$	0.043	&	1.032	$\pm$	0.002	&	0	&	4012	$\pm$	106&		0.25	$\pm$	0.12	&	0.35	$\pm$	0.17	\\
Gl 725 A		&	3.068	$\pm$	0.034	&	2.152	$\pm$	0.031	&	0.982	$\pm$	0.002	&	1	&	3680	$\pm$	18&		-0.34	$\pm$	0.12	&	-0.49	$\pm$	0.17	\\
Gl 412 A		&	3.431	$\pm$	0.091	&	2.618	$\pm$	0.085	&	0.983	$\pm$	0.003	&	1	&	3684	$\pm$	20&		-0.28	$\pm$	0.12	&	-0.40	$\pm$	0.17	\\
Gl 686		&	3.855	$\pm$	0.039	&	3.268	$\pm$	0.037	&	0.985	$\pm$	0.002	&	1	&	3693	$\pm$	20&		-0.20	$\pm$	0.12	&	-0.28	$\pm$	0.17	\\
Gl 752 AB		&	5.626	$\pm$	0.074	&	4.018	$\pm$	0.068	&	1.003	$\pm$	0.003	&	1	&	3789	$\pm$	20&		-0.03	$\pm$	0.12	&	-0.05	$\pm$	0.17	\\
Gl 649		&	4.966	$\pm$	0.087	&	4.534	$\pm$	0.033	&	0.993	$\pm$	0.002	&	1	&	3733	$\pm$	20&		-0.02	$\pm$	0.12	&	-0.04	$\pm$	0.17	\\
HIP 12961	&	5.175	$\pm$	0.105	&	4.963	$\pm$	0.105	&	1.014	$\pm$	0.004	&	1	&	3838	$\pm$	19&		0.01	$\pm$	0.12	&	0.01	$\pm$	0.17	\\
Gl 212		&	5.078	$\pm$	0.045	&	5.206	$\pm$	0.045	&	1.016	$\pm$	0.002	&	1	&	3851	$\pm$	17&		0.02	$\pm$	0.12	&	0.03	$\pm$	0.17	\\
HD 46375 B	&	7.136	$\pm$	0.035	&	5.407	$\pm$	0.034	&	0.978	$\pm$	0.002	&	1	&	3663	$\pm$	15&		0.21	$\pm$	0.12	&	0.29	$\pm$	0.17	\\
HIP 79431	&	8.948	$\pm$	0.041	&	5.436	$\pm$	0.039	&	0.984	$\pm$	0.002	&	1	&	3689	$\pm$	20&		0.33	$\pm$	0.12	&	0.46	$\pm$	0.17	\\
V* V1513 Cyg	&	1.508	$\pm$	0.061	&	1.964	$\pm$	0.059	&	0.935	$\pm$	0.002	&	2	&	3483	$\pm$	17&		-0.45	$\pm$	0.12	&	-0.64	$\pm$	0.17	\\
Gl 411		&	3.115	$\pm$	0.065	&	2.623	$\pm$	0.059	&	0.945	$\pm$	0.002	&	2	&	3526	$\pm$	18&		-0.28	$\pm$	0.12	&	-0.41	$\pm$	0.17	\\
Gl 526		&	3.646	$\pm$	0.015	&	3.207	$\pm$	0.014	&	0.973	$\pm$	0.001	&	2	&	3642	$\pm$	17&		-0.21	$\pm$	0.12	&	-0.30	$\pm$	0.17	\\
V* V547 Cas	&	3.603	$\pm$	0.075	&	3.246	$\pm$	0.068	&	0.945	$\pm$	0.003	&	2	&	3526	$\pm$	18&		-0.19	$\pm$	0.12	&	-0.28	$\pm$	0.17	\\
Gl 872 B		&	4.101	$\pm$	0.084	&	3.173	$\pm$	0.075	&	0.955	$\pm$	0.003	&	2	&	3569	$\pm$	20&		-0.17	$\pm$	0.12	&	-0.25	$\pm$	0.17	\\
Gl 797 B		&	4.298	$\pm$	0.054	&	3.182	$\pm$	0.045	&	0.955	$\pm$	0.002	&	2	&	3569	$\pm$	20&		-0.16	$\pm$	0.12	&	-0.23	$\pm$	0.17	\\
LHS 3577		&	4.704	$\pm$	0.095	&	3.919	$\pm$	0.085	&	0.968	$\pm$	0.003	&	2	&	3621	$\pm$	19&		-0.07	$\pm$	0.12	&	-0.11	$\pm$	0.17	\\
Gl 581		&	4.965	$\pm$	0.078	&	3.656	$\pm$	0.077	&	0.947	$\pm$	0.003	&	2	&	3534	$\pm$	18&		-0.06	$\pm$	0.12	&	-0.10	$\pm$	0.17	\\
Gl 408		&	4.755	$\pm$	0.015	&	3.895	$\pm$	0.013	&	0.945	$\pm$	0.001	&	2	&	3526	$\pm$	18&		-0.06	$\pm$	0.12	&	-0.09	$\pm$	0.17	\\
Gl 251		&	4.935	$\pm$	0.047	&	3.862	$\pm$	0.053	&	0.939	$\pm$	0.002	&	2	&	3500	$\pm$	17&		-0.04	$\pm$	0.12	&	-0.07	$\pm$	0.17	\\
Gl 297.2 B	&	5.154	$\pm$	0.107	&	4.055	$\pm$	0.125	&	0.959	$\pm$	0.003	&	2	&	3585	$\pm$	20&		-0.03	$\pm$	0.12	&	-0.04	$\pm$	0.17	\\
Gl 250 B		&	5.124	$\pm$	0.039	&	4.427	$\pm$	0.041	&	0.955	$\pm$	0.002	&	2	&	3569	$\pm$	20&		0.01	$\pm$	0.12	&	0.01	$\pm$	0.17	\\
Gl 176		&	6.155	$\pm$	0.042	&	4.806	$\pm$	0.042	&	0.958	$\pm$	0.002	&	2	&	3581	$\pm$	20&		0.11	$\pm$	0.12	&	0.15	$\pm$	0.17	\\
NLTT 14186	&	6.259	$\pm$	0.047	&	5.302	$\pm$	0.041	&	0.946	$\pm$	0.002	&	2	&	3530	$\pm$	18&		0.17	$\pm$	0.12	&	0.24	$\pm$	0.17	\\
Gl 849		&	7.198	$\pm$	0.055	&	5.382	$\pm$	0.049	&	0.963	$\pm$	0.002	&	2	&	3601	$\pm$	19&		0.23	$\pm$	0.12	&	0.31	$\pm$	0.17	\\
Gl 725 B		&	3.594	$\pm$	0.035	&	2.308	$\pm$	0.033	&	0.895	$\pm$	0.002	&	3	&	3288	$\pm$	27&		-0.25	$\pm$	0.12	&	-0.36	$\pm$	0.17	\\
Gl 661 AB		&	3.577	$\pm$	0.039	&	2.714	$\pm$	0.035	&	0.892	$\pm$	0.002	&	3	&	3272	$\pm$	28&		-0.21	$\pm$	0.12	&	-0.31	$\pm$	0.17	\\
G 262-29		&	3.474	$\pm$	0.072	&	3.056	$\pm$	0.065	&	0.929	$\pm$	0.002	&	3	&	3455	$\pm$	16&		-0.21	$\pm$	0.12	&	-0.30	$\pm$	0.17	\\
LHS 3605		&	3.946	$\pm$	0.098	&	2.879	$\pm$	0.086	&	0.925	$\pm$	0.003	&	3	&	3437	$\pm$	16&		-0.19	$\pm$	0.12	&	-0.28	$\pm$	0.17	\\
LHS 115		&	4.101	$\pm$	0.085	&	2.683	$\pm$	0.078	&	0.905	$\pm$	0.004	&	3	&	3339	$\pm$	23&		-0.19	$\pm$	0.12	&	-0.27	$\pm$	0.17	\\
Gl 625		&	4.112	$\pm$	0.049	&	3.144	$\pm$	0.046	&	0.931	$\pm$	0.002	&	3	&	3464	$\pm$	17&		-0.16	$\pm$	0.12	&	-0.23	$\pm$	0.17	\\
Gl 643		&	4.703	$\pm$	0.055	&	2.617	$\pm$	0.052	&	0.912	$\pm$	0.002	&	3	&	3376	$\pm$	20&		-0.15	$\pm$	0.12	&	-0.22	$\pm$	0.17	\\
Gl 273		&	4.667	$\pm$	0.085	&	2.915	$\pm$	0.104	&	0.896	$\pm$	0.003	&	3	&	3293	$\pm$	26&		-0.12	$\pm$	0.12	&	-0.17	$\pm$	0.17	\\
LHS 3591		&	4.777	$\pm$	0.104	&	3.416	$\pm$	0.092	&	0.923	$\pm$	0.003	&	3	&	3428	$\pm$	16&		-0.08	$\pm$	0.12	&	-0.12	$\pm$	0.17	\\
Gl 860 AB		&	5.075	$\pm$	0.097	&	3.348	$\pm$	0.088	&	0.931	$\pm$	0.003	&	3	&	3464	$\pm$	17&		-0.07	$\pm$	0.12	&	-0.11	$\pm$	0.17	\\
Gl 687		&	4.643	$\pm$	0.057	&	3.712	$\pm$	0.027	&	0.916	$\pm$	0.002	&	3	&	3395	$\pm$	18&		-0.06	$\pm$	0.12	&	-0.09	$\pm$	0.17	\\
Gl 628		&	5.548	$\pm$	0.036	&	3.508	$\pm$	0.034	&	0.913	$\pm$	0.002	&	3	&	3380	$\pm$	20&		-0.01	$\pm$	0.12	&	-0.02	$\pm$	0.17	\\
Gl 873		&	5.573	$\pm$	0.072	&	3.642	$\pm$	0.068	&	0.917	$\pm$	0.003	&	3	&	3400	$\pm$	18&		-0.00	$\pm$	0.12	&	-0.01	$\pm$	0.17	\\
LHS 3558		&	4.927	$\pm$	0.048	&	4.312	$\pm$	0.042	&	0.925	$\pm$	0.002	&	3	&	3437	$\pm$	16&		0.01	$\pm$	0.12	&	0.01	$\pm$	0.17	\\
G 168-24		&	5.266	$\pm$	0.043	&	4.029	$\pm$	0.037	&	0.923	$\pm$	0.002	&	3	&	3428	$\pm$	16&		0.01	$\pm$	0.12	&	0.01	$\pm$	0.17	\\
HD 222582 B	&	5.622	$\pm$	0.076	&	3.607	$\pm$	0.075	&	0.895	$\pm$	0.003	&	3	&	3288	$\pm$	27&		0.02	$\pm$	0.12	&	0.02	$\pm$	0.17	\\
Gl 436		&	5.373	$\pm$	0.029	&	4.302	$\pm$	0.024	&	0.932	$\pm$	0.002	&	3	&	3469	$\pm$	17&		0.03	$\pm$	0.12	&	0.04	$\pm$	0.17	\\
HIP 57050	&	5.659	$\pm$	0.053	&	3.806	$\pm$	0.051	&	0.895	$\pm$	0.002	&	3	&	3288	$\pm$	27&		0.04	$\pm$	0.12	&	0.05	$\pm$	0.17	\\
LP 816-60	&	6.185	$\pm$	0.034	&	3.725	$\pm$	0.031	&	0.918	$\pm$	0.002	&	3	&	3405	$\pm$	17&		0.05	$\pm$	0.12	&	0.06	$\pm$	0.17	\\
Gl 896 A		&	5.861	$\pm$	0.032	&	4.515	$\pm$	0.026	&	0.904	$\pm$	0.002	&	3	&	3334	$\pm$	23&		0.11	$\pm$	0.12	&	0.15	$\pm$	0.17	\\
Gl 876		&	6.777	$\pm$	0.049	&	4.363	$\pm$	0.047	&	0.933	$\pm$	0.002	&	3	&	3473	$\pm$	17&		0.14	$\pm$	0.12	&	0.19	$\pm$	0.17	\\
Gl 402		&	7.144	$\pm$	0.021	&	3.882	$\pm$	0.017	&	0.904	$\pm$	0.001	&	3	&	3334	$\pm$	23&		0.15	$\pm$	0.12	&	0.20	$\pm$	0.17	\\
Gl 53.1 B		&	7.006	$\pm$	0.055	&	3.954	$\pm$	0.055	&	0.894	$\pm$	0.002	&	3	&	3282	$\pm$	27&		0.15	$\pm$	0.12	&	0.21	$\pm$	0.17	\\
Gl 555		&	7.051	$\pm$	0.041	&	3.992	$\pm$	0.039	&	0.895	$\pm$	0.002	&	3	&	3288	$\pm$	27&		0.16	$\pm$	0.12	&	0.22	$\pm$	0.17	\\
Gl 179		&	6.793	$\pm$	0.105	&	4.554	$\pm$	0.098	&	0.922	$\pm$	0.003	&	3	&	3424	$\pm$	16&		0.17	$\pm$	0.12	&	0.23	$\pm$	0.17	\\
LHS 494		&	6.477	$\pm$	0.061	&	4.619	$\pm$	0.054	&	0.895	$\pm$	0.002	&	3	&	3288	$\pm$	27&		0.17	$\pm$	0.12	&	0.24	$\pm$	0.17	\\
Gl 388		&	6.523	$\pm$	0.107	&	5.105	$\pm$	0.098	&	0.915	$\pm$	0.003	&	3	&	3390	$\pm$	19&		0.20	$\pm$	0.12	&	0.28	$\pm$	0.17	\\
Gl 169.1 A	&	8.323	$\pm$	0.019	&	4.058	$\pm$	0.018	&	0.892	$\pm$	0.001	&	3	&	3272	$\pm$	28&		0.26	$\pm$	0.12	&	0.36	$\pm$	0.17	\\
LHS 3409		&	3.254	$\pm$	0.235	&	1.313	$\pm$	0.109	&	0.875	$\pm$	0.003	&	4	&	3178	$\pm$	41&		-0.36	$\pm$	0.12	&	-0.52	$\pm$	0.18	\\
Gl 699		&	3.745	$\pm$	0.043	&	1.967	$\pm$	0.037	&	0.891	$\pm$	0.002	&	4	&	3266	$\pm$	29&		-0.27	$\pm$	0.12	&	-0.39	$\pm$	0.17	\\
LHS 220		&	3.905	$\pm$	0.094	&	1.966	$\pm$	0.091	&	0.863	$\pm$	0.003	&	4	&	3093	$\pm$	70&		-0.24	$\pm$	0.12	&	-0.35	$\pm$	0.17	\\
Gl 783.2 B	&	4.032	$\pm$	0.053	&	2.365	$\pm$	0.044	&	0.875	$\pm$	0.002	&	4	&	3178	$\pm$	41&		-0.20	$\pm$	0.12	&	-0.29	$\pm$	0.17	\\
Gl 445		&	3.884	$\pm$	0.145	&	2.825	$\pm$	0.135	&	0.878	$\pm$	0.004	&	4	&	3195	$\pm$	38&		-0.17	$\pm$	0.12	&	-0.25	$\pm$	0.17	\\
Gl 213		&	4.358	$\pm$	0.088	&	2.443	$\pm$	0.086	&	0.874	$\pm$	0.003	&	4	&	3167	$\pm$	47&		-0.17	$\pm$	0.12	&	-0.25	$\pm$	0.17	\\
Gl 544 B		&	5.263	$\pm$	0.037	&	2.748	$\pm$	0.038	&	0.851	$\pm$	0.002	&	4	&	3058	$\pm$	65&		-0.05	$\pm$	0.12	&	-0.09	$\pm$	0.17	\\
LHS 1723		&	5.463	$\pm$	0.051	&	2.837	$\pm$	0.047	&	0.855	$\pm$	0.002	&	4	&	3054	$\pm$	69&		-0.03	$\pm$	0.12	&	-0.06	$\pm$	0.17	\\
GJ 1224		&	6.032	$\pm$	0.035	&	2.702	$\pm$	0.033	&	0.887	$\pm$	0.002	&	4	&	3245	$\pm$	31&		-0.03	$\pm$	0.12	&	-0.05	$\pm$	0.17	\\
GJ 1119		&	5.599	$\pm$	0.032	&	3.091	$\pm$	0.028	&	0.882	$\pm$	0.002	&	4	&	3218	$\pm$	34&		-0.02	$\pm$	0.12	&	-0.04	$\pm$	0.17	\\
G 041-014	&	5.324	$\pm$	0.033	&	3.309	$\pm$	0.031	&	0.875	$\pm$	0.002	&	4	&	3178	$\pm$	41&		-0.02	$\pm$	0.12	&	-0.03	$\pm$	0.17	\\
GJ 3348 B	&	5.438	$\pm$	0.073	&	3.323	$\pm$	0.064	&	0.877	$\pm$	0.002	&	4	&	3189	$\pm$	39&		-0.01	$\pm$	0.12	&	-0.02	$\pm$	0.17	\\
LHS 1809		&	6.476	$\pm$	0.037	&	2.359	$\pm$	0.036	&	0.855	$\pm$	0.002	&	4	&	3054	$\pm$	69&		-0.00	$\pm$	0.12	&	-0.01	$\pm$	0.17	\\
Gl 447		&	5.655	$\pm$	0.092	&	3.006	$\pm$	0.088	&	0.853	$\pm$	0.003	&	4	&	3065	$\pm$	69&		-0.00	$\pm$	0.12	&	-0.01	$\pm$	0.17	\\
LHS 1066		&	5.459	$\pm$	0.049	&	3.244	$\pm$	0.048	&	0.851	$\pm$	0.002	&	4	&	3058	$\pm$	65&		0.01	$\pm$	0.12	&	0.00	$\pm$	0.17	\\
GJ 3134		&	5.562	$\pm$	0.058	&	3.274	$\pm$	0.055	&	0.861	$\pm$	0.002	&	4	&	3080	$\pm$	74&		0.01	$\pm$	0.12	&	0.01	$\pm$	0.17	\\
Gl 231.1 B	&	5.602	$\pm$	0.034	&	3.458	$\pm$	0.032	&	0.877	$\pm$	0.002	&	4	&	3189	$\pm$	39&		0.02	$\pm$	0.12	&	0.02	$\pm$	0.17	\\
LHS 3593		&	6.126	$\pm$	0.125	&	3.095	$\pm$	0.123	&	0.868	$\pm$	0.004	&	4	&	3129	$\pm$	58&		0.03	$\pm$	0.12	&	0.03	$\pm$	0.17	\\
GJ 3379		&	6.106	$\pm$	0.042	&	3.502	$\pm$	0.045	&	0.875	$\pm$	0.002	&	4	&	3178	$\pm$	41&		0.06	$\pm$	0.12	&	0.08	$\pm$	0.17	\\
Gl 268 AB		&	6.753	$\pm$	0.047	&	3.175	$\pm$	0.054	&	0.876	$\pm$	0.002	&	4	&	3184	$\pm$	40&		0.08	$\pm$	0.12	&	0.10	$\pm$	0.17	\\
Gl 768.1 B	&	5.847	$\pm$	0.056	&	4.005	$\pm$	0.051	&	0.887	$\pm$	0.002	&	4	&	3245	$\pm$	31&		0.08	$\pm$	0.12	&	0.10	$\pm$	0.17	\\
NLTT 25869	&	6.452	$\pm$	0.051	&	3.229	$\pm$	0.049	&	0.853	$\pm$	0.002	&	4	&	3065	$\pm$	69&		0.08	$\pm$	0.12	&	0.11	$\pm$	0.17	\\
LHS 6007		&	6.546	$\pm$	0.062	&	3.951	$\pm$	0.062	&	0.891	$\pm$	0.002	&	4	&	3266	$\pm$	29&		0.12	$\pm$	0.12	&	0.16	$\pm$	0.17	\\
Gl 905		&	7.796	$\pm$	0.035	&	2.732	$\pm$	0.035	&	0.851	$\pm$	0.002	&	4	&	3058	$\pm$	65&		0.14	$\pm$	0.12	&	0.19	$\pm$	0.17	\\
LHS 495		&	7.259	$\pm$	0.115	&	3.354	$\pm$	0.104	&	0.861	$\pm$	0.003	&	4	&	3080	$\pm$	74&		0.15	$\pm$	0.12	&	0.20	$\pm$	0.17	\\
GJ 1214		&	7.753	$\pm$	0.096	&	3.275	$\pm$	0.067	&	0.887	$\pm$	0.003	&	4	&	3245	$\pm$	31&		0.15	$\pm$	0.12	&	0.20	$\pm$	0.17	\\
G 246-33		&	6.468	$\pm$	0.051	&	4.232	$\pm$	0.052	&	0.861	$\pm$	0.002	&	4	&	3080	$\pm$	74&		0.17	$\pm$	0.12	&	0.24	$\pm$	0.17	\\
Gl 324 B		&	7.818	$\pm$	0.035	&	4.019	$\pm$	0.034	&	0.889	$\pm$	0.002	&	4	&	3256	$\pm$	29&		0.22	$\pm$	0.12	&	0.31	$\pm$	0.17	\\
G 203-47		&	7.978	$\pm$	0.045	&	3.969	$\pm$	0.043	&	0.882	$\pm$	0.002	&	4	&	3218	$\pm$	34&		0.24	$\pm$	0.12	&	0.33	$\pm$	0.17	\\
Gl 299		&	3.354	$\pm$	0.035	&	1.456	$\pm$	0.032	&	0.839	$\pm$	0.002	&	5	&	3021	$\pm$	49&		-0.32	$\pm$	0.12	&	-0.46	$\pm$	0.17	\\
LHS 224		&	2.905	$\pm$	0.063	&	1.762	$\pm$	0.056	&	0.826	$\pm$	0.002	&	5	&	2982	$\pm$	34&		-0.32	$\pm$	0.12	&	-0.46	$\pm$	0.17	\\
Gl 611 B		&	3.277	$\pm$	0.037	&	1.635	$\pm$	0.029	&	0.849	$\pm$	0.002	&	5	&	3051	$\pm$	61&		-0.32	$\pm$	0.12	&	-0.45	$\pm$	0.17	\\
Gl 630.1 A	&	3.295	$\pm$	0.071	&	1.978	$\pm$	0.067	&	0.829	$\pm$	0.002	&	5	&	2991	$\pm$	37&		-0.27	$\pm$	0.12	&	-0.39	$\pm$	0.17	\\
NSV 13261	&	4.375	$\pm$	0.075	&	1.707	$\pm$	0.075	&	0.812	$\pm$	0.003	&	5	&	2939	$\pm$	20&		-0.20	$\pm$	0.12	&	-0.29	$\pm$	0.17	\\
NLTT 15867	&	4.513	$\pm$	0.087	&	2.242	$\pm$	0.086	&	0.821	$\pm$	0.003	&	5	&	2967	$\pm$	28&		-0.14	$\pm$	0.12	&	-0.21	$\pm$	0.17	\\
Gl 166 C		&	4.522	$\pm$	0.046	&	2.658	$\pm$	0.049	&	0.825	$\pm$	0.002	&	5	&	2979	$\pm$	32&		-0.10	$\pm$	0.12	&	-0.15	$\pm$	0.17	\\
LHS 3376		&	5.288	$\pm$	0.042	&	2.333	$\pm$	0.039	&	0.836	$\pm$	0.002	&	5	&	3012	$\pm$	45&		-0.08	$\pm$	0.12	&	-0.12	$\pm$	0.17	\\
GJ 3253		&	5.315	$\pm$	0.068	&	2.723	$\pm$	0.065	&	0.842	$\pm$	0.003	&	5	&	3030	$\pm$	52&		-0.05	$\pm$	0.12	&	-0.07	$\pm$	0.17	\\
GJ 3069		&	5.675	$\pm$	0.069	&	2.448	$\pm$	0.065	&	0.839	$\pm$	0.003	&	5	&	3021	$\pm$	49&		-0.04	$\pm$	0.12	&	-0.07	$\pm$	0.17	\\
GJ 1286		&	6.408	$\pm$	0.126	&	2.083	$\pm$	0.129	&	0.836	$\pm$	0.004	&	5	&	3012	$\pm$	45&		-0.02	$\pm$	0.12	&	-0.04	$\pm$	0.17	\\
Gl 164		&	5.956	$\pm$	0.055	&	2.927	$\pm$	0.053	&	0.849	$\pm$	0.002	&	5	&	3051	$\pm$	61&		0.02	$\pm$	0.12	&	0.02	$\pm$	0.17	\\
Gl 777 B		&	5.857	$\pm$	0.055	&	3.005	$\pm$	0.048	&	0.839	$\pm$	0.002	&	5	&	3021	$\pm$	49&		0.03	$\pm$	0.12	&	0.03	$\pm$	0.17	\\
Gl 866		&	6.938	$\pm$	0.071	&	2.085	$\pm$	0.065	&	0.816	$\pm$	0.002	&	5	&	2952	$\pm$	23&		0.04	$\pm$	0.12	&	0.05	$\pm$	0.17	\\
Gl 473 AB		&	6.368	$\pm$	0.072	&	2.652	$\pm$	0.085	&	0.825	$\pm$	0.003	&	5	&	2979	$\pm$	32&		0.05	$\pm$	0.12	&	0.05	$\pm$	0.17	\\
Gl 234 AB		&	6.419	$\pm$	0.038	&	3.348	$\pm$	0.039	&	0.845	$\pm$	0.002	&	5	&	3039	$\pm$	56&		0.10	$\pm$	0.12	&	0.13	$\pm$	0.17	\\
LHS 3549		&	7.097	$\pm$	0.087	&	2.958	$\pm$	0.059	&	0.836	$\pm$	0.002	&	5	&	3012	$\pm$	45&		0.12	$\pm$	0.12	&	0.16	$\pm$	0.17	\\
LHS 1706		&	7.681	$\pm$	0.075	&	2.822	$\pm$	0.079	&	0.834	$\pm$	0.003	&	5	&	3006	$\pm$	43&		0.16	$\pm$	0.12	&	0.21	$\pm$	0.17	\\
V* V388 Cas	&	7.326	$\pm$	0.067	&	3.737	$\pm$	0.066	&	0.845	$\pm$	0.002	&	5	&	3039	$\pm$	56&		0.21	$\pm$	0.12	&	0.28	$\pm$	0.17	\\
LHS 3799		&	8.478	$\pm$	0.075	&	3.508	$\pm$	0.068	&	0.845	$\pm$	0.003	&	5	&	3039	$\pm$	56&		0.28	$\pm$	0.12	&	0.38	$\pm$	0.17	\\
Gl 285		&	7.343	$\pm$	0.019	&	4.545	$\pm$	0.019	&	0.847	$\pm$	0.002	&	5	&	3045	$\pm$	59&		0.29	$\pm$	0.12	&	0.40	$\pm$	0.17	\\
LHS 18 		&	3.285	$\pm$	0.095	&	1.213	$\pm$	0.071	&	0.802	$\pm$	0.002	&	6	&	2909	$\pm$	17&		-0.34	$\pm$	0.12	&	-0.48	$\pm$	0.17	\\
Gl 412 B		&	5.138	$\pm$	0.076	&	0.385	$\pm$	0.074	&	0.804	$\pm$	0.003	&	6	&	2915	$\pm$	18&		-0.27	$\pm$	0.12	&	-0.39	$\pm$	0.17	\\
LHS 1901		&	4.322	$\pm$	0.054	&	1.231	$\pm$	0.053	&	0.775	$\pm$	0.003	&	6	&	2826	$\pm$	21&		-0.23	$\pm$	0.12	&	-0.34	$\pm$	0.17	\\
LSPM J0011\tablenotemark{a}	&	5.109	$\pm$	0.049	&	1.285	$\pm$	0.051	&	0.805	$\pm$	0.003	&	6	&	2918	$\pm$	18&		-0.18	$\pm$	0.12	&	-0.27	$\pm$	0.17	\\
GJ 1245 AC	&	5.595	$\pm$	0.092	&	1.675	$\pm$	0.108	&	0.795	$\pm$	0.002	&	6	&	2890	$\pm$	19&		-0.09	$\pm$	0.12	&	-0.14	$\pm$	0.17	\\
GJ 1245 B	&	5.695	$\pm$	0.068	&	1.615	$\pm$	0.065	&	0.792	$\pm$	0.003	&	6	&	2881	$\pm$	20&		-0.09	$\pm$	0.12	&	-0.13	$\pm$	0.17	\\
GJ 1116 AB	&	5.735	$\pm$	0.028	&	1.709	$\pm$	0.025	&	0.797	$\pm$	0.002	&	6	&	2896	$\pm$	18&		-0.08	$\pm$	0.12	&	-0.12	$\pm$	0.17	\\
GJ 3146		&	6.066	$\pm$	0.062	&	1.457	$\pm$	0.058	&	0.788	$\pm$	0.002	&	6	&	2868	$\pm$	21&		-0.07	$\pm$	0.12	&	-0.11	$\pm$	0.17	\\
LHS 252		&	6.004	$\pm$	0.036	&	2.287	$\pm$	0.036	&	0.798	$\pm$	0.002	&	6	&	2898	$\pm$	17&		0.00	$\pm$	0.12	&	-0.01	$\pm$	0.17	\\
Gl 376 B		&	7.725	$\pm$	0.077	&	1.687	$\pm$	0.081	&	0.776	$\pm$	0.003	&	6	&	2829	$\pm$	21&		0.11	$\pm$	0.12	&	0.14	$\pm$	0.17	\\
Gl 1156		&	6.645	$\pm$	0.058	&	2.765	$\pm$	0.057	&	0.792	$\pm$	0.002	&	6	&	2881	$\pm$	20&		0.11	$\pm$	0.12	&	0.15	$\pm$	0.17	\\
Gl 406		&	7.994	$\pm$	0.015	&	1.977	$\pm$	0.015	&	0.794	$\pm$	0.001	&	6	&	2887	$\pm$	20&		0.14	$\pm$	0.12	&	0.18	$\pm$	0.17	\\
GJ 3147		&	7.022	$\pm$	0.058	&	2.773	$\pm$	0.054	&	0.779	$\pm$	0.002	&	6	&	2839	$\pm$	21&		0.16	$\pm$	0.12	&	0.21	$\pm$	0.17	\\
LHS 292		&	4.791	$\pm$	0.035	&	0.294	$\pm$	0.035	&	0.759	$\pm$	0.002	&	7	&	2772	$\pm$	25&		-0.28	$\pm$	0.12	&	-0.41	$\pm$	0.17	\\
Gl 644 C		&	5.117	$\pm$	0.054	&	0.571	$\pm$	0.051	&	0.745	$\pm$	0.002	&	7	&	2727	$\pm$	29&		-0.21	$\pm$	0.12	&	-0.32	$\pm$	0.17	\\
GJ 1111		&	5.661	$\pm$	0.039	&	1.039	$\pm$	0.038	&	0.758	$\pm$	0.002	&	7	&	2769	$\pm$	26&		-0.12	$\pm$	0.12	&	-0.19	$\pm$	0.17	\\
LHS 2090		&	6.242	$\pm$	0.032	&	1.405	$\pm$	0.032	&	0.758	$\pm$	0.002	&	7	&	2769	$\pm$	26&		-0.03	$\pm$	0.12	&	-0.06	$\pm$	0.17	\\
V* V492 Lyr	&	6.406	$\pm$	0.075	&	1.225	$\pm$	0.078	&	0.739	$\pm$	0.003	&	7	&	2709	$\pm$	28&		-0.02	$\pm$	0.12	&	-0.04	$\pm$	0.17	\\
Teegarden$Õ$s	&	3.429	$\pm$	0.039	&	0.269	$\pm$	0.032	&	0.713	$\pm$	0.003	&	8	&	2637	$\pm$	30&		-0.38	$\pm$	0.12	&	-0.55	$\pm$	0.17	\\
2MASS J1835\tablenotemark{b}	&	3.886	$\pm$	0.075	&	0.721	$\pm$	0.073	&	0.692	$\pm$	0.003	&	8	&	2578	$\pm$	39&		-0.27	$\pm$	0.12	&	-0.40	$\pm$	0.17	\\
LHS 2065		&	7.252	$\pm$	0.104	&	1.316	$\pm$	0.099	&	0.687	$\pm$	0.003	&	8	&	2564	$\pm$	40&		0.12	$\pm$	0.12	&	0.16	$\pm$	0.17	\\
LHS 2924		&	4.995	$\pm$	0.049	&	0.845	$\pm$	0.053	&	0.661	$\pm$	0.002	&	9	&	2492	$\pm$	43&		-0.13	$\pm$	0.12	&	-0.19	$\pm$	0.17	\\
Gl 908\tablenotemark{c}	&	2.153	$\pm$	0.115	&	2.275	$\pm$	0.113	&	1.035	$\pm$	0.004	&	0	&	3995	$\pm$	47&		-0.41	$\pm$	0.12	&	-0.59	$\pm$	0.17	\\
Gl 809\tablenotemark{c}		&	4.085	$\pm$	0.043	&	4.193	$\pm$	0.041	&	1.057	$\pm$	0.002	&	0	&	\nodata&		-0.15	$\pm$	0.12	&	-0.21	$\pm$	0.17	\\
Gl 644 AB\tablenotemark{c}		&	3.623	$\pm$	0.048	&	2.769	$\pm$	0.045	&	1.015	$\pm$	0.003	&	1	&	3845	$\pm$	19&		-0.27	$\pm$	0.12	&	-0.39	$\pm$	0.17	\\
Gl 829 AB\tablenotemark{c}		&	4.696	$\pm$	0.123	&	3.265	$\pm$	0.129	&	0.907	$\pm$	0.004	&	3	&	3349	$\pm$	22&		-0.09	$\pm$	0.12	&	-0.13	$\pm$	0.17	\\

\hline
\enddata
\tablenotetext{a}{ LSPM J0011+5908}
\tablenotetext{b}{2MASS J18353790+3259545}
\tablenotetext{c}{Stars with low quality K band spectra}
\tablecomments{}
\label{spectral_table}
\end{deluxetable}
\end{landscape}

\begin{deluxetable}{lccccc}
\tabletypesize{\scriptsize}
\tablewidth{0pt}
\tablecaption{BT-Settl H$_2$O-K2 indices}
\tablehead{
\colhead{T$_{eff}$}    & \multicolumn{5}{c}{[M/H]} \\
\colhead{} & \colhead{-1.0} & \colhead{-0.5} & \colhead{0.0} & \colhead{+0.3} & \colhead{+0.5} }   
\startdata
4400 & 1.0633894  &  1.0551468  &  1.0459396  &  1.0346350  &  \nodata \\
4300 & 1.0610328  &  1.0530623  &  1.0447932  &  1.0325842  &   \nodata \\
4200 & 1.0555427  &  1.0486027  &  1.0440688  &  1.0326834  &  \nodata \\ 
4100 & 1.0496983  &  1.0415743  &  1.0399236  &  1.0308537  &  \nodata \\
4000 & 1.0419007  &  1.0330761  &  1.0336640  &  1.0265948  &  1.0123681\\
3900 & 1.0302266  &  1.0223378  &  1.0250625  &  1.0202581  &  1.0082819\\
3800 & 1.0083129  &  1.0047255  &  1.0108148  &  1.0072626  &  0.99948109\\
3700 & 0.98369432  &  0.98394169  &  0.98987729  &  0.99077235  &  0.98483467\\
3600 & 0.95944974  &  0.95519382  &  0.96753477  &  0.96615374  &  0.96534533\\
3500 & 0.93851486  &  0.93301470  &  0.94241768  &  0.93860547  &  0.94239536\\
3400 & 0.92003132  &  0.91276339  &  0.91955948  &  0.91273634  &  0.91865987\\
3300 & 0.90412690  &  0.89518129  &  0.89743695  &  0.89064011  &  0.89750592\\
3200 & 0.88679857  &  0.87540384  &  0.87727299  &  0.86834456  &  0.88154377\\
3100 & 0.87051604  &  0.85641549  &  0.85691544  &  0.84896337  &  0.86984104\\
3000 & 0.85463719  &  0.83051899  &  0.83022602  &  0.82209357  &  0.86153062\\
2900 & 0.84301951  &  0.87462956  &  0.80480058  &  0.79553348  &  0.79291515\\
2800 & 0.83379220  &  0.86284561  &  0.77417570  &  0.76562934  &  0.76116386\\
2700 & 0.83353582  &  0.85048313  &  0.74699251  &  0.73209011  &  0.72973959\\
2600 & 0.83127154  &  0.84092836  &  0.71229056  &  0.69373495  &  0.68974967\\
2500 & 0.84359432  &  0.83413574  &  0.68145572  &  0.65795920  &  0.65185812\\
2400 & 0.83627202  &  0.82175367  &  0.64586819  &  0.61983636  &  0.61785883\\
2300 & 0.84981127  &  0.81318717  &  0.62799732  &  0.60425485  &  0.59720128\\
2200 & 0.85792452  &  0.79609664  &  0.58683221  &  0.60165298  &  0.57924869\\
\hline 
\enddata
\tablecomments{}
\label{H2O-K2Table}
\end{deluxetable}

\begin{deluxetable}{lccccc}
\tabletypesize{\scriptsize}
\tablewidth{0pt}
\tablecaption{M dwarf Metallicity Calibration Sample}
\tablehead{
\colhead{Name}    & \colhead{Sp.T.} &\multicolumn{2}{c}{SPOCS} & \multicolumn{2}{c}{Predicted}  \\
\colhead{} &\colhead{}    & \colhead{[M/H]} & \colhead{[Fe/H]} & \colhead{[M/H]} & \colhead{[Fe/H]} }   
\startdata
Gl 212		&		M1	&	0.16	&	0.19	&	0.02	&	0.03	\\
HD 46375 B	&		M1	&	0.20	&	0.24	&	0.21	&	0.29	\\
Gl 872 B		&		M2	&	-0.16	&	-0.22	&	-0.17	&	-0.25	\\
Gl 797 B		&		M2	&	-0.09	&	-0.09	&	-0.16	&	-0.23	\\
Gl 250 B		&		M2	&	-0.01	&	0.14	&	0.01	&	0.01	\\
NLTT 14186	&		M2	&	0.04	&	0.05	&	0.17	&	0.24	\\
HD 222582 B	&		M3	&	-0.02	&	-0.03	&	0.02	&	0.02	\\
Gl 324 B		&		M4	&	0.25	&	0.31	&	0.22	&	0.31	\\
Gl 768.1 B	&		M4	&	0.12	&	0.16	&	0.08	&	0.10	\\
Gl 231.1 B	&		M4	&	-0.08	&	-0.04	&	0.02	&	0.02	\\
GJ 3348 B	&		M4	&	-0.10	&	-0.22	&	-0.01	&	-0.02	\\
Gl 783.2 B	&		M4	&	-0.09	&	-0.15	&	-0.20	&	-0.29	\\
Gl 544 B		&		M4	&	-0.15	&	-0.18	&	-0.05	&	-0.09	\\
Gl 611 B		&		M5	&	-0.49	&	-0.69	&	-0.32	&	-0.45	\\
Gl 777 B		&		M5	&	0.19	&	0.21	&	0.03	&	0.03	\\
Gl 166 C		&		M5	&	-0.08	&	-0.28	&	-0.10	&	-0.15	\\
NLTT 15867	&		M5	&	-0.05	&	-0.10	&	-0.14	&	-0.21	\\
Gl 376 B		&		M6	&	0.11	&	0.20	&	0.11	&	0.14	\\

\hline 
\enddata
\tablecomments{}
\label{caltable}
\end{deluxetable}


\begin{deluxetable}{lcccccc}
\tabletypesize{\scriptsize}
\tablewidth{0pt}
\tablecaption{Statistics of the [Fe/H] prediction models}
\tablehead{
\colhead{Calibration}     & \colhead{p}  &\colhead{N}  &\colhead{rms} &\colhead{RMSp} & \colhead{$R^2_{ap}$}  &   \colhead{Source}}
\startdata

B05		&		5	&48	&0.483$\pm$0.047	&0.037$\pm$0.006    & 0.858$\pm$0.028   &  Table 4$^a$ \\
JA09	&		1	&6	&0.082$\pm$0.016	&0.004$\pm$0.002    & 0.472$\pm$1.224   &  Table 1$^b$  \\                                           
SL10	&		1	&19	&0.144$\pm$0.017	&0.022$\pm$0.005    & 0.489$\pm$0.188   &  Table 1$^c$  \\ 
N11		&		2	&23	&0.167$\pm$0.035	&0.034$\pm$0.010    & 0.426$\pm$0.238   &  Table 2$^d$  \\ 
RA11	&		3	&18	&0.208$\pm$0.030	&0.020$\pm$0.004    & 0.670$\pm$0.161   &  Table \ref{caltable} \\

\hline 
\enddata
\tablecomments{a) Bonfils et al. (2005), b) Johnson \& Apps (2009), c) Schlaufman \& Laughlin (2010), d) Neves et al. (2011)}
\label{rms_r2a}
\end{deluxetable}


\begin{deluxetable}{lcccccc}
\tabletypesize{\scriptsize}
\tablecolumns{7}
\tablewidth{0pt}
\tablecaption{[Fe/H] Comparison Sample}
\tablehead{
\colhead{}    &  \colhead{B05} &   \colhead{JA09}   & \colhead{SL10} &   \colhead{W09}   &\colhead{RA11} & \colhead{L07}  \\
\colhead{Name} & \colhead{[Fe/H]} & \colhead{[Fe/H]}   & \colhead{[Fe/H]}    & \colhead{[Fe/H]} & \colhead{[Fe/H]} &
\colhead{Class} }   
\startdata

Gl 699	&	\nodata	&	\nodata	&	-0.76	&	\nodata	&	-0.39	&	dM\\
Gl 411	&	-0.34	&	\nodata	&	\nodata	&	-0.14	&	-0.41	&	dM\\
Gl 905	&	\nodata	&	0.17		&	0.14		&	\nodata	&	0.19		&	dM\\
Gl 447	&	\nodata	&	\nodata	&	-0.28	&	\nodata	&	-0.01	&	dM\\
Gl 725 B	&	-0.37	&	\nodata	&	-0.44	&	-0.04	&	-0.36	&	dM\\
Gl 725 A	&	-0.32	&	\nodata	&	-0.29	&	0.01		&	-0.49	&	dM\\
GJ 1111	&	\nodata	&	\nodata	&	0.25		&	\nodata	&	-0.19	&	dM\\
Gl 273	&	-0.15	&	0.07		&	-0.06	&	0.02		&	-0.17	&	dM\\
Gl 860 AB	&	-0.18	&	0.03	&	-0.10	&	-0.15	&	-0.11	&	dM\\
Gl 234 AB	&	-0.00	&	0.32	&	0.22	&	\nodata	&	0.13	&	dM\\
Gl 628	&	-0.12	&	0.12	&	-0.01	&	-0.08	&	-0.02	&	dM\\
Gl 687	&	-0.13	&	0.12	&	-0.02	&	0.04	&	-0.09	&	dM\\
LHS 292	&	\nodata	&	\nodata	&	0.20	&	\nodata	&	-0.41	&	dM\\
GJ 1245 AC	&	\nodata	&	0.20	&	0.20	&	\nodata	&	-0.14	&	dM\\
GJ 1245 B	&	\nodata	&	-0.07	&	-0.18	&	\nodata	&	-0.13	&	dM\\
Gl 876	&	0.01	&	0.38	&	0.22	&	0.02	&	0.18	&	dM\\
Gl 412 A	&	-0.52	&	\nodata	&	-0.50	&	-0.33	&	-0.40	&	dM\\
Gl 388	&	0.03	&	0.37	&	0.20	&	-0.21	&	0.28	&	dM\\
Gl 873	&	-0.10	&	0.17	&	0.03	&	-0.21	&	-0.01	&	dM\\
GJ 1116 AB	&	\nodata	&	\nodata	&	0.89	&	\nodata	&	-0.12	&	dM\\
GJ 3379	&	-0.12	&	0.08	&	-0.04	&	\nodata	&	0.08	&	dM\\
Gl 445	&	-0.30	&	\nodata	&	-0.34	&	\nodata	&	-0.25	&	dM\\
Gl 526	&	-0.26	&	-0.05	&	-0.17	&	0.06	&	-0.30	&	dM\\
LP 816-60	&	-0.17	&	-0.05	&	-0.16	&	\nodata	&	0.06	&	\nodata\\
Gl 251	&	-0.23	&	-0.03	&	-0.16	&	-0.13	&	-0.07	&	dM\\
Gl 169.1 A	&	-0.01	&	0.34	&	0.20	&	0.06	&	0.36	&	dM\\
Gl 402	&	\nodata	&	-0.06	&	-0.18	&	-0.04	&	0.20	&	dM\\
Gl 205	&	-0.10	&	0.12	&	\nodata	&	-0.13	&	0.35	&	dM\\
Gl 213	&	\nodata	&	-0.12	&	-0.23	&	\nodata	&	-0.25	&	dM\\
Gl 752 AB	&	-0.03	&	0.23	&	0.09	&	0.00	&	-0.05	&	dM\\
Gl 285	&	0.09	&	0.55	&	0.40	&	\nodata	&	0.40	&	dM\\
Gl 555	&	0.00	&	0.36	&	0.23	&	0.03	&	0.22	&	dM\\
Gl 338 A	&	-0.19	&	\nodata	&	\nodata	&	-0.42	&	-0.18	&	sdM\\
Gl 581	&	-0.25	&	-0.09	&	-0.21	&	-0.10	&	-0.10	&	dM\\
Gl 338 B	&	-0.34	&	\nodata	&	\nodata	&	-0.29	&	-0.15	&	dM\\
Gl 268 AB	&	0.17	&	0.76	&	0.65	&	\nodata	&	0.10	&	dM\\
Gl 643	&	\nodata	&	\nodata	&	-0.41	&	-0.07	&	-0.22	&	dM\\
Gl 1156	&	\nodata	&	0.07	&	-0.01	&	\nodata	&	0.15	&	dM\\
Gl 625	&	-0.48	&	\nodata	&	-0.50	&	-0.35	&	-0.23	&	sdM\\
Gl 408	&	-0.22	&	-0.02	&	-0.14	&	-0.09	&	-0.09	&	dM\\
GJ 1286	&	\nodata	&	-0.01	&	-0.10	&	\nodata	&	-0.04	&	dM\\
G 203-47	&	-0.08	&	0.19	&	0.06	&	\nodata	&	0.33	&	dM\\
LHS 3376	&	\nodata	&	\nodata	&	-0.68	&	\nodata	&	-0.12	&	dM\\
LHS 3799	&	\nodata	&	0.20	&	0.14	&	\nodata	&	0.38	&	dM\\
GJ 1224	&	\nodata	&	\nodata	&	-0.29	&	\nodata	&	-0.05	&	dM\\
LHS 3558	&	-0.18	&	0.04	&	-0.10	&	-0.14	&	0.01	&	dM\\
Gl 686	&	-0.39	&	\nodata	&	-0.33	&	-0.25	&	-0.28	&	dM\\
GJ 3146	&	\nodata	&	\nodata	&	\nodata	&	\nodata	&	-0.11	&	dM\\
LHS 2065	&	\nodata	&	\nodata	&	\nodata	&	\nodata	&	0.16	&	dM\\
Gl 849	&	0.15	&	0.53	&	0.36	&	\nodata	&	0.31	&	dM\\
LHS 224	&	\nodata	&	-0.06	&	-0.18	&	\nodata	&	-0.46	&	dM\\
LHS 1809	&	\nodata	&	-0.10	&	-0.25	&	\nodata	&	-0.01	&	dM\\
LHS 3549	&	\nodata	&	0.05	&	-0.05	&	\nodata	&	0.16	&	dM\\
Gl 176	&	-0.06	&	0.19	&	0.06	&	-0.16	&	0.15	&	dM\\
LHS 495	&	\nodata	&	0.13	&	0.03	&	\nodata	&	0.20	&	dM\\
LHS 252	&	\nodata	&	0.06	&	-0.01	&	\nodata	&	-0.01	&	dM\\
V* V547 Cas	&	-0.26	&	-0.05	&	-0.18	&	-0.18	&	-0.28	&	dM\\
Gl 436	&	-0.03	&	0.26	&	0.10	&	-0.05	&	0.04	&	dM\\
GJ 1119	&	\nodata	&	0.14	&	0.03	&	\nodata	&	-0.04	&	dM\\
Gl 649	&	-0.15	&	0.07	&	-0.05	&	-0.13	&	-0.04	&	dM\\
GJ 3147	&	\nodata	&	\nodata	&	\nodata	&	\nodata	&	0.21	&	dM\\
LHS 2924	&	\nodata	&	\nodata	&	\nodata	&	\nodata	&	-0.19	&	dM\\
HIP 57050	&	-0.02	&	0.31	&	0.15	&	0.07	&	0.05	&	dM\\
Gl 777 B	&	\nodata	&	\nodata	&	-0.39	&	\nodata	&	0.03	&	dM\\
Gl 164	&	-0.04	&	0.23	&	0.12	&	\nodata	&	0.02	&	dM\\
LHS 18	&	\nodata	&	\nodata	&	\nodata	&	\nodata	&	-0.48	&	dM\\
Gl 179	&	-0.01	&	0.32	&	0.16	&	0.02	&	0.23	&	dM\\
Gl 212	&	-0.04	&	0.18	&	0.07	&	-0.24	&	0.03	&	dM\\
LHS 3577	&	-0.26	&	-0.05	&	-0.18	&	-0.22	&	-0.11	&	dM\\
GJ 1214	&	\nodata	&	0.04	&	-0.07	&	\nodata	&	0.20	&	dM\\
LHS 220	&	\nodata	&	\nodata	&	-0.72	&	\nodata	&	-0.35	&	dM\\
LHS 3605	&	-0.51	&	\nodata	&	-0.57	&	-0.19	&	-0.28	&	dM\\
Gl 611 B	&	\nodata	&	\nodata	&	\nodata	&	\nodata	&	-0.45	&	dM\\
LHS 3593	&	\nodata	&	0.03	&	-0.08	&	\nodata	&	0.03	&	dM\\
LHS 1706	&	\nodata	&	0.23	&	\nodata	&	\nodata	&	0.21	&	dM\\
V* V492 Lyr	&	\nodata	&	\nodata	&	\nodata	&	\nodata	&	-0.04	&	dM\\
GJ 3253	&	0.05	&	0.47	&	0.38	&	\nodata	&	-0.07	&	dM\\
Gl 630.1 A	&	-0.11	&	0.15	&	0.02	&	\nodata	&	-0.39	&	dM\\
HIP 79431	&	0.19	&	0.56	&	0.39	&	\nodata	&	0.46	&	\nodata\\
V* V1513 Cyg	&	-0.89	&	\nodata	&	-1.02	&	-0.80	&	-0.64	&	sdM\\
LHS 494	&	0.08	&	0.49	&	0.32	&	\nodata	&	0.24	&	dM\\
NSV 13261	&	\nodata	&	0.31	&	0.26	&	\nodata	&	-0.29	&	dM\\
G 168-24	&	-0.29	&	\nodata	&	-0.26	&	-0.09	&	0.01	&	dM\\
LHS 3576	&	-0.13	&	\nodata	&	-0.02	&	-0.20	&	-0.19	&	dM\\
LHS 3409	&	\nodata	&	\nodata	&	\nodata	&	\nodata	&	-0.52	&	sdM\\
LHS 6007	&	-0.04	&	0.27	&	0.14	&	\nodata	&	0.16	&	dM\\
G 210-45	&	-0.16	&	0.06	&	-0.05	&	-0.12	&	-0.05	&	dM\\
HIP 12961	&	-0.20	&	\nodata	&	-0.07	&	\nodata	&	0.01	&	\nodata\\
LHS 3591	&	0.06	&	0.35	&	0.21	&	\nodata	&	-0.12	&	\nodata\\
G 262-29	&	0.16	&	0.38	&	0.28	&	-0.19	&	-0.30	&	dM\\
\hline 
\enddata
\label{comptable}
\end{deluxetable}


\begin{deluxetable}{lccccccc}
\tabletypesize{\scriptsize}
\tablecolumns{8}
\tablewidth{0pt}
\tablecaption{M dwarf Wide Binaries [Fe/H] Comparison}
\tablehead{
\colhead{}   & \colhead{}  &\multicolumn{2}{c}{This Work} &  \colhead{B05} &   \colhead{JA09}   & \colhead{SL10} &   \colhead{W09}    \\
\colhead{Name} & \colhead{Sp.Type} & \colhead{[M/H]}   & \colhead{[Fe/H]}    & \colhead{[Fe/H]} & \colhead{[Fe/H]} & \colhead{[Fe/H]} &
\colhead{[Fe/H]} }   
\startdata
GJ 1245 AC	&M6	&-0.09	&-0.14	&\nodata	&0.20	&0.20	&\nodata	\\
GJ 1245 B	&M6	&-0.09	&-0.13	&\nodata	&-0.07	&-0.18	&\nodata	\\
Diff:			&	&0.00	&0.01	&	&0.26	&0.3	&     \\
&&&&&&&\\
Gl 338 A		&M0	&-0.13	&-0.18	&-0.19	&\nodata	&\nodata	&-0.42	\\
Gl 338 B		&M0	&-0.11	&-0.15	&-0.34	&\nodata	&\nodata	&-0.29	\\
Diff:			&	&0.02	&0.03	&0.15	&	&	&0.13	\\
&&&&&&&\\
Gl 725 A		&M1	&-0.34	&-0.49	&-0.32	&\nodata	&-0.29	&0.01	\\
Gl 725 B		&M3	&-0.25	&-0.36	&-0.37	&\nodata	&-0.44	&-0.04	\\
Diff:			&	&0.09	&0.13	&0.05	&	&0.15	&0.05	\\
&&&&&&&\\
V$*$ V547 Cas	&M2	&-0.19	&-0.28	&-0.26	&-0.05	&-0.18	&-0.18	\\
LHS 115		&M3	&-0.19	&-0.27	&\nodata	&\nodata	&\nodata	&\nodata	\\
Diff:			&	&0.00	&0.01	&	&	&	&	\\
&&&&&&&\\
Gl 412 A		&M1	&-0.28	&-0.40	&-0.52	&\nodata	&-0.50	&-0.33	\\
Gl 412 B		&M6	&-0.27	&-0.39	&\nodata	&\nodata	&\nodata	&\nodata	\\
Diff:				&	&0.01	&0.01	&	&	&	&	\\
&&&&&&&\\
Gl 643		&M3	&-0.15	&-0.22	&\nodata	&\nodata	&-0.41	&-0.07	\\
Gl 644 C		&M7	&-0.21	&-0.32	&\nodata	&\nodata	&\nodata	&\nodata	\\
Diff:			&	&0.06	&0.10	&	&	&	& \\
	
\hline 
\enddata
\label{tab_binaries}
\end{deluxetable}


\begin{deluxetable}{lccccccccc}
\tabletypesize{\scriptsize}
\tablecolumns{10}
\tablewidth{0pt}
\tablecaption{Best M dwarf Targets for Planet Searches}
\tablehead{
\colhead{Name}    &  \colhead{PMSU}   & \colhead{Sp.T.} & \colhead{V} & \colhead{K$_s$} & \colhead{[M/H]} & \colhead{[Fe/H]} & \colhead{\ha\tablenotemark{1}} & \colhead{\vsini\tablenotemark{2}} &\colhead{Pop}  }   
\startdata


Gl 581		&2420	 	&M2		&10.57	&5.84	&-0.06	&-0.10   	&-0.359	&$\leq$2.1\tablenotemark{a}	&Thin\\
Gl 687		&2797	 	&M3		&9.15	&4.55	&-0.06	&-0.09	&-0.34	&$\leq$2.8\tablenotemark{a}	& Thin\\
Gl 408		&1709	 	&M2		&10.03	&5.50	&-0.06	&-0.09	&-0.307	& $\leq$2.3\tablenotemark{a}	& Thin\\
Gl 251		&1094		&M2		&9.9		&5.3		&-0.04	&-0.07	&-0.272	&$\leq$2.4\tablenotemark{a}	&Thin\\
Gl 752 AB		&3037	 	&M1		&9.12	&4.67	&-0.03	&-0.05 	&-0.42	&$<$2.5\tablenotemark{b}	&Thin/Thick\\
Gl 649		&2673		&M1		&9.7		&5.6		&-0.02	&-0.04	&-0.49	&$\leq$1.9\tablenotemark{c}	&Thin\\
Gl 628		&2599	 	&M3		&10.1	&5.08	&-0.01	&-0.02 	&-0.234	& 1.1\tablenotemark{d}	&Thin\\
Gl 447		&1849	 	&M4		&11.12	&5.65	&-0.00	&-0.01 	&-0.181	& $\leq$2\tablenotemark{e}	& Thin\\
LHS 1809		&1004		&M4		&14.5	&8.4		&-0.00	&-0.01	&-0.231	&4.3\tablenotemark{f}		&Thin\\
LHS 252		&1355	 	&M6		&15.05	&8.67	&0.00	&-0.01 	&0.312	&4\tablenotemark{f}	&Thin/Thick\\
LHS 3558		&3203		&M3		&10.4	&5.9		&0.01	&0.01	&-0.039	&$\leq$3.2\tablenotemark{a}	&Thin\\
Gl250 B		&1092		&M2		&10.1	&5.7		&0.01	&0.01	&-0.292	&$<$2.5\tablenotemark{b}	&Thin\\
Gl 212		&956		&M1		&9.9		&5.8		&0.02	&0.03	&-0.391	&$<$2.5\tablenotemark{b}	&Thin\\
Gl 436		&1830		&M3		&10.7	&6.1		&0.03	&0.04	&-0.37	&$\leq$1\tablenotemark{c}		&Thin/Thick\\
LHS 3549		&3195		&M5		&14.0	&8.1		&0.12	&0.16	&-0.131	&$\leq$4.5\tablenotemark{f}	&Thin/Thick\\
Gl 905		&3743		&M4		&12.3	&5.9		&0.14	&0.19	&0.092	&$\leq$1.2\tablenotemark{a}	&Thick\\
Gl 876		&3604		&M2		&10.2	&5.0		&0.14	&0.19	&-0.2	&$\leq$2\tablenotemark{e}		&Thin\\
Gl 402		&1687		&M3		&11.7	&6.4		&0.15	&0.20	&-0.28	&$\leq$2.3\tablenotemark{e}	&Thin\\
Gl 555		&2310		&M3		&11.3	&5.9		&0.16	&0.22	&-0.177	&2.7	\tablenotemark{e}	&Thin\\
Gl 324 B		&1383		&M4		&13.2	&7.7		&0.22	&0.31	&-0.249	&$<$2.5\tablenotemark{b}	&Thin\\
Gl 849		&3478		&M2		&10.4	&5.6		&0.23	&0.31	&-0.384	&$<$2.5\tablenotemark{b}	&Thin\\
Gl 205		&930		&M0		&8.0		&4.0		&0.25	&0.35	&-0.401	&1\tablenotemark{d}		&Thin/Thick\\
Gl 169.1 A	&780		&M3		&11.1	&5.7		&0.26	&0.36	&-0.174	&1.9\tablenotemark{e}		&Thin/Thick\\
\hline
\enddata
\tablenotetext{2}{Reference of \ha: \citet{2002AJ....123.3356G}}
\tablenotetext{2}{Reference of $v$ sin $i$: (a) \citet{1998A&A...331..581D}, (b) \citet{2010AJ....139..504B}, (c) \citet{1992ApJ...390..550M}, (d) \citet{2007A&A...467..259R}, (e) \citet{2003ApJ...583..451M}, and (f) \citet{2009ApJ...704..975J}}
\label{desirabletargets}
\end{deluxetable}

\begin{landscape}
\begin{deluxetable}{lcccccccccccc}
\tabletypesize{\scriptsize}
\tablecolumns{12}
\tablewidth{0pt}
\tablecaption{M dwarf Planet Hosts and Metallicity Calibration Sample}
\tablehead{
\colhead{}    &  
\multicolumn{3}{c}{This Work} & \colhead{KHM}   & \colhead{B05} & \colhead{JA09} & \colhead{W09} & \colhead{SL10} & \colhead{R10} & \colhead{M sin(i)} & \colhead{} \\
\cline{2-4} 
\colhead{Name} & 
\colhead{Sp. Type}    & \colhead{[M/H]} & \colhead{[Fe/H]} & \colhead{Sp. Type}  & \colhead{[Fe/H]} & \colhead{[Fe/H]} & \colhead{[Fe/H]} &\colhead{[Fe/H]} & \colhead{[Fe/H]}  &\colhead{M$_J$}   & \colhead{Planet Notes} }   
\startdata
HIP 79431 		  		&M1 		&+0.33 		&+0.46 	&M3 		&+0.16	&+0.52	&\nodata	&+0.35	&+0.60 	&2.1  	&Jupiter\\
Gl 849 			 		&M2 		&+0.23 		&+0.31 	&M3		&+0.14	&+0.58	&+0.20	&+0.41	&+0.49 	&0.82	&Jupiter\\
Gl 179			 		&M3 		&+0.17		&+0.23	&M3.5	&\nodata	&+0.30	&+0.02	&+0.20	&\nodata	&0.82	&Jupiter\\	
GJ 1214 			 		&M4 		&+0.15		&+0.20	&M4.5	&\nodata	&+0.03	&\nodata	&+0.28	&+0.39	&0.0179 	&super-Earth\\
Gl 876			 		&M3 		&+0.14		&+0.19	&M4	 	&+0.03	&+0.37	&+0.02	&+0.23	&+0.43	&2.64	&2 Jupiters + Neptune + super-Earth\\
Gl 176			 		&M2 		&+0.11		&+0.15	&M2.5	&\nodata	&+0.18	&-0.16	&+0.06	&\nodata	&0.0265	&super-Earth\\
Gl 436 			 		&M3 		&+0.04		&+0.05	&M2		&-0.03	&+0.25	&-0.05	&+0.10	&-0.00	&0.072  	&Neptune\\
HIP 57050 		 		&M3 		&+0.03		&+0.04	&M4 		&-0.02	&+0.32	&+0.07	&\nodata	&+0.12	&0.298 	&Neptune\\
HIP 12961 		 		&M1 		&+0.01		&+0.01	&M0		&\nodata	&\nodata	&\nodata	&-0.07	&\nodata	&0.35  	&Neptune\\
Gl 649 			 		&M1 		&-0.02		&-0.04	&M1 		&-0.18	&+0.04	&-0.13	&-0.03	&+0.14	&0.328 	&Neptune\\
Gl 581 			 		&M2 		&-0.06		&-0.10	&M3 		&-0.25	&-0.10	&-0.10	&-0.22	&-0.02	&0.0492  	&Neptune + 3 super-Earths\\
\hline
\enddata
\label{planettable2}
\end{deluxetable}
\end{landscape}

\end{document}